\documentclass[structabstract]{aa}
\usepackage{txfonts}
\usepackage{aas_macros}
\usepackage{natbib}
\bibpunct{(}{)}{;}{a}{}{,} % to follow the A&A style
\usepackage{version}
\begin{document}
\title{Variations on a theme -- the evolution of hydrocarbon solids:}
\subtitle{II. Optical property modelling -- the optEC$_{\rm (s)}$ model \\[0.2cm] 
{\small Combined version of the published papers: A\&A, 540, A2 (2012) -- DOI: 10.1051/0004-6361/201117624 and  \\ \hspace*{5.7cm} A\&A, 545, C2 (2012) -- DOI: 10.1051/0004-6361/201117624e\\[0.2cm] 
Data files are available at the CDS via \\[-0.1cm] 
%anonymous ftp to cdsarc.u-strasbg.fr (130.79.128.5) or via 
http://cdsarc.u-strasbg.fr/viz-bin/qcat?J/A+A/545/C2 and http://cdsarc.u-strasbg.fr/viz-bin/qcat?J/A+A/545/C3}}

\author{A.P. Jones \inst{1,2}}

    \institute{Institut d'Astrophysique Spatiale, CNRS/Universit\'e Paris Sud, UMR\,8617, Universit\'e Paris-Saclay, Orsay F-91405, France\\
               \email{ Anthony.Jones@ias.u-psud.fr}              }

    \date{Received 4 July, 2011 / Accepted 30 October 2011}

   \abstract
% context
{The properties of hydrogenated amorphous carbon (a-C:H) dust are known to evolve in response to the local conditions.}
%aims
{We present an adaptable model for the determination of the optical properties of  low-temperature, interstellar a-C:H grains that is based on the fundamental physics of their composition.}
% methods
{The imaginary part of the refractive index, $k$, for a-C:H materials, from 50\,eV to cm wavelengths, is derived and the real part, $n$, of the refractive index is then calculated using the Kramers-Kronig relations.}
% results
{The formulated optEC$_{\rm (s)}$ model allows a determination of the complex dielectric function, $\epsilon$, and refractive index, $m(n,k)$,  for a-C:H materials as a continuous function the band gap, $E_{\rm g}$, which is shown to lie in the range $\simeq -0.1$ to $2.7$\,eV. We provide expressions that enable a determination of their optical constants and tabulate $m(n,k,E_{\rm g})$ for 14 different values of $E_{\rm g}$. We explore the evolution of the likely extinction and emission behaviours of a-C:H grains and estimate the relevant transformation time-scales.}
% conclusions
{With the optEC$_{\rm (s)}$ model we are able to predict how the optical properties of an a-C:H dust component in the interstellar medium will evolve in response to, principally, the local interstellar radiation field. The evolution of a-C:H materials appears to be consistent with many dust extinction, absorption, scattering and emission properties, and also with H$_2$ molecule, daughter `PAH' and hydrocarbon molecule formation resulting from its photo-driven decomposition.}


   \maketitle

%------------------------------------------------------------------
\section{Introduction}
%------------------------------------------------------------------

Studies of low-temperature interstellar dust and the interpretation of the dust extinction and emission rely upon the availability of laboratory data, and in particular optical property data, on dust analogues formed and analysed at low temperatures. Currently, such data on the optical properties of suitable dust analogue materials, formed from the gas phase at low temperatures, are rather scarce \citep[{\it e.g.},][for silicates and hydrogenated amorphous carbons, respectively]{1989LPSC...19..565N,2005A&A...432..895D}. Conversely, most experimental analogue materials are formed via a variety of techniques far from those appropriate to interstellar media, such as: aqueous precipitation, electron or laser ablation or the quenching of high-temperature melts. There is therefore clearly a need for new optical constant data appropriate to the astrophysical domain even if, as necessity currently dictates, those models are to be considered as ``straw men'' that will later be torn down and reconstructed. 

The optical properties of interstellar dust and how its properties vary with wavelength and respond to local conditions is absolutely critical for our understanding of dust evolution in interstellar, circumstellar and Solar System media. Ideally, we need a coherent determination of the complex dielectric function, and therefore the refractive index, of low-temperature, interstellar dust analogues from at least far-UV to mm wavelengths. Currently, such data is rather fragmentary and incomplete for this suite of complex materials. In this respect the carbon component of cosmic dust presents a real challenge because its behaviour in the interstellar medium (ISM) is probably far from being completely  understood. 
There are a number of laboratory-based determinations of the optical constants for carbonaceous dust ranging over the whole gamut from graphite to (nano)diamond   \citep[{\it e.g.},][]{1984ApJ...285...89D,1989Natur.339..117L,1995ApJS..100..149M,1991ApJ...377..526R,1996MNRAS.282.1321Z}. There has also been a theoretical determination of the optical properties, in the visible-UV wavelength range, of a very wide range of amorphous carbons by \cite{2007DiamondaRM...16.1813K}. However, despite the wealth of these data, they do not coherently cover the entire range of compositional interest over the full energy range required, {\it i.e.}, from the extreme-UV (EUV) to cm wavelengths. The problem here lies in the inherently and `infinitely' complex nature of carbonaceous matter, a material that, even in a pure carbon form, is widely varying in its nature. These pure carbon solids cover materials such as graphite, diamond, glassy carbon, amorphous carbon, graphene, fullerenes and nano-tubes. The addition of only hydrogen hetero-atoms into the structure opens up a whole new world of complexity for interstellar dust studies \cite[{\it e.g.},][]{2009ASPC..414..473J,jones2011A}. 

Laboratory amorphous hydrocarbon solids (a-C:H) are known to darken upon exposure to ultraviolet (UV) light and in response to thermal annealing \citep[{\it e.g.},][]{1985OpEffinAS.120..258I,1984JAP....55..764S}. This photo-darkening of a-C:H materials leads to a decrease in the band gap or optical gap energy, $E_{\rm g}$. It is this property that lies at the heart of the inherent variability in their optical properties and that will be of prime importance in unravelling the evolutionary histories of hydrocarbon grains in the ISM \cite[{\it e.g.},][]{1996MNRAS.283..343D,2009ASPC..414..473J,jones2011A}.

In contrast to microcrystalline and glassy carbon, which are essentially metallic, the range of amorphous carbon (a-C) and hydrogenated amorphous carbon (a-C:H) materials are truly amorphous and exhibit a semiconducting band gap, $E_{\rm g}$, which renders their structures less classifiable \citep[{\it e.g.},][]{1986AdPhy..35..317R}. 

Here, and following on from \cite{jones2011A} hereafter called paper~I, we focus on the optical properties of the carbonaceous dust component, and in particular on a-C (H-poor materials, $E_{\rm g} \simeq 0.4-0.7$\,eV) and a-C:H (H-rich materials, $E_{\rm g} \simeq 1.2-2.5$\,eV), and how their properties evolve. In the following we will use the designation a-C(:H) to imply the whole family of materials but use the separate terms where strictly relevant. The key parameter in determining the inherent variations in these properties is the band gap or, as we show here, the derived Tauc gap \citep[{\it e.g.}.][]{1973AmorphSemicond,1979ElProcinNonCrystMat}. 

The energetics of the thermal processes leading to hydrogen loss and aromatisation in hydrogenated amorphous carbon (HAC) materials has been considered in detail by \cite{1996MNRAS.283..343D}. Following \citep[][and references therein]{1996MNRAS.283..343D} the principal stages in HAC thermal evolution, for temperatures $\geq$\,600\,K, can be summarised as follows:
\begin{enumerate}
  \item $600-700$\,K: a loss of H, a decrease in the C atom $sp^3/sp^2$ ratio and a closing of the band gap. 
  \item $700-800$\,K: the formation of aromatic clusters with sizes of the order of $1-2$\,nm with dangling bonds on their edges. 
  \item $> 1200$\,K: growth of the aromatic clusters, their alignment and the eventual graphitisation of the solid. 
\end{enumerate}

In paper~I we derived the compositional and spectral properties of this suite of materials and here we extend this study to the calculation of their optical properties, the complex dielectric functions, $\epsilon(\epsilon_1,\epsilon_2,E_{\rm g})$, and refractive indices, $m(n,k,E_{\rm g})$, over the EUV-cm wavelength range. As in paper~I, we refer to an amorphous hydrocarbon particle as any finite-sized,  macroscopically-structured network of atoms, {\it i.e.} a {\it contiguous}, solid-state material consisting solely of carbon and hydrogen atoms (see paper~I, and references therein, for a description of their measured and modelled properties).  We rely heavily on this wealth of literature in constructing  a model for the evolution of hydrocarbon grains in the interstellar medium (ISM).

This paper is organised as follow: in 
\S\,\ref{sect_lab_data} and \S\,\ref{opt_props} we discuss the laboratory data, and modelled data derived from it, that we use in our analysis, 
in \S\,\ref{sect_optEC_model} we present our new model, optEC$_{(s)}$ ({\bf o}ptical property {\bf p}rediction {\bf t}ool for the {\bf E}volution of {\bf C}arbonaceous {\bf (s)}olids), for the determination of the complex refractive index of a-C(:H) materials,
%\footnote{Tables of the a-C(:H) band-gap dependent optical properties $n$ and $k$ are provided in the accompanying ASCII files.}  
in \S\,\ref{sect_astro_implications} the astrophysical implications for this new model are considered, 
in \S\,\ref{sect_predictions} we summarise the model predictions, 
in \S\,\ref{sect_limitations} we discuss its limitations 
and in
\S\,\ref{sect_conclusion} we end with a concluding summary of the principal results of this work.

%------------------------------------------------------------------
\section{The measured optical properties of amorphous hydrocarbons from the optical to the UV}
\label{sect_lab_data}
%------------------------------------------------------------------

\begin{figure}
 %\resizebox{\hsize}{!}{\includegraphics{the_model/ZZ_alpha_vs_wavelength_2011.ps}}
 \resizebox{\hsize}{!}{\includegraphics{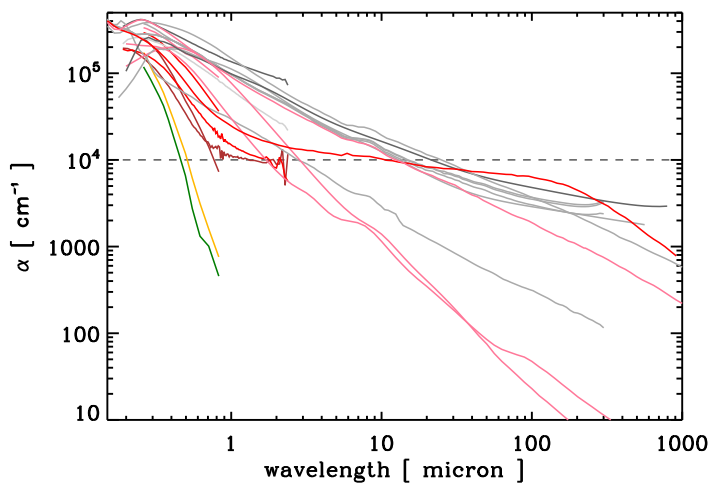}}
 \caption{Some available (hydrogenated) amorphous carbon optical data presented as the absorption coefficient, $\alpha$. The data are colour-coded by the band gap $E_{\rm g}$ (small gap - upper, grey lines; large gap - lower left, yellow and green lines ). See the text for details.}
 \label{fig_alpha}
\end{figure}

\begin{figure}
 %\resizebox{\hsize}{!}{\includegraphics{the_model/ZZ_sqrt_alphaE_vs_E_2011.ps}}
 \resizebox{\hsize}{!}{\includegraphics{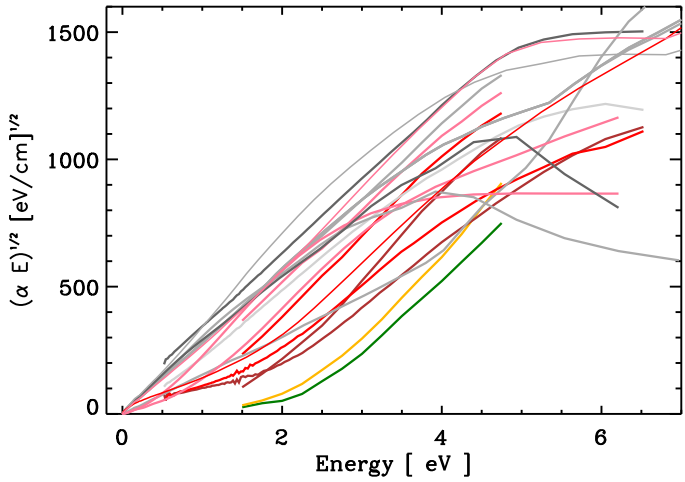}}
 \caption{The same data, and the same colour-coding scheme, as in Fig.~\ref{fig_alpha} but presented in the form of a Tauc plot, {\it i.e.}, $(\alpha E)^{0.5}$ {\it vs.} energy. The band gap increases from the upper to lower curves. See the text for details.}
 \label{fig_sqrt_alphaE}
\end{figure}

We base our study on the optical property data of (hydrogenated) amorphous carbons available from 
\cite{1984JAP....55..764S},  
\cite{1991ApJ...377..526R}, 
\cite{1995ApJS..100..149M}, 
\cite{1996MNRAS.282.1321Z} and the 
Jena group's ``Databases of Dust Optical Properties'' (DDOP).  \footnote{http://www.astro.uni-jena.de/Laboratory/Database/databases.html} These data represent experimentally-derived optical properties (in some cases the data have been extended by modelling) for a range of solid (hydro)carbon materials as a function of the preparation method and/or the material annealing temperature. We use these extensive data to constrain and compare our model-derived optical constants for amorphous (hydro)carbons with `reality'. For illustrative purposes, in Figs.~\ref{fig_alpha} and \ref{fig_sqrt_alphaE} we present all of these data in the form of the absorption coefficient, $\alpha = 4 \pi k / \lambda$, as a function of wavelength, and $(\alpha E)^{0.5}$ {\it vs.} energy Tauc plots, respectively.  In these plots the data are qualitatively colour-coded by  $E_{\rm g}$, where: grey - indicate values close to 0\,eV, red and pink - values in the range 0.5--1.5\,eV, and yellow and green - values close to 2\,eV. 

An extrapolation of experimentally-derived optical data, the linear portion of a $(\alpha E)^{1/2}$ {\it vs.} energy plot, can be used to determine the optical or Tauc gap, $E_{\rm g}$, for  a material; this is known as the ``Tauc relation''  
\citep{1966PSSBR..15..627T}. We have chosen to use as wide a range of available data as possible in order to empirically study the optical property variations of a-C(:H) materials using Tauc plots (see \S~\ref{sect_Tauc_fits} and Appendix~\ref{app_full_Tauc_analysis}). 

% FIGURE 1 *********************************************************
% *********************************************************
% FIGURE 2 *********************************************************
% *********************************************************

Perhaps the first thing to remark upon in Figs.~\ref{fig_alpha} and \ref{fig_sqrt_alphaE} is the rather large dispersion in the optical properties of hydrogenated amorphous carbons. This intrinsic variation reflects the not-too-surprising compositional complexity of a `simple' material containing only C and H atoms, which is after all the fundamental basis for life. However, and upon closer inspection, there appear to be systematic differences in the optical properties and it is these systematics that have been extensively exploited in order to understand the behaviour of these materials \citep[{e.g.},][]{1986AdPhy..35..317R}. 

In Fig.~\ref{fig_alpha} the horizontal dashed line at $\alpha = 10^4$\,cm$^{-1}$ can be used to determine the $E_{04}$ optical gap, this is the energy at which the absorption coefficient $\alpha$ has a value of $10^4$\,cm$^{-1}$. 
Note that in the Tauc plot (Fig.~\ref{fig_sqrt_alphaE}) these data indicate a linear dependence at near IR to near UV wavelengths, with the position of the linear portion depending on the material, and a rather narrow range of slopes. This reflects that fact that we are looking at a suite of related materials with the same functional, chemical building blocks (see paper~I). It is also clear from Fig.~\ref{fig_sqrt_alphaE} that an extrapolation of the linear portions of these data to the energy axis indicate values of the Tauc gap, $E_{\rm g}$, in the  range $-0.2$ to 2.5\,eV.

In \S~\ref{sect_Tauc_fits} we discuss the Tauc plot fitting and in Appendix~\ref{app_full_Tauc_analysis} we present a full Tauc analysis of these measured and calculated data and derive the optical or Tauc gaps for each material. The derived parameters for all of these data are presented in Table~\ref{table_Tauc_params} as a function of the deposition and/or annealing temperature.

%------------------------------------------------------------------
\section{The optical properties of hydrogenated amorphous carbons}
\label{opt_props}
%------------------------------------------------------------------

% FIGURE 3 *********************************************************
\begin{figure}
 \resizebox{\hsize}{!}{\includegraphics{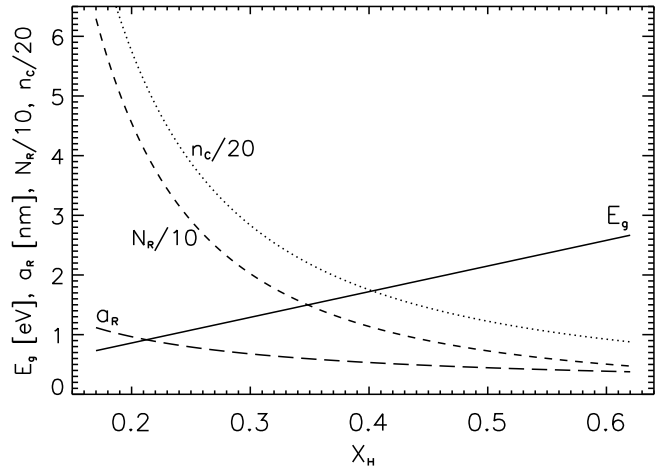}}
\caption{Band gap ($E_{\rm g}$, Eq.~\ref{Eg4XH}), number of aromatic rings per aromatic domain ($N_{\rm R}$, Eq.~\ref{eq_NR_Eg}), number of carbon atoms per aromatic domain ($n_{\rm C}$, Eq.~\ref{eq_nC_NR}) and aromatic domain radius ($a_{\rm R}$, Eq.~\ref{eq_aR}), as a function of the fractional atomic hydrogen content $X_{\rm H}$ for a-C:H using the expressions given by \cite{1988JVST....6.1778A} and \cite{1988Sci...241..913A}.}
\label{EgNRavsXH}
\end{figure}
% *********************************************************

In the hydrogen-poorer, a-C(:H) solids ({\it i.e.}, a-C) the valence and conduction band edges are due to $\pi$ states (conjugated olefinic and aromatic $\pi$-bonded structures). However, in the case of  the hydrogen-richer, a-C:H solids, as the $sp^2$ atom fraction, $X_{sp^2}$ (see paper~I), decreases the $\pi$ density-of-states decreases and the $\pi$ and $\pi^\ast$ bands become localised. For these truly amorphous materials it turns out that it is medium-range, rather than short-range, order that determines the band gap \citep{1986AdPhy..35..317R}. Laboratory studies of amorphous hydrocarbons show that there is a linear relationship between the band gap of these materials and their hydrocarbon content \citep{1990JAP....67.1007T}
\begin{equation}
E_{\rm g}({\rm eV}) \simeq 4.3 X_{\rm H},
\label{Eg4XH}
\end{equation}
and that the number of aromatic rings, $N_{\rm R}$, in an aromatic domain or cluster is a function of $E_g$  \citep{1987PhRvB..35.2946R}, {\it i.e.,} 
\begin{equation}
N_{\rm R} = \left[ \frac{5.8}{E_{\rm g}({\rm eV})} \right]^2. 
\label{eq_NR_Eg}
\end{equation}
The number of carbon atoms $n_{\rm C}$ per cluster, in terms of $N_{\rm R}$, is   
\begin{equation}
n_{\rm C} = 2 N_{\rm R} + 3.5 \surd N_{\rm R} + 0.5 
\label{eq_nC_NR}
\end{equation}
and the radius of the most compact aromatic domains is 
\begin{equation}
a_{\rm R} = 0.09 [2 N_{\rm R} + \surd N_{\rm R} + 0.5]^{0.5} \ \ {\rm nm} 
\label{eq_aR}
\end{equation}
(see paper~I for the details of these expressions). The aromatic coherence length, $L_a$, another measure of the aromatic domain size, is also a function of $E_{\rm g}$ \citep{1991PSSC..21..199R}, {\it i.e.,} 
\begin{equation}
L_{\rm a}({\rm nm}) = \left[ \frac{0.77}{E_{\rm g}({\rm eV})} \right].
\end{equation}
Fig.~\ref{EgNRavsXH} illustrates how the quantities $E_{\rm g}$, $N_{\rm R}$, $n_{\rm C}$ and $a_{\rm R}$ vary as a function of $X_{\rm H}$ as predicted by the above simple equations. Note that in Fig.~\ref{EgNRavsXH} the $N_{\rm R}$ and $n_{\rm C}$ values have been divided by 10 and 20, respectively, to fit on this plot. 

We note that for finite-sized particles the grain size itself imposes limits on the minimum possible band gap, {\it {\it i.e.}}, the aromatic cluster cannot be larger than the particle size. This effect is discussed in detail in a follow-up paper. 

\begin{figure}
 %\resizebox{\hsize}{!}{\includegraphics{the_model/alphaE.ps}}
 \resizebox{\hsize}{!}{\includegraphics{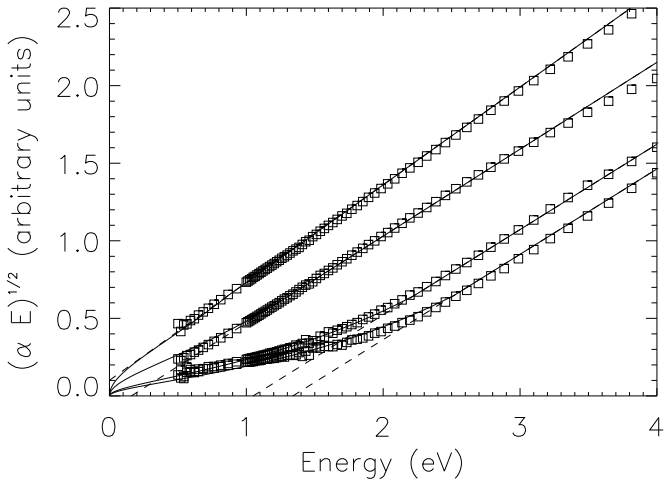}}
 \caption{A schematic Tauc plot for bulk hydrocarbon data \citep{1995ApJS..100..149M}. From bottom to top the deposition annealing temperature increases from 20 to 800\,$^\circ$C. The dashed lines show the extrapolation of the linear portions to derive the Tauc gap $E_g$.}
 \label{fig_alphaE}
\end{figure}

For disordered materials it has been shown that there is a linear dependence of the absorption coefficient, $\alpha$, of the material for photon energies, $E$, in the optical and UV range \citep{1973AmorphSemicond,1979ElProcinNonCrystMat} that is given by;
\begin{equation}
(\alpha E)^{1/2} = B^{1/2} (E-E_{\rm g}), 
\label{tauc}
\end{equation}
where $B$ is a constant that depends upon the degree to which the hydrocarbon is processed, and this expression is usually valid for $\alpha \geq 10^4$ cm$^{-1}$, corresponding to transitions across the absorption edge \citep{1988PhysAppAmorphSemicond}. In disordered materials the behaviour of the absorption becomes exponential at low energies and is referred to as the `Urbach tail' \citep{1953PhRv...92.1324U} where the energy dependence of the absorption coefficient is given by 
\begin{equation}
(\alpha E)^{1/2}  = \left[ E \alpha_0(E_1)\, {\rm exp}  \left( \frac{E-E_1}{\sigma_{\rm U}} \right) \right]^{1/2}, 
\label{urbach}
\end{equation}
which is valid in the range $10^2 \leq \alpha \leq 10^4$ cm$^{-1}$. This tail is thought to be due to structural disorder within the solid \citep{1986AdPhy..35..317R}.  Indeed, as \cite{1986AdPhy..35..317R} points out, aromatic and olefinic clusters have an energy spectrum symmetric about the Fermi level, $E_{\rm F}$, and that (less stable) clusters with an odd number of C atoms must have states at $E_{\rm F}$ and are considered as defects. Consequently, the band edges of amorphous carbons ought to be roughly symmetric about the mid-gap.  In Eq.~(\ref{urbach}) $\sigma_{\rm U}$ is a measure of the disorder and determines the width of the exponential tail and $\alpha_0(E_1)$ is a normalisation value.   The constants  $\sigma_{\rm U}$ and $E_1$ are determined empirically, $\sigma_{\rm U}$ is generally close to unity and $E_1$ can be considered as a measure of the onset of the exponential tail. Fig.~\ref{fig_alphaE} shows an example fit to data, using the linear and Urbach tail expressions, and how the extrapolation of the linear portion yields the Tauc gap (the energy-axis intercept). Note that as the annealing temperature increases that the absorption increases and the band gap diminishes.

This ``Tauc relation'' formalism has been used to investigate the fundamental wavelength dependence of interstellar extinction with some success \citep{1992MNRAS.255..243D}.

% FIGURE  4 *********************************************************
%  *********************************************************

%------------------------------------------------------------------
\subsection{Analytical fits to Tauc plots}
\label{sect_Tauc_fits}
%------------------------------------------------------------------

Eqs.~(\ref{tauc}) and (\ref{urbach}) can be used to derive analytical expressions for the absorption coefficient, and hence the imaginary part of the refractive index ($k = \alpha \lambda / 4 \pi$) as required later (see \S~\ref{sect_optEC_model}), of hydrocarbon materials over the optical and infrared wavelength range, {\it {\it i.e.}} re-arranging Eqs.~(\ref{tauc}) and (\ref{urbach}) we have,
\begin{equation}
\alpha(E) = B \frac{(E-E_{\rm g})^2}{E}, 
\label{alphaE_fit_0}
\end{equation}
\begin{equation}
\alpha(E)  = \alpha_0(E_1)\, {\rm exp} \left( \frac{E-E_1}{\sigma_{\rm U}} \right).
\label{eq_Urbach} 
\end{equation}
In order to ensure a smooth transition between the linear and exponential portions of $(\alpha E)^{1/2}$ in the Tauc plot we can combine Eqs.~(\ref{tauc}) and (\ref{urbach}) and solve for $\alpha_0(E_1)$, which then defines the transition to the exponential tail, or alternatively, the exponential tail occurs for $E < E_1$. Therefore, 
\begin{equation}
\alpha_0(E_1) = B \frac{(E_1-E_{\rm g})^2}{E_1}.
\end{equation} For the \cite{1984JAP....55..764S}, DDOP database and \cite{1996MNRAS.282.1321Z} laboratory data we find that $E_1$ can be well-matched empirically by the expression 
\begin{equation}
E_1 = E_{\rm g} + \frac{x}{\surd B} 
\end{equation}
with $x = 275, 275$ and 400 for these data sets, respectively. However, for the \cite{1995ApJS..100..149M} and \cite{1991ApJ...377..526R} laboratory data we were able to derive an analytical band gap-dependent fit to $E_1$, which can be expressed as
\begin{equation}
E_1 = E_{\rm g} + \frac{0.575}{\surd [B^\prime(E_{\rm g})]} 
\end{equation}
where the slope parameter $B^\prime(E_{\rm g})$ is given by
\begin{equation}
B^\prime(E_{\rm g}) = \frac{0.01}{E_{\rm g} + 0.25} + 0.29 
\label{alphaE_fit_5}
\end{equation}
We note that these fits to the linear portion of the Tauc plot are not particularly sensitive to the values of the adopted constants. 

Additionally, we find that very satisfactory fits to the laboratory data Urbach tails can be obtained by setting the disorder parameter $\sigma_{\rm U}$  to unity. However, a more detailed fitting indicates that allowing Urbach tail disorder parameter, $\sigma_{\rm U}$, to vary over the range $0.4-1.1$ leads to more satisfying fits (see Appendix~\ref{app_full_Tauc_analysis}).

In Fig.~\ref{fig_Smith_Tauc} we show the Tauc plots for the \cite{1984JAP....55..764S} data (lines with data points) and fits to these data (solid red lines), including an Urbach tail component. The dashed red lines in the upper figure show the extrapolated linear fit portion used to determine $E_{\rm g}$. In this figure, and for `esthetic' purposes only, we have added an `inversed-Gaussian' wing (the right hand term in Eq.~\ref{eq_gaussian_UV_tail} below) to empirically fit the UV `turnover', {\it i.e.}, 
\begin{equation}
(\alpha E)^{1/2}  = \surd B(E-E_{\rm g}) - \left( B\ E\ {\rm exp} \left[ \frac{-(E-E_{\rm UV})^2}{2 \sigma_{\rm UV}^2} \right] \, \right)^{1/2}, 
\label{eq_gaussian_UV_tail}
\end{equation}
where the left hand term is the linear contribution. Here, for the \cite{1984JAP....55..764S} data,  we take $E_{\rm UV} = 10$\,eV and $\sigma_{\rm UV}$ in the range $1.0-1.7$. In order to `fit' all of the laboratory data UV `tails' we find that we can restrict the  parameter $E_{\rm UV}$ ($\sigma_{\rm UV}$) to the range $7-10$\,eV ($0.9-1.9$). The fits to the laboratory data used here, along with a table of the relevant fit parameters, can be found in Appendix~\ref{app_full_Tauc_analysis}. 

% TABLE
\begin{table}
\caption{The optEC$_{\rm (s)}$ material band gap, $E_{\rm g}$, and hydrogen atom fraction, $X_{\rm H}$, colour coding scheme.}
\begin{center}
\begin{tabular}{llll}
                                         &                           &                 &                 \\[-0.35cm]
\hline
\hline
                                         &                          &              &                 \\[-0.35cm]
 $E_{\rm g}$  [ eV ]          &     $X_{\rm H}$  &   colour  &                 \\[0.05cm]
\hline
                                        &                           &                        &                 \\[-0.35cm]
   \hspace*{-0.3cm} $-0.1$  [$E_{g-}$]  &    0.00   &   black              &                 \\
    0.0                               &        0.00            &   dark grey        &                 \\
    0.1                               &        0.02             &   mid grey        &                 \\
    0.25                             &        0.05             &   light grey        &   $|$ typical             \\
    0.5                               &        0.11             &   pink               &   $|$   of           \\
    0.75                             &        0.17             &   red                &   $|$  a-C            \\
    1.0                               &        0.23             &   brown            &                 \\
    1.25                             &        0.29             &   orange           &   $|$              \\
    1.5                               &        0.35             &   yellow            &   $|$              \\
    1.75                             &        0.41             &   green            &   $|$  typical            \\
    2.0                               &        0.47             &   blue              &    $|$  of           \\
    2.25                             &        0.52             &   cobalt            &    $|$  a-C:H           \\
    2.5                               &        0.58             &   violet             &    $|$             \\
  2.67 [$E_{g+}$]             &        0.62             &   purple           &     $|$            \\\hline
\hline
                                        &                            &                        &                 \\[-0.25cm]
\end{tabular}
\end{center}
\label{table_colour_code}
\end{table}

The $E_{\rm g}$ colour-coded lines (with purple indicating high $E_{\rm g}$ and grey low $E_{\rm g}$ materials -- see Table~\ref{table_colour_code} for the full colour-coding scheme) in the Fig~\ref{fig_Smith_Tauc} show the full $E_{\rm g}$-dependent a-C(:H) model data presented later in \S~\ref{sect_optEC_model}.

The fits to the Tauc plots can be used to analytically-derive the optical properties for hydrocarbon materials as a function of annealing  temperature using Eqs.~(\ref{alphaE_fit_0}) to (\ref{alphaE_fit_5}), which will then allow us to determine the optical properties of any hydrogenated amorphous carbon material, from ultraviolet to visible wavelengths, as a function of $E_{\rm g}$, or equivalently using Eq.~(\ref{Eg4XH}), as a function of $X_{\rm H}$ ($\simeq E_{\rm g}/4.3$) and the annealing conditions. 

% FIGURE 5 *********************************************************
\begin{figure}
 %\resizebox{\hsize}{!}{\includegraphics{the_model/ZZ_Tauc_plots_Smith_energy_2011.ps}}
 %\resizebox{\hsize}{!}{\includegraphics{the_model/ZZ_Tauc_plots_Smith_wavelength_2011.ps}}
 \resizebox{\hsize}{!}{\includegraphics{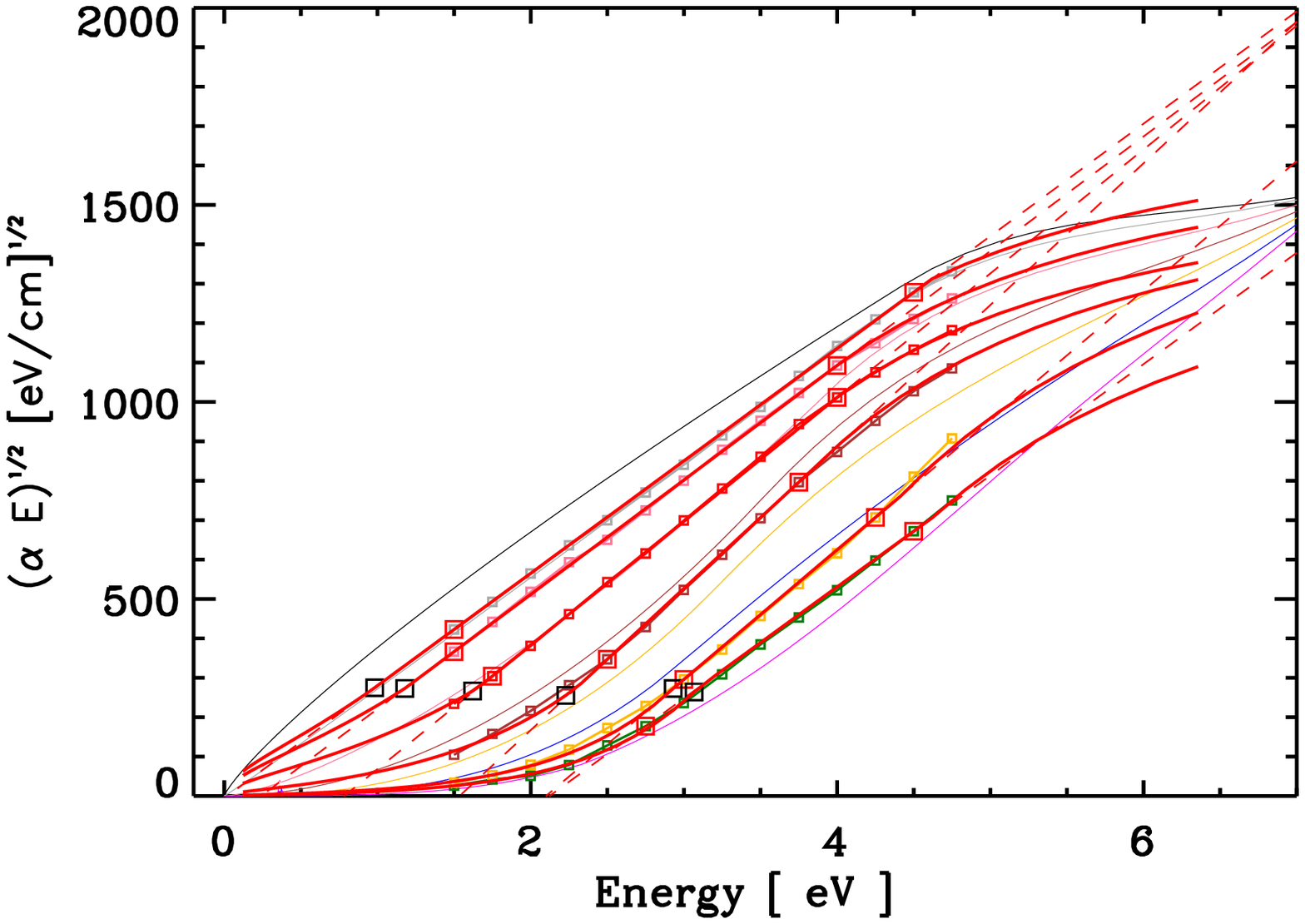}}
 \resizebox{\hsize}{!}{\includegraphics{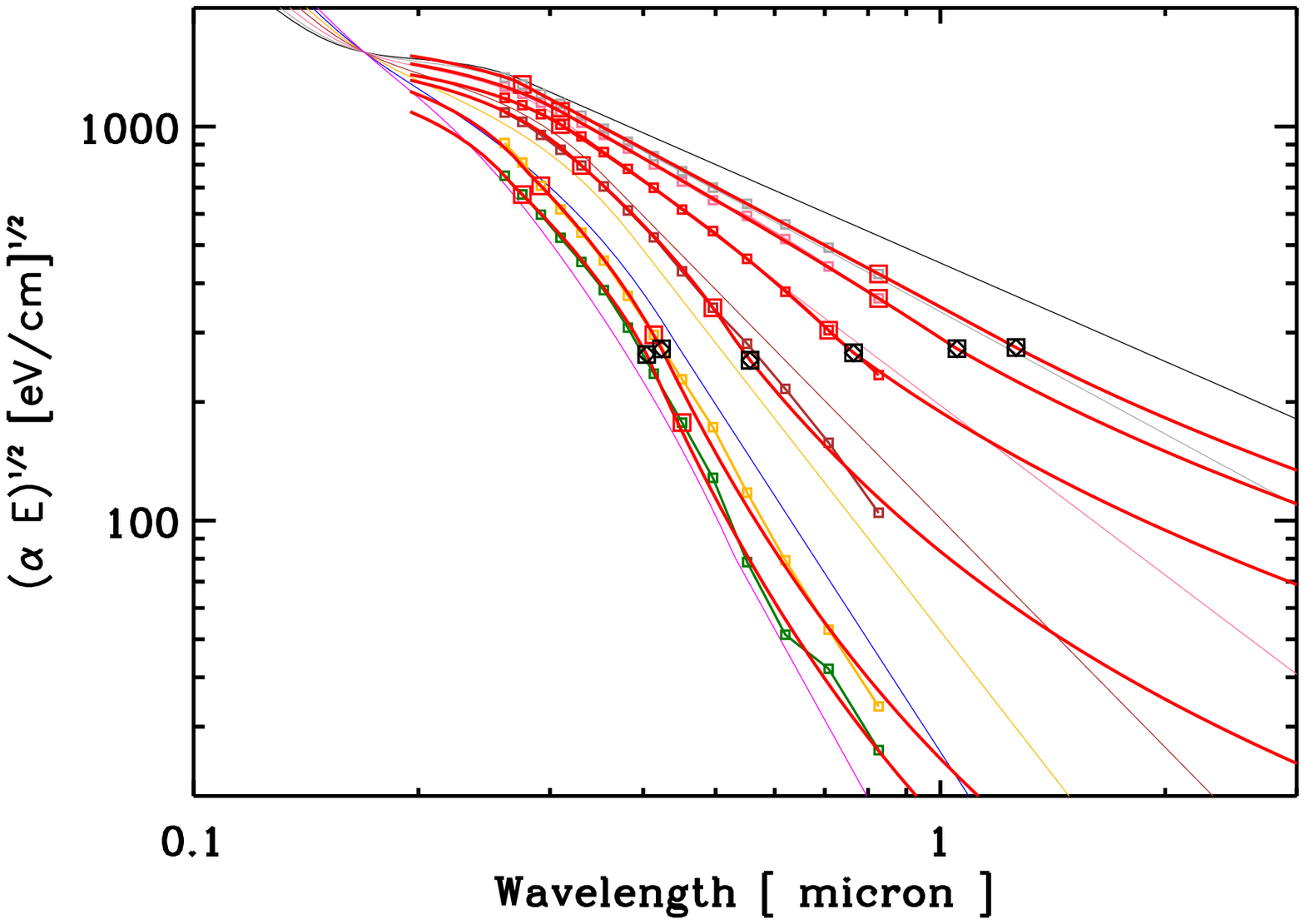}}
 \caption{The laboratory data from \cite{1984JAP....55..764S} plotted as $(\alpha E)^{0.5}$ {\it vs.} energy (upper plot), coloured curves with small data points, {\it i.e.}, a Tauc plot, and also versus wavelength (lower plot). The red lines show the modelled fits to the data using Eqs.~(\ref{alphaE_fit_0}) and (\ref{eq_Urbach}), with an `inversed-Gaussian' wing at high energies. The red squares delineate the portion used to determine $B$ and the extrapolation to the Tauc gap, $E_{\rm g}$. The black squares indicate the transition energies to the Urbach tail, $E_1$.}
 \label{fig_Smith_Tauc}
\end{figure}
% *********************************************************

%------------------------------------------------------------------
\subsection{The temperature dependence of $E_{\rm g}$}
\label{sect_T_dep_Eg}
%------------------------------------------------------------------

% FIGURE  6 *********************************************************
\begin{figure}
 %\resizebox{\hsize}{!}{\includegraphics{the_model/ZZ_Eg_vs_T_2011.ps}}
 \resizebox{\hsize}{!}{\includegraphics{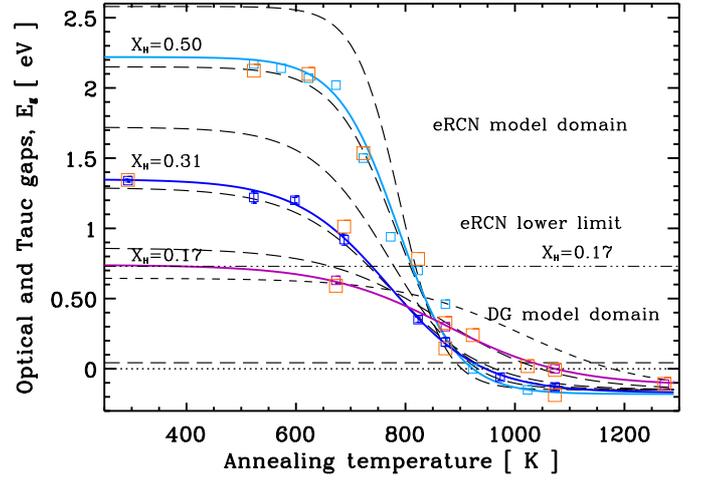}}
 \caption{The Tauc gap, $E_g$, for (hydro)carbon materials as a function of the annealing temperature.
   Light blue squares:  \cite{1984JAP....55..764S} data. Blue squares: \cite{1995ApJS..100..149M} data. Purple squares: DDOP data.  
   The orange squares show the derived Tauc gaps (see Appendix~\ref{app_full_Tauc_analysis}).
   The solid lines show the analytical fits to each data set using  Eq.~(\ref{EgT_fit}). 
   The long-dashed lines show the analytical fits for $X_{\rm H} = 0.1, 0.6\ (0.1)$, from top to bottom, and the short dashed line shows the fit for $X_{\rm H} = 0.15$, using Eq.~(\ref{EgT_fit_gen}). 
   The horizontal dashed-triple dotted line shows the lower limit to the validity of the eRCN model.}
 \label{EgvsT_2}
\end{figure}
%  *********************************************************

The thermal annealing of a-C:H at temperatures $\gtrsim 550$--600\,K leads to the loss of hydrogen from the structure and a decrease in the material band gap \cite[{\it e.g.},][]{1983SolidStComm..48..105D,1983ApPhL..42..636D,1996MNRAS.283..343D,2005A&A...432..895D}. In this section we explore and quantify this effect in terms of the relationship between the material band gap and the annealing temperature. 

The \cite{1984JAP....55..764S}, \cite{1995ApJS..100..149M} and DDOP data are particularly interesting because they can be used to derive the Tauc gap, $E_{\rm g}$,  for hydrocarbon materials as a function of the annealing temperature (see Fig.~\ref{EgvsT_2}). The optical gap for the initial hydrocarbon material is different in each case, due to the different sample preparation and deposition conditions in the experiments, and therefore the different hydrogen atom contents, $X_{\rm H}$.  For the \cite{1984JAP....55..764S} (\cite{1995ApJS..100..149M}), [DDOP] data the initial gap at the quoted temperature of 523\,K (293\,K), [673\,K] is 1.34\,eV or equivalent to $X_H =  0.31$ (2.17\,eV $\equiv X_H =  0.50$), [0.59\,eV $\equiv X_H =  0.14$], and then declines to -0.13\.eV (-0.15\,eV), [-0.11\,eV] after annealing to temperatures in excess of 1070\,K. 

In order to make use of the \cite{1984JAP....55..764S}, \cite{1995ApJS..100..149M} and DDOP laboratory data we have analytically fitted the temperature-dependent optical gap, $E_{\rm g}(T)$, transition with a suitably smooth (tanh) function that allows the determination of $E_{\rm g}$ at any temperature. This function, $E_{\rm g}(T)$, takes the form 
\begin{equation}
E_{\rm g}(T) = E_{\rm g-} +(E_{\rm g+}-E_{\rm g-}) \ \{ 1 + {\rm tanh} \left(-[T-T_{\rm m}]/2 \delta \right)  \} / 2
\label{EgT_fit}
\end{equation}
where $E_{\rm g+}$ is the band gap of the high-$E_{\rm g}$ material deposited at low temperature, $E_{\rm g-}$ is the band gap of the low-$E_{\rm g}$ material after annealing to high temperatures and where $\delta$ determines the width of the transition between the extreme values $E_{\rm g+}$ and $E_{\rm g-}$. Table~\ref{tab_EgT_fit_params} gives the values of the fit parameters for the \cite{1984JAP....55..764S}, \cite{1995ApJS..100..149M}, DDOP data. 

% TABLE
\begin{table}
\caption{The $E_{\rm g}(T)$ parameters used in Eq.~(\ref{EgT_fit}) to fit the \cite[][S84]{1984JAP....55..764S}, \cite[][M95]{1995ApJS..100..149M} and DDOP data.}
\begin{center}
\begin{tabular}{lccc}
                                         &                             &                   &                    \\[-0.35cm]
\hline
\hline
                                         &                             &                     &                  \\[-0.35cm]
  Parameter              &      S84         &      M95       &      DDOP      \\[0.05cm]
\hline
                                &                             &                    &                   \\[-0.35cm]
$E_{\rm g+}$ [eV]    &     2.22       &       1.35     &       0.74   \\
$E_{\rm g-}$  [eV]    &     -0.18     &      -0.17     &     -0.12    \\
$T_{\rm m}$  [K]      &      785       &      770       &      873    \\
$\delta$  [K]             &      110       &      170       &      220   \\             
\hline
\hline
                       &                             &                   &                    \\[-0.25cm]
\end{tabular}
\end{center}
\label{tab_EgT_fit_params}
\end{table}
In Fig.~\ref{EgvsT_2} we plot these laboratory data and our analytical fits to these data using Eq.~(\ref{EgT_fit}) and the fit parameters in Table~\ref{tab_EgT_fit_params}. 

In fact the fit given by Eq.~(\ref{EgT_fit}) can be generalised to any initial band gap material using the constraints on $X_{\rm H}$, {\it {\it i.e.}}, 
\begin{equation}
E_{\rm g}(T) =  -0.15 + 0.5(4.3X_{{\rm H},i} + 0.15)  \{ 1 +  {\rm tanh} \left(-[T-T_{\rm m}]/ 2 \delta \right) \} \label{EgT_fit_gen}
\end{equation}
where $X_{{\rm H},i}$ is the initial hydrogen atom fraction. We have here assumed that for the high-temperature-annealed materials the band gap is independent of the initial value of $X_{\rm H}$. We have therefore set $E_{\rm g-} = -0.15$ and have substituted  for $E_{\rm g+}$ from Eq.~(\ref{Eg4XH}). In Fig.~\ref{EgvsT_2}, and based on fits to the laboratory data, we adopt the values for $T_{\rm m}$ given by the expression 
\begin{equation}
T_{\rm m} = 788 + \{ 0.49 [ {\rm exp}( 2.7 - E_{\rm g+})]^{3.2} \}  - [ 30 (2.7 - E_{\rm g+})^2]
\label{eq_Tm_fit }
\end{equation}
and values for $\delta$ given by
\begin{equation}
\delta = 137.0 - 37.5 E_{\rm g+}. 
\label{ eq_delta_fit}
\end{equation}
In Fig.~\ref{EgvsT_2} we show a series of evolutionary tracks for materials with different initial hydrogen atom fractions,  $X_{{\rm H},i} =$ 0.1 to 0.6 (in 0.1 steps, and also the intermediate case for 0.15). Eq.~(\ref{EgT_fit_gen}) and Fig.~\ref{EgvsT_2} can now be used to determine the evolution of the optical gap of any  hydrocarbon  material annealed to any temperature. 

We note that Fig.~\ref{EgvsT_2} indicates that the temperature-dependent annealing profile is steeper for the more hydrogen-rich materials, {\it i.e.}, these materials can apparently reach a smaller band gap state at lower temperatures than for H-poorer materials. We speculate that such an effect could be due to a sort of band gap-locking effect where the smaller gap of the H-poorer materials, which is determined by the largest aromatic clusters, somewhat inhibits or `blocks' evolution to smaller band gap. The H-rich materials, being of lower density, could give the material more `room to manoeuvre' in the formation of larger aromatic clusters, than is possible in the H-poorer case, and therefore lead to larger aromatic clusters at lower annealing temperatures. Possibly related to this, and as pointed out by \cite{2004PhilTransRSocLondA..362.2477F}, is that there can be a ``memory'' effect in some nano-structured carbon films, whereby precursor clusters determine the subsequent evolution ({\it e.g.}, possibly as a result of hydrogen atom loss-induced re-structuring).  It appears that small clusters tend to be chain-like while larger clusters are rather three-dimensional, cage-like, $sp^2$ structures \citep[{\it e.g.},][]{2004PhilTransRSocLondA..362.2477F}.

%------------------------------------------------------------------
\section{A model for the optical properties of a-C(:H) - the optEC$_{\rm (s)}$ model}
\label{sect_optEC_model}
%------------------------------------------------------------------

In this section we develop an {\bf o}ptical property {\bf p}rediction {\bf t}ool for the {\bf E}volution of {\bf C}arbonaceous solids ({\bf optEC$_{\rm (s)}$}). This model should be considered as something of a ``straw man'' that can be partially, or even totally, torn down and re-constructed as and when dictated by the availability of more-constraining laboratory data on low-temperature formed-and-analysed hydrogenated carbonaceous materials. 

A model, based on two Tauc-Lorentz oscillators (2-TL), has been developed for the optical properties of amorphous carbons in the visible-UV wavelength domain by \cite{2007DiamondaRM...16.1813K}. However, and as we show later in this paper, the need for models and spectroscopic data, over a wide range of hydrogen contents and over a much broader wavelength range, especially at FIR-mm wavelengths, is particularly critical. 

The range of amorphous carbons encompassed by the optEC$_{\rm (s)}$ model includes the large band gap ($E_{\rm g} \simeq 3$\,eV), hydrogenated amorphous carbons (a-C:H) through to the hydrogen poorest, smallest band gap ($E_{\rm g} \simeq 0$\,eV) amorphous carbons (a-C).  The model predicts their optical properties over more than seven  orders of magnitude in energy, $2.5 \times 10^{-6}$ -- $56$\,eV (i.e., $0.022\,\mu$m to 50\,cm in wavelength), and more than eight orders (one order) of magnitude in the imaginary (real) part of the refractive index $k$ ($n$).  The aim of this tool is to provide a set of self-consistent data that can be used to test to the effects of large (radii $\geq 100$\,nm) carbonaceous grain evolution within the astrophysical context.  To this aim we provide a pair of ASCII $n$ and $k$ data files that give the optEC$_{\rm (s)}$ model real and imaginary parts of the complex index of refraction, $m(n,k)$ for {\em bulk} a-C(:H) solids. These data are provided as a function of the  band gap ($E_{\rm g} \equiv X_{\rm H}$), which is taken to be the single, most important, material-characterising parameter. 

In a follow-up paper we extend this work, to include the particle size dependence, and give pairs of ASCII data files, for $n$ and $k$, as a function of the band gap ($E_{\rm g} \equiv X_{\rm H}$) for a range of particle radii $a$ 
({\it i.e.}, for $a = 100, 30, 10, 3, 1, 0.5$ and 0.33\,nm).  

The optEC$_{\rm (s)}$ model has its foundations firmly planted in the nature of amorphous carbon, so comprehensively-reviewed by \cite{1986AdPhy..35..317R}. As is described by \cite{1986AdPhy..35..317R} the wide-band optical spectra of amorphous carbons (a-C) and hydrogenated amorphous carbons (a-C:H) show two clear and separated peaks: a $\pi-\pi^\ast$ peak at $\sim 4$\, eV and $\sigma-\sigma^\ast$ at $\sim 13$\, eV. \cite{2007DiamondaRM...16.1813K} noted that the energy positions of the $\pi-\pi^\ast$ and $\sigma-\sigma^\ast$ transitions do not change significantly with hydrogen atom content and therefore fixed the band positions in their optical property determination.  An additional energy peak at $\sim 6.5$\, eV has been attributed to C$_6$, `benzene-like' aromatic clusters in the structure. With thermal annealing, or equivalently with photo-darkening, the $\pi-\pi^\ast$ peak strengthens with respect to both the $\sigma-\sigma^\ast$ and $\sim 6.5$\, eV (C$_6$) peaks. These trends are the result of increasing aromaticity with increased annealing. Here we assume these three characteristic band energies ({\it i.e.}, 4.0\,eV for $\pi-\pi^\ast$, 6.5\,eV for C$_6$ and 13.0\,eV for $\sigma-\sigma^\ast$), and their annealing-dependent behaviour, as the fundamental and underlying basis for the optEC$_{\rm (s)}$ model. \cite{2007Carbon.45.1542L} and \cite{2011A&A...528A..56G} have undertaken a similar decompositional analysis for hydrogenated amorphous carbon materials but they allow for more constituent bands and also allow their band positions to vary in their fits to their laboratory data. 

With the quantitative annealing behaviour determined (see \S~\ref{sect_T_dep_Eg}), we find that the optical properties, from UV to cm wavelengths, depend on only the band gap or, equivalently, the H atom fraction of the a-C(:H) material. The presence of an optical gap in these materials reflects the fact that the conjugated aromatic (and olefinic) clusters are discontinuous and isolated, {\it i.e.}, they do not percolate. Note that the non-percolation of the clusters is implicitly assumed in the eRCN model (paper~I). The $sp^2$-bonded clusters are therefore islands, even when most of the carbon atoms in the structure are $sp^2$ hybridised \citep{1986AdPhy..35..317R}.

We now derive a single-parameter model for the evolution of the imaginary part, $k(E)$, of the complex refractive index, $m(E)=n(E)+ik(E)$, as a function of energy, where the critical characterising parameter is the band gap of the material, $E_{\rm g}$. We then use the derived $k(E,E_{\rm g})$ to calculate $n(E,E_{\rm g})$ using the Kramers-Kronig Fortran toolbox (KKTOOL) provided by V\"olker Ossenkopf \footnote{Available for download from the following website http://hera.ph1.uni-koeln.de/~ossk/Jena/pubcodes.html \label{footnote2}}. The KKTOOL fortran code was lightly updated and modified to allow for more wavelength coverage and more materials but was otherwise used {\it as-is}.

%------------------------------------------------------------------
\subsection{Band profiles}
%------------------------------------------------------------------

In order to cover the full optical behaviour of carbonaceous materials we use the properties of diamond and graphite, in addition to the well-determined a-C:H and a-C properties, as a guide to the cross-sections for the characterising $\pi-\pi^\ast$,  C$_6$ and $\sigma-\sigma^\ast$ bands. We find that these bands in a-C(:H) can be empirically well-fit with a log-normal profile in energy, $g_i(E)$, of the form 
\begin{equation}
g_i(E) = {\rm exp} \Bigg\{ - \left[ {\rm ln} \left( \frac{E}{E_{0,i}} \right) \right]^2 \frac{1}{2 \sigma_i^2} \Bigg\}, 
\label{eq_log_normal}
\end{equation}
where we assume a width, $\sigma_i = 0.3$\,eV, for each of the bands and where $E_{0,i}$ is the band-centre energy (seeTable~\ref{optECs_params}). At high and low energies the adopted log normal band profiles need to be modified to take account of the appropriate energy-dependencies of the material optical properties. 

% TABLE
\begin{table}
\caption{The optEC$_{\rm (s)}$ model input parameters.}
\begin{center}
\begin{tabular}{lccc}
                                         &                             &                   &                    \\[-0.35cm]
\hline
\hline
                                         &                             &                     &                  \\[-0.35cm]
  Parameter                      &  $\pi-\pi^\ast$       &  C$_6$         &  $\sigma-\sigma^\ast$      \\[0.05cm]
\hline
                                         &                             &                    &                   \\[-0.35cm]
$i$                                    &             1               &          2          &           3           \\
E$_{0,i}$ [ eV ]                 &             4.0            &        6.5          &        13.0        \\
$\sigma_{i}$ \, [ eV ]        &             0.3            &        0.3          &        0.3        \\
$S_i(E_{\rm g+})$             &             0              &       0.30         &         1.00        \\
$S_i(E_{\rm g-})/0.47$     &          1.65             &       0.60         &         1.00        \\
%$\delta_i$                         &           0.30           &        0.30         &         0.30         \\\hline
\hline
                       &                             &                   &                    \\[-0.25cm]
\end{tabular}
\end{center}
\label{optECs_params}
\end{table}

%------------------------------------------------------------------
\subsubsection{High energy behaviour}
%------------------------------------------------------------------

For energies beyond 16\,eV ($E_{\rm EUV}$) we modify and extend the $\sigma-\sigma^\ast$ band using a power law behaviour determined by the expected photo-electron emission cross-section at EUV to x-ray wavelengths,  $k(E,E_{\rm g}) \propto E^{-2.5}$, {\it i.e.}, 
\begin{equation}
g_i^\prime(E) = g_i(E) \ \times \Bigg\{ \frac{ E }{ E_{\rm EUV} } \Bigg\}^{-2.5} \ \ \ \ \ \ \ \ \ \ \ \ {\rm for} \ \ E \geq E_{\rm EUV}. 
\label{eq_high_E_fit} 
\end{equation}
This results in asymmetric $\sigma-\sigma^\ast$ band profiles essentially identical in form to the 2-TL (two Tauc-Lorentz oscillators) formalism used by \cite{2007DiamondaRM...16.1813K} in their optical property derivation. 

We extend the photo-electron-emission energy-dependence out to energies well beyond 60\,eV in order to allow a proper determination of $k(E,E_{\rm g})$, using KKTOOL, for $E \leq 55$\,eV. In the tabulations of $n$ and $k$ we therefore only present complex refractive index data for energies $\leq 56$\,eV (wavelengths $\geq 22$\,nm), {\it i.e.}, from the EUV to the dm domain. 

%------------------------------------------------------------------
\subsubsection{Low energy behaviour}
%------------------------------------------------------------------

At low energies and long wavelengths Fig.~\ref{fig_alpha} indicates that the optical property behaviours of practically all of the laboratory-measured a-C:H and a-C materials show approximately linear trends. For $E < E_1^\prime$, where we set 
\begin{equation}
E_1^\prime = 4.5-(E_{\rm g}{\rm [eV]}/1.2) \ \ {\rm eV},
\label{eq_k_lowE_E1p}
\end{equation} 
we therefore assume that the optical properties show only an underlying linear trend (with superimposed bands, see paper~I and \S\,\ref{sect_add_IR_bands}). We can then empirically fit the experimentally-determined wavelength- and $E_{\rm g}$-dependent values of $k(E,E_{\rm g})$ with the following simple expression 
\begin{equation}
k(E,E_{\rm g}) \propto E^\gamma \ \ \ \ \ \ \ \ \ \ \ \ {\rm where} \ \ \gamma  = 2(E_{\rm g}{\rm [eV]}-0.07). 
\label{eq_k_lowE}
\end{equation}
As we show below, this underlying linear trend in the $E_{\rm g}$-dependent behaviour, which is most evident for the low $E_{\rm g}$ materials with little contribution from the IR bands (for $\lambda \geqslant 0.5\,\mu$m), is quite remarkable in that it allows us to predict the imaginary part of the complex index of refraction, $k$, of a-C(:H) materials over more than three orders of magnitude in wavelength; the value of $k$ itself varies by more that eight orders of magnitude in the mm wavelength range. 

It is clear that the linear behaviour adopted here for the long-wavelength optical properties is entirely empirical and it is likely that no `real' dust analogue material will show such a simple, {\em linear} dependence. However, current data do not allow better constraints on the modelling of low-temperature-formed carbonaceous dust analogues. Primarily this arises because the long-wavelength properties are affected by bands, such as those seen in the DDOP data, which have not yet been measured for a sufficiently large range of materials, especially so for the high hydrogen content a-C:Hs, to allow for the characterisation and inclusion of long-wavelength bands in the model.

%------------------------------------------------------------------
\subsection{Derivation of the imaginary part of the refractive index}
%------------------------------------------------------------------

We find that a very satisfactory fit to the laboratory-measured optical properties, at visible to UV wavelengths, of the a-C and a-C:H end-members, {\it i.e.}, $E_{\rm g-} \simeq -0.15$\,eV  and $E_{\rm g+} \simeq 2.6$\,eV, is possible using a simple scaling of the 4, 6.5 and 13\,eV band strengths, $S_i$, where $i=1,2$ or 3. We choose to normalise these scaling factors to the 13\,eV $\sigma-\sigma^\ast$ band strength of the $E_{\rm g+}$ material and adopt $S_i$ values of 1.65, 0.60 and 1.00 (equivalently, 0.00, 0.30 and 1.00 for the $E_{\rm g-}$ material), respectively. We also apply a single scaling factor of 0.47 to all of the low $E_{\rm g}$ material $S_i$ band values, with respect to the H-rich end-member. All of the determined parameter values are shown in Table~\ref{optECs_params}. Using these data we now construct the imaginary part of the refractive index for any a-C(:H) material as a function of energy, $E$, and the single `free' parameter, $E_{\rm g}$, as a linear combination of the end-member compositions, {\it i.e.}, 
\begin{equation}
k(E) =  \sum_{i=1}^3 \Bigg\{ f(E_{\rm g+}) S_i(E_{\rm g+}) + f(E_{\rm g-}) S_i(E_{\rm g-}) \Bigg\} g_i(E)
\label{eq_optEC_1 }
\end{equation}
where $g_i(E)$ is the log normal band profile defined in Eq.~(\ref{eq_log_normal}). The fractions of the high [low] band gap material $f(E_{\rm g+})$ [f($E_{\rm g-}$)] are given by a simple linear interpolation between the limiting values for the band gap, {\it i.e.}, 
\begin{equation}
f(E_{\rm g+}) = \frac{(E_{\rm g}-E_{\rm g-})}{(E_{\rm g+}-E_{\rm g-})}, \ \ \ \ \ f(E_{\rm g-}) = 1-f(E_{\rm g+}). 
\label{eq_optEC_2 }
\end{equation}

\begin{figure} 
 %\resizebox{\hsize}{!}{\includegraphics{the_model/ZZ_eps2_vs_energy_basics_2011.ps}}
 \resizebox{\hsize}{!}{\includegraphics{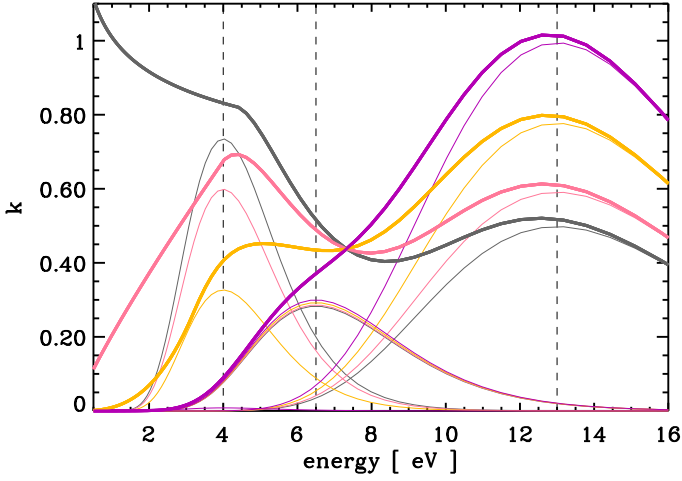}}
 \caption{The imaginary part of the refractive index, $k$, for the optEC$_{(\rm s)}$ model materials with $E_{\rm g} = 0, 0.5, 1.5$ and 2.67\,eV (dark grey, pink, yellow and purple lines, respectively). The vertical dashed lines mark the band centres of the $\pi-\pi^\ast$, C$_6$ `benzene ring' and $\sigma-\sigma^\ast$ bands at 4.0, 6.5 and 13\,eV, respectively.}
 \label{fig_k_basics}
\end{figure}

In this work we present data for a range of a-C(:H) materials defined by their band gaps. The $E_{\rm g}$ values that we adopt, and the colour coding scheme that we use in all of the plots, are shown in Table~\ref{table_colour_code}. In Fig.~\ref{fig_k_basics} we show the energy dependence of $k$ for several a-C(:H) materials, including the `close-to' limiting cases where $E_{\rm g} =0$ and 2.67\,eV. In the figure we also show the decomposition into the separate $\pi-\pi^\ast$,  C$_6$ and $\sigma-\sigma^\ast$ bands along with their band centres. The calibration of these data against the available laboratory and modelled data is discussed in \S~\ref{sect_look}. 

% FIGURE 7 *********************************************************
% plot of eps_2 vs E for the basics
% *********************************************************

%------------------------------------------------------------------
\subsection{The addition of IR bands into the $k$ determination}
\label{sect_add_IR_bands}
%------------------------------------------------------------------

\begin{figure} 
 \resizebox{\hsize}{!}{\includegraphics{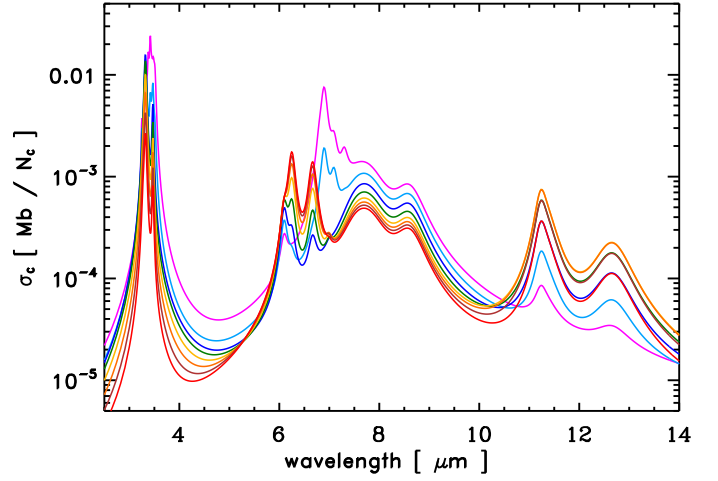}}
 \caption{The adopted wavelength-dependent IR cross-section per carbon atom, $\sigma_{\rm C}$, in Mb (1\,Mb $= 10^{-18}$\,cm$^2$) for the eRCN model.
 The upper curves (purple) are for the H-rich, wide band gap materials and the cross-sections, in the 8\,$\mu$m region, decrease with decreasing $X_{\rm H}\equiv E_{\rm g}$ from top to bottom  -- $E_{\rm g} = 2.5$ (violet), 2.25 (cobalt), 2.0 (blue), 1.75 (green), 1.5 (yellow), 1.25 (orange), 1.0 (brown) and 0.75\,eV (red).}
  \label{fig_sigmaC}
\end{figure}

\begin{figure} 
 \resizebox{\hsize}{!}{\includegraphics{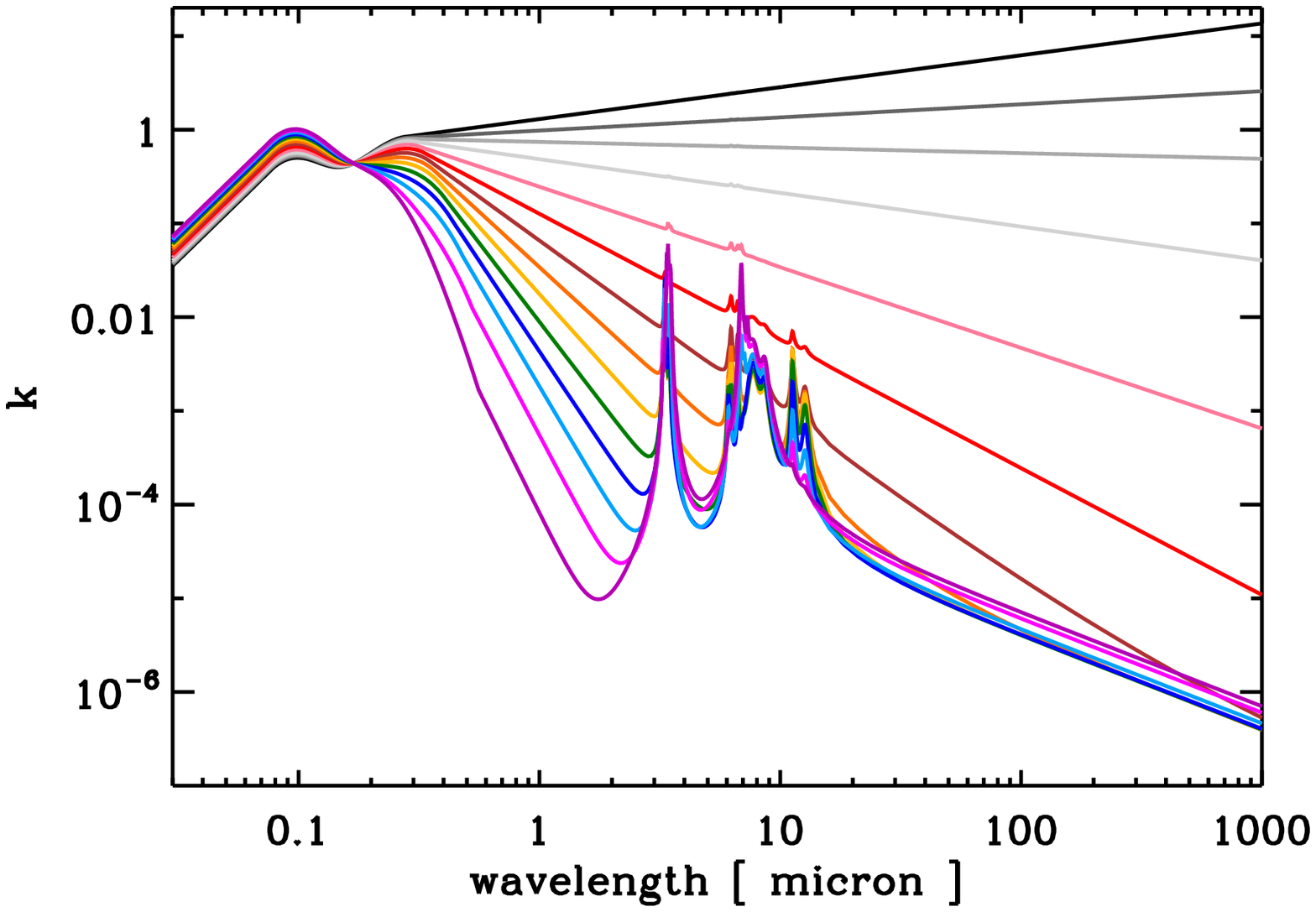}}
 \resizebox{\hsize}{!}{\includegraphics{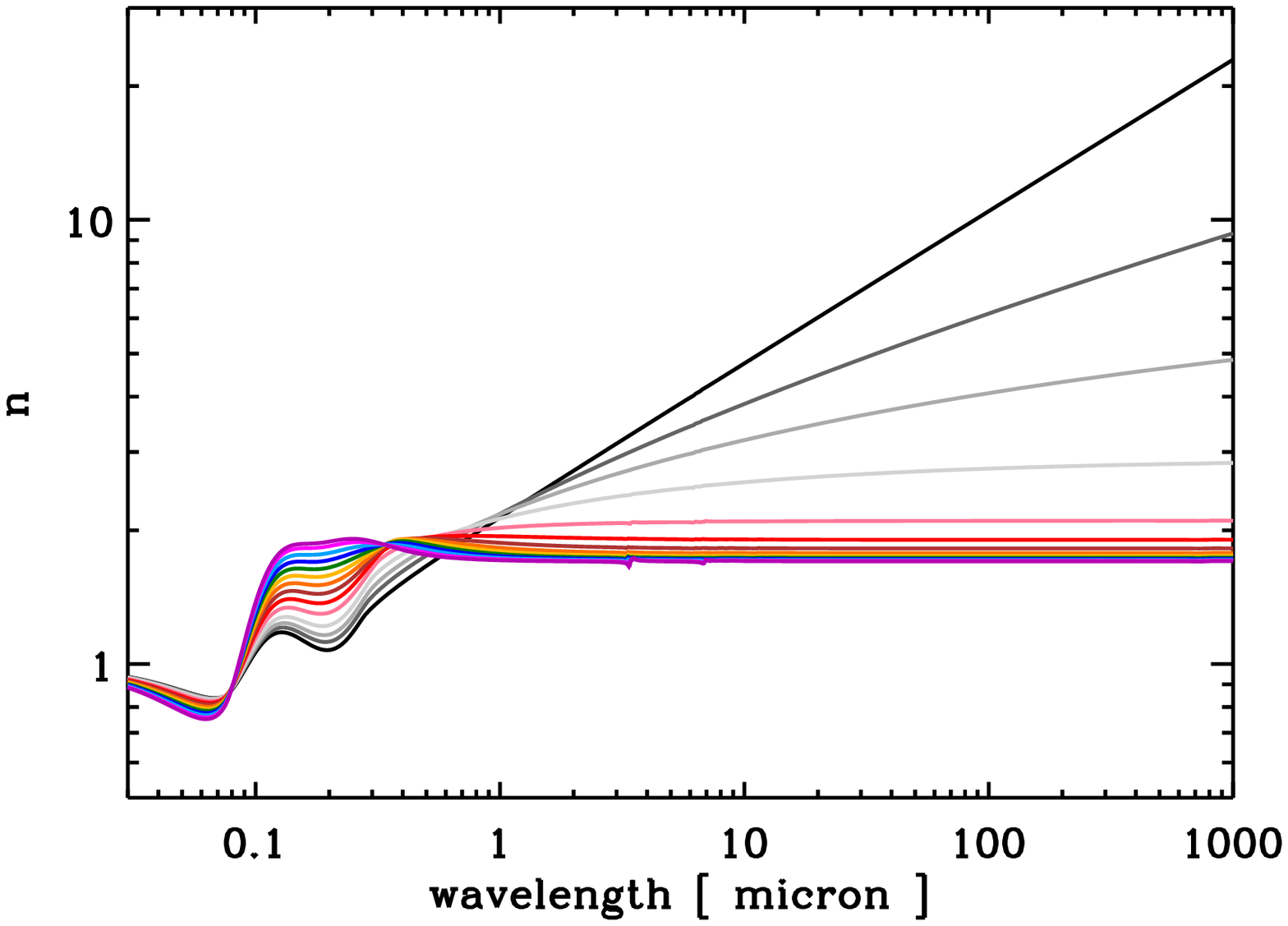}}
 \caption{The imaginary part, $k$ (upper), and derived real part, $n$ (lower), of the refractive index data for the suite of a-C(:H) materials predicted by the optEC$_{(s)}$ model as a function of $E_{\rm g}$ (see Table~\ref{table_colour_code} for the line colour-coding).} 
 \label{fig_optECs_nk_only}
\end{figure}

In paper~I we derived the characteristic infrared $X_{\rm H}$-dependent band profiles for a-C(:H) materials using the eRCN and DG models presented there. These same bands, with their characteristic positions, widths and intensities have coherently been added into the determination of $k$, where  $k = \sigma_{\rm C} \, N_{\rm C} \, \lambda / (4 \pi) = h\, c\, \sigma_{\rm C} \, N_{\rm C}  / (4 \pi \, E) $.  This is done by summing over the $n$ contributing bands, {i.e.}, 
\begin{equation}
k_{\rm IR}(E,E_{\rm g} ) = k(E,E_{\rm g} ) + N_{\rm C}(E_{\rm g})  \sum_{j=1}^n \Bigg\{ \frac{ h\  c\ \sigma_{{\rm C},j}(E)}{ 4 \pi E } \Bigg\}, 
\label{eq_IR_bands_k}
\end{equation}
where $N_{\rm C}$ is the number of carbon atoms per unit volume in the material under consideration and is given by 
\begin{equation}
N_{\rm C}(E_{\rm g}) = \frac{ \rho(X_{\rm H}) } {m_{\rm C}} \Bigg\{ 1 + \frac{X_{\rm H}}{12 (1-X_{\rm H})} \Bigg\}^{-1} ,
\label{eq_N_C}  
\end{equation}
where the a-C(:H) material specific mass density is given by $\rho(X_{\rm H}) {\rm [g\,cm^{-3}]} \approx 1.3  + 0.4 \, {\rm exp}[ -( E_{\rm g} + 0.2 ) ]$ and the right hand term in \{brackets\} `removes' the H atom contribution. $m_{\rm C}$ is the carbon atom mass. Fig.~\ref{fig_sigmaC} shows the adopted band cross-sections in Mb, where 1\,Mb $= 10^{-18}$\,cm$^2$, which for the $j^{\, \rm th}$ band is determined from
\begin{equation}
\sigma_{{\rm C},j}(E) = \sigma_{0,j}(E) \ X_j \ g_j(E)
\label{eq_sigma_C}  
\end{equation}
 cross-section $\sigma_{0,j}$ is derived from the integrated cross-sections (see Table~2 and \S~2.2.4 in paper~I), $X_j$ is the C atom fractional abundance for the participating C$_p$H$_q$ ($p\geq1$, $q\geq0$) functional group. For the band shapes, $g_j(E)$, we assume Drude profiles (see paper~I) in order to give the correct long wavelength behaviour. The effects of adding the infrared bands to the refractive index data are clearly evident in Fig.~\ref{fig_optECs_nk_only} (also see later \S\,\ref{sect_look} and Fig.~\ref{fig_all_nk}). 

The refractive index data shown in Fig.~\ref{fig_optECs_nk_only} clearly indicate the effects of the electronic transitions shortward of 1\,$\mu$m, the IR properties in the $1-30\,\mu$m region and the FIR-mm behaviour, which is here determined by the wings of the Drude-profile IR bands for hydrogen-rich materials. Thus, it is clear that a `full' knowledge of all of the IR bands and their characteristics is absolutely essential for an understanding of the FIR-mm behaviour. Currently our knowledge is incomplete and hence the current model must be considered as something of a `straw man' model. 
% *********************************************************

% *********************************************************
% plot of optECs n and k data only - {\bf Paper~II, Fig.~9:} 
% *********************************************************

%------------------------------------------------------------------
\subsection{The conductivity of a-C:H and a-C: implications for the long-wavelength optical properties}
%------------------------------------------------------------------

The electrical conductivity of interstellar dust analogue materials is a key issue for their long-wavelength behaviour but within the astrophysical context it is not completely evident how important this will be for small, isolated, hydrogen-rich semi-conductor materials such as a-C:H. In this section, and based upon the available experimental evidence,  we attempt to assess and evaluate the effects of electrical conductivity on the FIR-mm properties of a-C(:H) at the low temperatures appropriate to interstellar carbonaceous dust. 

The H-poorer a-C members of the a-C(:H) family exhibit higher electrical conductivities as a result of the increased aromatic character, {\it i.e.}, enhanced but locally delocalised (within the $\pi$ cluster) electrons resulting from the $\pi-\pi^\ast$ band contribution. However, in these materials the aromatic structures that provide the conduction (electrons) are surrounded by insulating aliphatic and olefinic structures, and are therefore {\em not contiguous} as is the case for graphite. The onset of electrical conductivity is therefore guarded by an activation energy, for the transition between `nearby' aromatic sites, and the conductivity must then be thermally- or photo-activated. As \cite{1986AdPhy..35..317R} points out, there is a continuous distribution of these defect or localised (aromatic and olefinic) states across the pseudogap and their density increases with heat treatment. 

At long wavelengths the behaviour of the imaginary parts of the dielectric function and the refractive index for conducting materials depend upon their temperature-dependent conductivity and we therefore need to consider this in our calculation of the optical properties. The long wavelength behaviour is expected to follow the form $a \, \lambda^{-1} + b\, \lambda \, \sigma_{\rm ec}(T)$, where the right hand term includes the temperature-dependent electrical conductivity, $\sigma_{\rm ec}(T)$, which results in an `upturn' in the behaviour of the imaginary parts of $k$ and $\epsilon_2$. 

\begin{figure} 
 %\resizebox{\hsize}{!}{\includegraphics{the_model/aCH_IR_model_NEW/ZZZ_conductivity_band_shapes/ZZ_conductivity_vs_1000ovT_2011.ps}}
 %\resizebox{\hsize}{!}{\includegraphics{the_model/aCH_IR_model_NEW/ZZZ_conductivity_band_shapes/ZZ_conductivity_vs_logT_2011.ps}}
 \resizebox{\hsize}{!}{\includegraphics{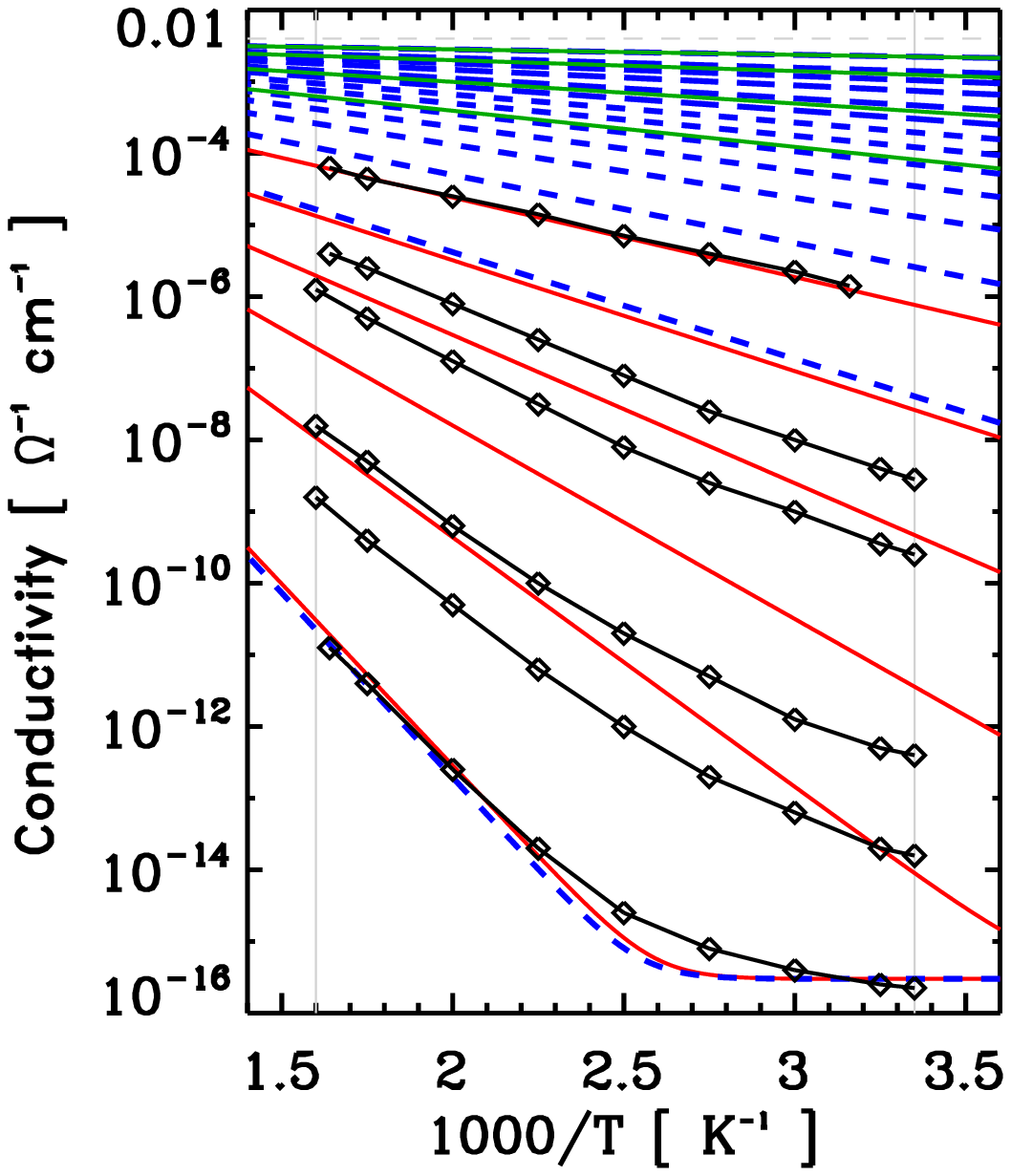}}
 \resizebox{\hsize}{!}{\includegraphics{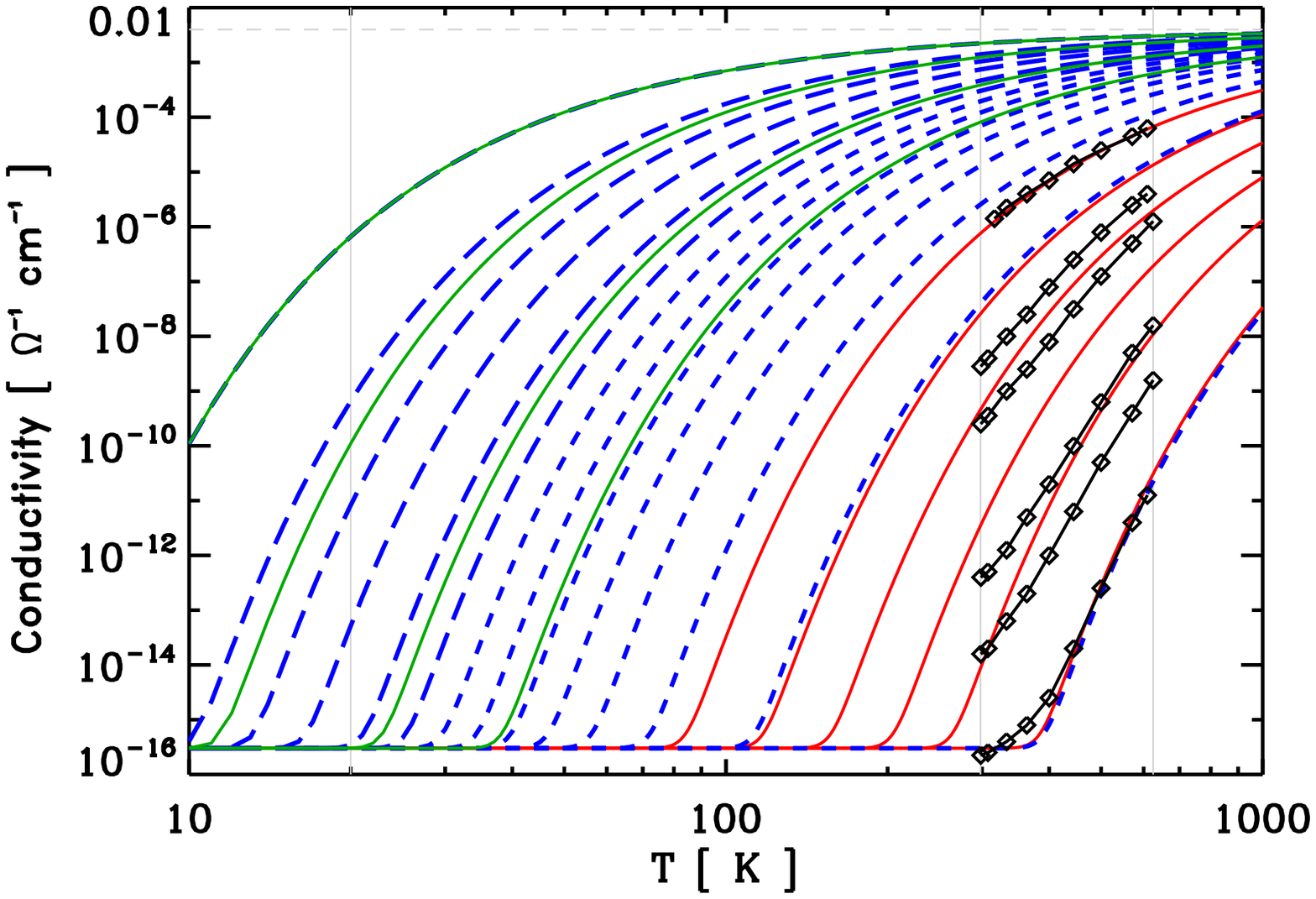}}
 \caption{The conductivity of a-CH and a-C materials (a-C:H - black lines with data points, as per Fig.~33 of \cite{1986AdPhy..35..317R}; red lines - fits to these a-C:H data  using Eq.~(\ref{eq_conductivity_fit}); green lines - fits for a-C,  using the \cite{2001Scriptamater.44..1191D} activation energies) compared to those predicted for the eRCN a-C:H/a-C (short-dashed blue) and DG a-C models (long-dashed blue) by Eq~(\ref{eq_conductivity_fit}).} 
 \label{fig_conductivity}
\end{figure}

The thermally-activated electrical conductivity of a-C and a-C:H materials is reviewed by \cite[][and references cited therein]{1986AdPhy..35..317R} and has been studied more recently by \cite{2001Scriptamater.44..1191D}. From these studies it has been shown that $\sigma_{\rm ec}(T)$ follows a simple thermally-activated behaviour at high temperatures,
\begin{equation}
\sigma_{\rm ec}(T) = \sigma_0 \, {\rm exp} \{ -\Delta E / kT \}
\label{eq_conductivity}
\end{equation}
where, for a-C and a-C:H materials, it was shown that $\Delta E = 0.02-1.5$\,eV and $\sigma_0$ lies in the range $10^{-4}-10^{-1}$\,$\Omega^{-1}$\,cm$^{-1}$.  In Fig.~\ref{fig_conductivity} we show the data from \citet[][Fig.~33]{1986AdPhy..35..317R} and \cite{2001Scriptamater.44..1191D}. The upper part of this figure shows the data as a function of 1000/$T$ [K$^{-1}$] over a limited temperature range (as per Fig.~33 of \cite{1986AdPhy..35..317R}) and the lower part shows the electrical conductivity over the full temperature range. Also shown in Fig.~\ref{fig_conductivity} are our adopted fits to these data over the temperature range $10-1000$\,K, where we set $\sigma_0 = 4 \times 10^{-3}$\,$\Omega^{-1}$\,cm$^{-1}$ and 
\begin{equation}
\Delta E =  0.075 \Bigg\{ \left( \frac{1000}{T_{\rm ann}} \right)^{\frac{5}{2}} - 0.0265 \Bigg\}
\label{eq_conductivity_fit}
\end{equation}
Here $T_{ann}$ is the annealing temperature ({\it e.g.}, see Fig.~\ref{EgvsT_2}). Note that our fits to the electrical conductivity of these materials cover over 15 orders of magnitude in $\sigma_{\rm ec}(T)$ and two orders of magnitude in temperature. 

From this analysis, and from our detailed fitting to the laboratory data, we find that the electrical conductivities of a-C and a-C:H materials are  small ($\leq 10^{-2}$\,$\Omega^{-1}$\,cm$^{-1}$ at $T \leq 1000$\,K) and that, at the temperatures of interest for interstellar and solar system carbonaceous dust ({\it viz.}, $T \simeq 10-300$\,K), the electrical conductivity $\sigma_{\rm ec}(T) \lesssim  10^{-2}$\,$\Omega^{-1}$\,cm$^{-1}$. For ISM dust studies $T_{\rm dust}$ is generally $\lesssim 50$\,K, which implies electrical conductivities $\ll 10^{-4}$\,$\Omega^{-1}$\,cm$^{-1}$ for a-C and $\ll 10^{-10}$\,$\Omega^{-1}$\,cm$^{-1}$ for a-C:H. 

We therefore conclude from our analysis that that we can safely ignore any contribution of the electrical conductivity to the optical properties of a-C and a-C:H materials at long wavelengths ($\lambda > 100$\,$\mu$m) and low temperatures ($T < 300$\,K). However, we do note that in some of the laboratory data ({e.g.}, see Fig.~\ref{fig_all_nk}) that there is an `upturn' in some of the $k$ data for the more H-poor materials and we suspect that this may be due to the difficulty in completely isolating the particles in these experiments, resulting in some `macroscopic' conduction between connected or clustered particles. This effect has been taken into account in the analyses of \cite{1991ApJ...377..526R} and \cite{1996MNRAS.282.1321Z}, using a variety of approaches, and it is to be noted that there does appear to be a linear trend in  $k$ for the BE and ACAR data of \cite{1996MNRAS.282.1321Z} once the clustering effect has been compensated for. 

The fact that the electrical conductivity of these materials is rather low, and especially so at low temperatures, implies that their thermal conductivities must also be very low, which may have interesting consequences for the a-C(:H) dust thermal conductivities and hence for the peak temperatures attained by small stochastically-heated carbonaceous particles in the ISM. 

% FIGURE 10 *********************************************************
% plot of a-C:H / a-C conductivity
% *********************************************************

%------------------------------------------------------------------
\subsection{A detailed look at the optEC$_{\rm (s)}$ data}
\label{sect_look}
%------------------------------------------------------------------

\begin{figure} 
 \resizebox{\hsize}{!}{\includegraphics{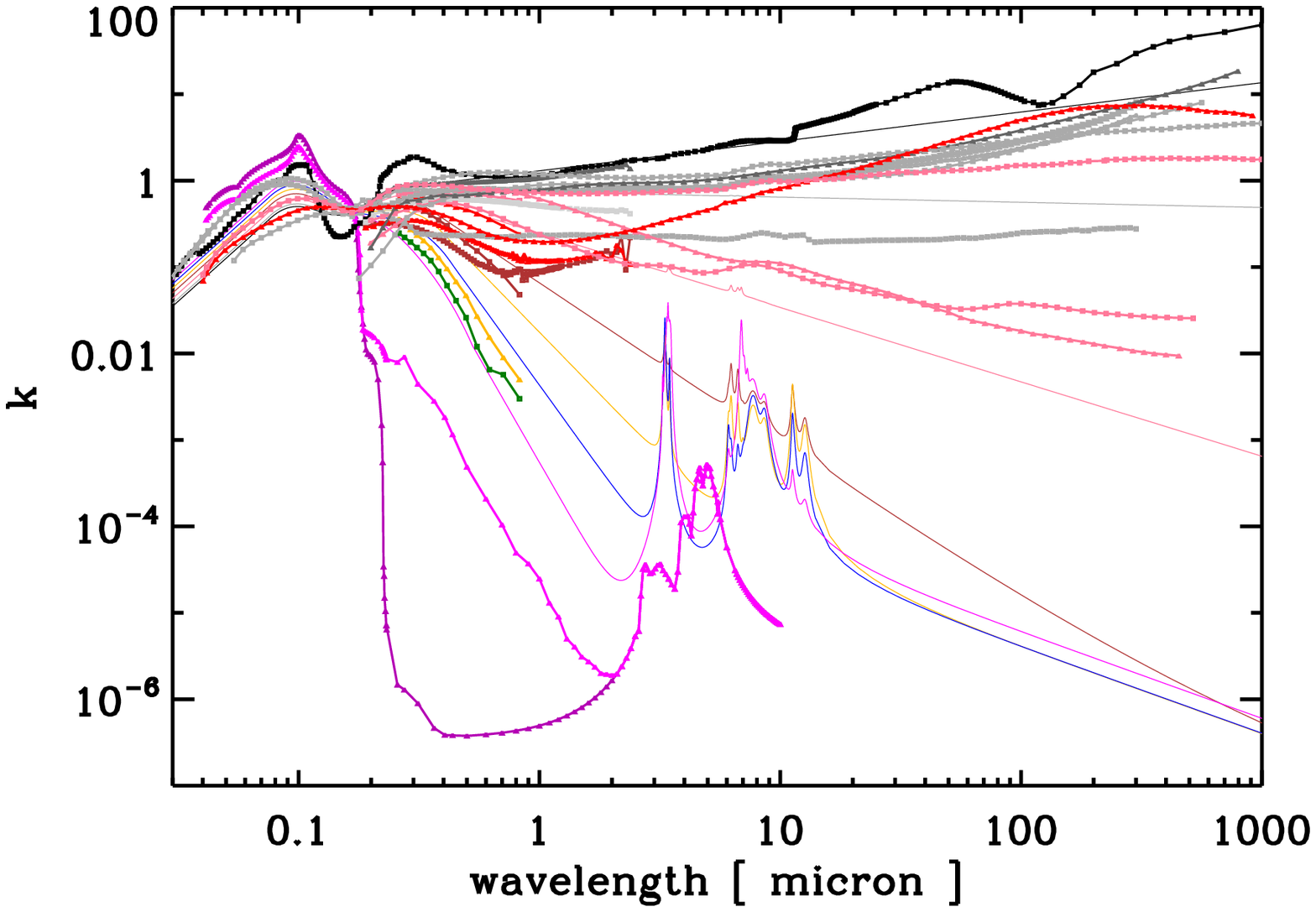}}
 \resizebox{\hsize}{!}{\includegraphics{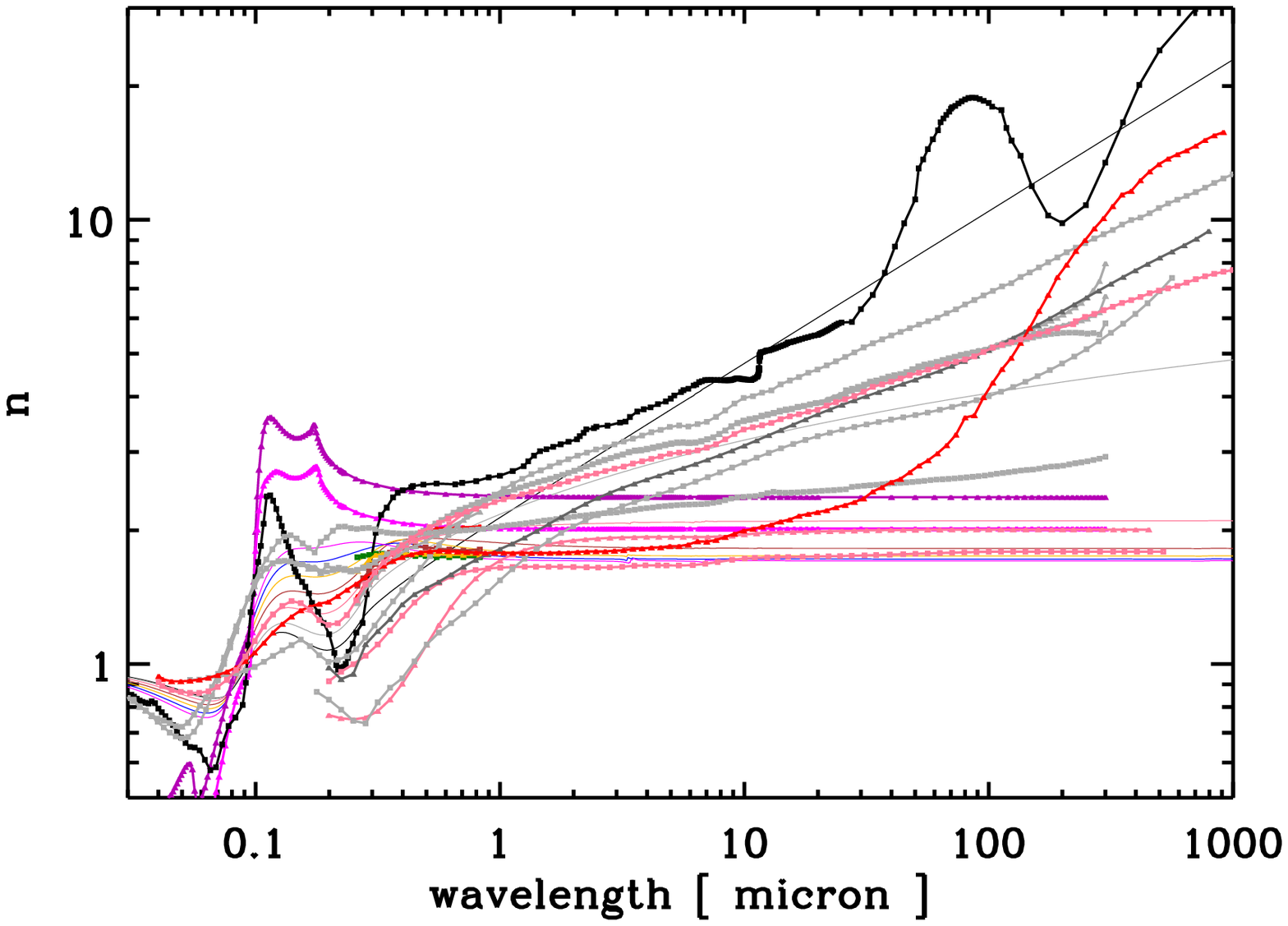}}
 \caption{The smooth, coloured lines show the optEC$_{\rm (s)}$ model-derived imaginary (top) and real (bottom) parts of the refractive index (large band gap, purple, to low band gap, grey). The lines with data points are the laboratory-measured and model-derived data (see references earlier for the sources). The upper black lines with data points show the data for graphite \citep{1984ApJ...285...89D} and the purple and violet lines with data points show the data for diamond \citep{1985HandbookOptConst...665,1989Natur.339..117L}. }
 \label{fig_all_nk}
\end{figure}

\begin{figure} 
 %\resizebox{\hsize}{!}{\includegraphics{the_model/ZZ_eps2_vs_energy_ALL_2011.ps}}
 \resizebox{\hsize}{!}{\includegraphics{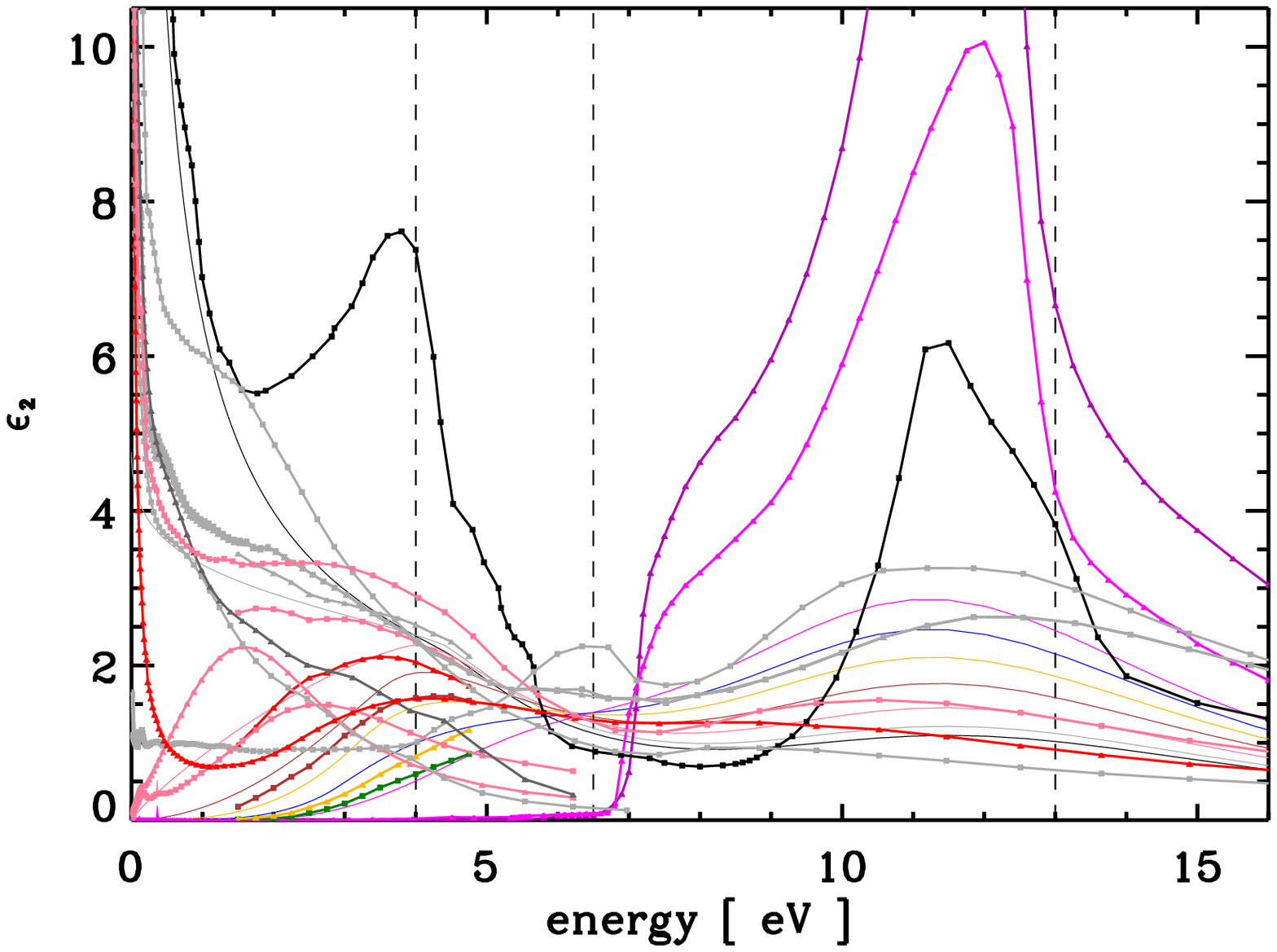}}
 \caption{The imaginary part of the dielectric function, $\epsilon_2$, for the optEC$_{(\rm s)}$ model (lines without data points) compared to laboratory and other model data (lines with data points). The diamond (purple and violet lines) and graphite data (black) are shown for comparison. The vertical dashed lines mark the band centres of the $\pi-\pi^\ast$, C$_6$ and $\sigma-\sigma^\ast$ bands at 4.0, 6.5 and 13\,eV, respectively.}
 \label{fig_e2_all}
\end{figure}

The derived data, from 50\,eV to 1\,mm, are again shown in Fig.~\ref{fig_all_nk} ($k$ and $n$) and Fig.\ref{fig_e2_all} ($\epsilon_2$ for $E = 0$ -- 16\,eV), where we compare our data to the modelled and laboratory-derived data. The smooth lines show the optEC$_{(\rm s)}$ model data and the lines with data points indicate the laboratory data from \cite{1984JAP....55..764S}, \cite{1995ApJS..100..149M} and DDOP, the modelled data from \cite{1991ApJ...377..526R} and \cite{1996MNRAS.282.1321Z}, and for comparison the graphite data \citep[][with the usual $\frac{1}{3} \epsilon_{||} + \frac{2}{3} \epsilon_{\bot}$ combination]{1984ApJ...285...89D} and diamond data \citep{1985HandbookOptConst...665,1989Natur.339..117L}. The colour coding here is as indicated in Table~\ref{table_colour_code}. The complementary values of the real part of the refractive index, $n$, shown in the lower part of Fig.~\ref{fig_all_nk}, were calculated using the Kramers-Kronig toolbox KKTOOL (see footnote~\ref{footnote2}). For high $E_{\rm g}$ a-C(:H) materials, for $\lambda \geqslant 2$\,mm, we have assumed a linear extrapolation of $n$ from the KKTOOL-determined 1 and 2\,mm data points. We find this is necessary because in our experience the KKTOOL does not converge well for $\lambda > 3$\,mm due to the large dynamic range (greater that four orders of magnitude difference between $k$ and $n$). Conversely, for low $E_{\rm g}$ a-C(:H) $k$ and $n$ differ by about only an order of magnitude and the KKTOOL calculation converges well out to a wavelength of $\approx 1$\,cm.

%% FIGURE 11 ********************************************************* ORIGINAL
%\begin{figure} 
% %\resizebox{\hsize}{!}{\includegraphics{the_model/ZZ_k_vs_wavelength_ALL_2011.ps}}
% %\resizebox{\hsize}{!}{\includegraphics{the_model/ZZ_n_vs_wavelength_ALL_2011.ps}}
% \resizebox{\hsize}{!}{\includegraphics{ZZ_k_vs_wavelength_ALL_2011.ps}}
% \resizebox{\hsize}{!}{\includegraphics{ZZ_n_vs_wavelength_ALL_2011.ps}}
% \caption{The smooth, coloured lines show the optEC$_{\rm (s)}$ model-derived imaginary (top) and real (bottom) parts of the refractive index (large band gap, purple, to low band gap, grey). The lines with data points are the laboratory-measured and model-derived data (see references earlier for the sources). The upper black lines with data points show the data for graphite \citep{1984ApJ...285...89D} and the purple and violet lines with data points show the data for diamond \citep{1985HandbookOptConst...665,1989Natur.339..117L}. }
% \label{fig_all_nk}
%\end{figure}
%% *********************************************************
% ********************************************************* UPDATED VERSION 
% {\bf Paper~II, Fig.~11:} 
% *********************************************************

% FIGURE 12 *********************************************************
% plot of eps_2 vs E
% *********************************************************

\begin{figure} 
 %\resizebox{\hsize}{!}{\includegraphics{the_model/ZZ_eps2_vs_energy_aCH_2011.ps}}
 \resizebox{\hsize}{!}{\includegraphics{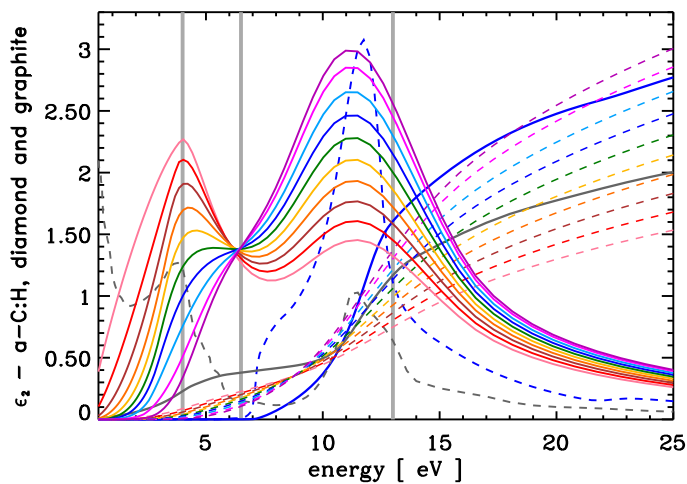}}
 \caption{As per Fig.~\ref{fig_e2_all}, the imaginary part of the dielectric function, $\epsilon_2$, for the a-C(:H) optEC$_{(\rm s)}$ model for comparison with Fig.~29 of \cite{1986AdPhy..35..317R}. Note that the diamond (short-dashed blue) and graphite (short-dashed grey) $\epsilon_2$ data have been scaled down here for comparison. The dashed lines rising towards high energies show the running integrals for $n_{\rm eff}$ (Eq.~\ref{eq_n_eff}) for the optEC$_{(\rm s)}$ data and also for diamond (blue) and graphite (dark grey). }
 \label{fig_e2_limited}
\end{figure}

Following the detailed methodology of \cite{1992A&A...266..513V} in Figs.~\ref{fig_e2_all} and \ref{fig_e2_limited} we plot  the imaginary part of the complex dielectric function for our derived optical data, {\it i.e.}, $\epsilon_2 = 2nk$. The data in these figures show that the  optEC$_{(\rm s)}$ model gives a very good quantitative fit to the available data for hydrogenated amorphous carbons. Note that the band positions, as defined above in determining $k$ (indicated as the vertical dashed and solid grey lines), do not match the positions of the bands in $\epsilon_2$ because of the additional dependence on $n$. 

Using the generated $\epsilon_2(E)$ we can now compare our data with those of the H-free, crystalline forms of solid carbon, {\it i.e.}, diamond \citep[all $sp^3$ C,][]{1985HandbookOptConst...665,1989Natur.339..117L} and graphite \citep[all $sp^2$ C,][]{1984ApJ...285...89D} by comparing the effective number of contributing electrons, $n_{\rm eff}$. 
Fig.~\ref{fig_e2_limited} shows the values of $n_{\rm eff}$ calculated as a running integral using the sum rule \citep[{\it e.g.}.][]{1986AdPhy..35..317R}, {\it i.e.}, 
\begin{equation}
n_{\rm eff} = \frac{m}{2 \pi^2 N_{\rm A} e^2 \hbar^2} \int^{E_2}_{E_1} E \epsilon_2(E) \ {\rm d}E, 
\label{eq_n_eff}
\end{equation}
where $m$ is the electron mass and $N_A$ is the number of atoms per unit volume. For the running integral we take $E_1$ as the lowest energy data point ($>$ mm wavelengths) and $E_2$ as the given energy, which therefore runs to EUV wavelengths on the extreme right of Fig.~\ref{fig_e2_limited}. 

We note that the data in Fig.~\ref{fig_e2_limited} compare well with that for a denser ta-C:H material  generated using a 2-TL model \citep[][see their Fig.~3]{2007DiamondaRM...16.1813K}.  The difference in the $\epsilon_2(E)$ peak intensities (in the $\sigma-\sigma^\ast$ band), of about a factor of two, is reflected in the specific density difference of the same order. However, and as they make clear, their 2-TL model does not apparently work so well for a-C:H materials, possibly due to the presence of $sp^1$ bonds or cumulenic chains, and so a direct comparison is difficult. 

\begin{figure} 
 %\resizebox{\hsize}{!}{\includegraphics{the_model/ZZ_sigmaC_vs_E_2011.ps}}
 \resizebox{\hsize}{!}{\includegraphics{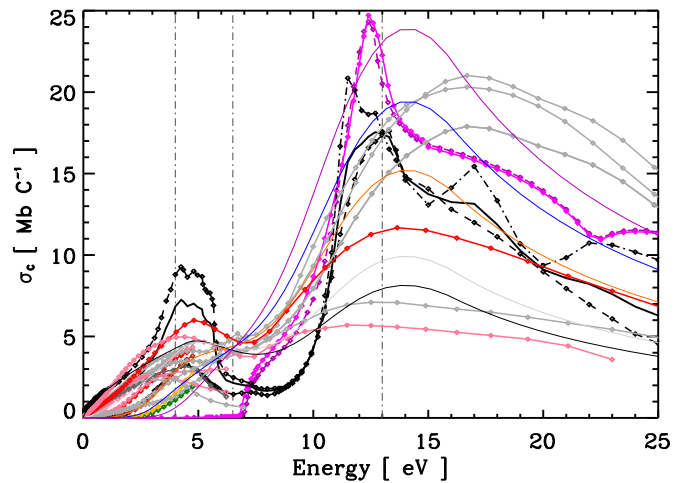}}
 \caption{The cross-section per carbon atom, $\sigma_{\rm C}$ (in Mb per C atom, 1\,Mb $= 10^{-18}$\,cm$^2$), for the optEC$_{(\rm s)}$ model (smooth curves peaking at $\sim 14$\,eV, for the extreme and three intermediate $E_{\rm g}$ cases), compared to laboratory data and other models. The vertical dot-dashed lines mark the band centres of the $\pi-\pi^\ast$, C$_6$ and $\sigma-\sigma^\ast$ bands at 4.0, 6.5 and 13\,eV, respectively.  The diamond (purple and violet lines) and graphite data (black) are shown  for comparison. }
 \label{fig_sigma_C_Mb}
\end{figure}

The behaviour of the contributing bands can also be seen in detail in Fig.~\ref{fig_sigma_C_Mb} where we plot the absorption cross-section per C atom, $\sigma_{\rm C}$ (in Mb per C atom where 1\,Mb $= 10^{-18}$\,cm$^2$), compared with the measured laboratory data and the modelled data (as per Fig.~\ref{fig_e2_all}). Overall we find that our optEC$_{(\rm s)}$ model matches the laboratory-measured data \citep[][and DDOP]{1984JAP....55..764S,1995ApJS..100..149M} and the previously-modelled data \citep{1991ApJ...377..526R,1996MNRAS.282.1321Z} extremely well. The optEC$_{(\rm s)}$ model data, and indeed all of the available laboratory and modelled data, are ``well-bracketed" by the data for the crystalline solids (diamond and graphite). Note that the optEC$_{(\rm s)}$ and other available data, in the limiting a-C:H and a-C cases, are not expected to match the diamond and graphite data, respectively, because these materials contain a significant fraction of H atoms (a-C:H) and a considerable degree of disorder in their structure (both a-C:H and a-C), which lead to much broader characteristic $\pi-\pi^\ast$ and $\sigma-\sigma^\ast$ features than in their distant crystalline cousins.  

% FIGURE 13 *********************************************************
% plot of eps_2 vs E
% *********************************************************

% FIGURE 14 *********************************************************
% plot of cross-section per C atom
% *********************************************************

%------------------------------------------------------------------
\section{Astrophysical implications}
\label{sect_astro_implications}
%------------------------------------------------------------------

Clearly the major indication, from the derived optEC$_{(\rm s)}$ model complex refractive index data, is that the evolution of amorphous hydrogenated carbon solids in the astrophysical media is a rather complicated issue. In fact the evolution of the physico-chemical properties of these materials depends upon where they come from ({\it i.e.}, what was their original band gap composition), where they have been ({\it i.e.}, what has happened to them in their past and since they were formed) and where they are ({\it i.e.}, what is happening to them and how are their properties evolving).

%------------------------------------------------------------------
\subsection{a-C(:H) processing time-scales}
A key question here is: what are the critical photo-processing time-scales that determine the evolution of a-C(:H) materials in the ISM? We now examine this is some detail but return to the issue again in a following paper where we consider the added complication of size effects. 

In Fig.~\ref{fig_dtau} we show the depth, $d_1$ (in nm), at which the optical depth, $\tau$, for photons of a given energy is unity ({\it i.e.}, $\tau = 1$), for the derived a-C(:H) optical properties. What this shows is that, for all the derived materials, photons with $E \gtrsim 7$\,eV are able to penetrate the entire particle volume for grains smaller than a few tens of nm in radius. This then also implies that the surfaces of larger grains can only be UV/EUV photo-processed to depths of the order of a few tens of nm by photons with energies greater than $\sim 7$\,eV. Such photons, above some threshold energy, will then lead to a-C(:H) grain photo-darkening (photo-processing).

% figure counter reset 
\setcounter{figure}{14}
% FIGURE 15 *********************************************************
\begin{figure} 
 \resizebox{\hsize}{!}{\includegraphics{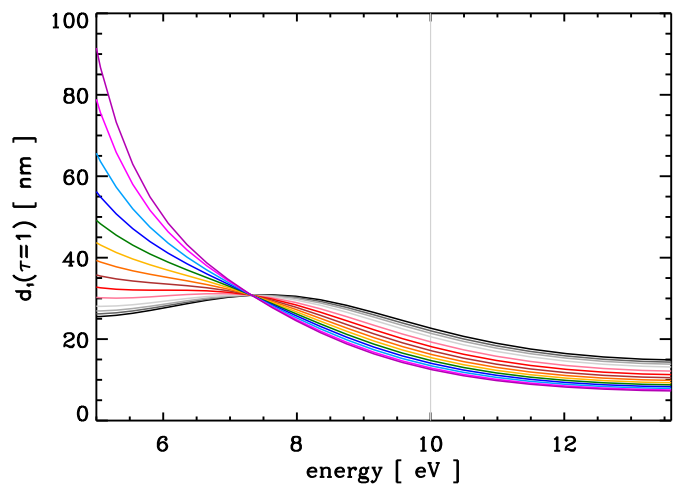}}
 \caption{The depth, $d_1$, at which $\tau =1$ for optEC$_{(\rm s)}$ model data. The vertical grey line marks the assumed lower limit for photons capable of photo-dissociating CH bonds in the assumed a-C(:H) materials.}
 \label{fig_dtau}
\end{figure}
% *********************************************************

We can estimate a UV/EUV photo-darkening rate, $\Lambda_{\rm UV,pd}$, for carbonaceous dust subject to a given radiation field using
% equation counter reset 
\setcounter{equation}{30}
\begin{equation}
\Lambda_{\rm UV,pd} = F_{\rm EUV} \ \sigma_{\rm CH-diss.} \ Q_{\rm abs}(a,E)  \ \epsilon,  
\label{eq_UV_dehyd_rate}
\end{equation}
where $F_{\rm EUV}$ is the dissociating photon flux, $\sigma_{\rm CH-diss.}$ is the CH bond photo-dissociation cross-section, $Q_{\rm abs}(a,E)$ is the particle absorption efficiency and  $\epsilon$ is a photo-darkening efficiency, which takes into account that photon absorption can also lead to heating and fluorescence \cite[{\it e.g.},][]{1990MNRAS.243..570S}. Based on CH$_4$ photo-dissociation cross-section studies in the EUV \citep{1972JChPh..57..286W,1994CPL...227..243G} we adopt a value of $\sigma_{\rm CH-diss.} = 10^{-19}$\,cm$^2$ centred at $\sim 107$\,nm (11.6\,eV) and (somewhat conservatively) assume a bandwidth of 33\,nm ($10-13.6$\,eV).  At UV/EUV wavelengths the photon flux, $F_{\rm EUV}$, appropriate to the local ISRF, can be approximated by $F_{\rm EUV} \approx 10^6$\,photons cm$^{-2}$ s$^{-1}$ nm$^{-1}$ \citep{2002ApJ...570..697H}. Integrated over the assumed 33\,nm bandwidth for the photo-dissociation cross-section, this then yields $F_{\rm EUV} \simeq 3 \times 10^7$\,photons cm$^{-2}$ s$^{-1}$. For large grains at these wavelengths $Q_{\rm abs}(a,EUV) \sim 1$ ({\it i.e.}, the short-wavelength, linear behaviour in the $\lambda Q_{\rm abs}(a,EUV)/a$ plot in  Fig.~\ref{fig_Qs_100_30nm} below). The photo-darkening (or aromatisation) time-scale for the outer $\simeq 20$\,nm of large a-C(:H) particles, $\tau_{\rm UV,pd}$, is then simply the inverse of the photo-darkening time-scale, {\it i.e.},
\begin{equation} 
\tau_{\rm UV,pd} \approx ( F_{\rm EUV} \ \sigma_{\rm CH-diss.} \ Q_{\rm abs}(a,\lambda) \ \epsilon )^{-1}
\approx \frac {10^4}{Q_{\rm abs}(a,\lambda) \ \epsilon} \ \ \ \ {\rm yr}.  
\label{eq_UV_dehyd_rate_1}
\end{equation}
If we assume a {\em per bond} photo-darkening efficiency of $\epsilon = 0.1$, rather than the {\em per grain} photofission efficiency of $10^{-3}$ adopted by \cite{1990MNRAS.243..570S}, this yields an $\approx 20$\,nm outer-layer photo-darkening time-scale of $\approx 10^5$\,yr for large a-C(:H) grains ($a > 20$\,nm) in the diffuse ISM. In photo-dissociation regions where the local radiation field can be orders of magnitude over that of the diffuse ISM ({\it e.g.}, a factor of $\simeq 10^4$ higher for the Orion PDR) these timescales will be significantly shorter. For example, in a PDR with a radiation field 100 ($10^4$) times that of the diffuse ISM, the outer-layer photo-darkening timescales reduce to $\approx 10^3$\,yr ($\approx10$\,yr) for large particles. However, in all of these cases the 
complete photo-processing of the particle is not possible because core material, at depths greater than a few tens of nm, is likely to be unaffected by photo-darkening. Thus, the large carbonaceous grains in the ISM can retain an H-rich interior, as indicated by observations \citep[{\it e.g.},][]{2004A&A...423..549D,2004A&A...423L..33D}. 

The above a-C:H UV processing can also perhaps be considered to engender a sort of band gap `velocity', {\it i.e.}, $dE_{\rm g}/dt$, which will principally depend upon the local ISRF in the diffuse ISM or given PDR and that only apply to the outer few tens of nm of large particles ($a > 20$\,nm). The band gap `velocity', $dE_{\rm g}/dt$, derived from the above, can be  expressed as  
\begin{equation}
\frac{dE_{\rm g}}{dt} \approx \Lambda_{\rm UV,pd}  \ E_{\rm g}(t) \approx 10^{-4} \ Q_{\rm abs}(a,\lambda) \ \epsilon \ E_{\rm g}(t) \  \  \  \ {\rm [\ eV\  yr^{-1}\ ]},  
\label{eq_dEgdt}
\end{equation}
which yields a band gap velocity of $\approx 10^{-4} \ \epsilon \ E_{\rm g}(t)$ eV yr$^{-1}$, where $E_{\rm g}(t)$ is the band gap at time $t$.   

However, we note that the above determination may only give an upper limit to the a-C(:H) aromatisation time-scale because it concerns only the direct photo-dissociation of CH bonds by EUV photons. To this must be added the thermal effects due to photon absorption leading to grain heating to temperatures sufficiently high for H atom loss by thermal annealing to occur. 

Thermal processing effects on amorphous hydrocarbons, and the associated kinetics, have been studied in detail by \cite{1996MNRAS.283..343D}. Using the data of \cite{1984JAP....55..764S} he showed that, for H atom loss leading to band gap closing, the reaction can be described in terms of a thermally activated process, {\it i.e.},
\begin{equation}
E_{\rm g}(t) = E^0_{\rm g} e^{-k_1 t},
\label{eq_Walt_thermal_act}
\end{equation}
where $E^0_{\rm g}$\,eV is the initial band gap (taken to be 2.2\,eV) and the rate constant $k_1 = Ae^{-\Delta H/T}$ (where $A=6.8$\,s$^{-1}$ and the activation energy $\Delta H = 8000\pm2000$\,K). With this approach \cite{1996MNRAS.283..343D} finds that, for a temperature of 350\,K (typical of the extended atmospheres of evolved stars such as IRC+10216), band gap closure ($\equiv$ aromatisation) can occur on a time-scale of $\approx 10$\,yr. Thus, it appears that a thermal processing effect could be important but only if sufficiently high temperatures can be maintained long enough for thermally-driven aromatisation to occur. For grains in the ISM or in PDRs, where grain temperatures are considerably lower than 350\,K for large particles in thermal equilibrium with the ISRF ($T \simeq 20$\,K), or can only be achieved for very short time-scales in stochastically-heated small particles (for periods of the order of a few seconds every month, {\it i.e.}, $\approx$ one millionth of their time), the aromatisation time-scales will be considerably longer than 10\,yr and probably of the order of, at least, several millions of years. 
Given that thermal processing is probably only going to be important for stochastically-heated, small a-C(:H) grains we will re-examine this process when we specifically consider grain size effects.

%------------------------------------------------------------------
\subsection{a-C(:H) evolution,  decomposition and H$_2$ formation}
%------------------------------------------------------------------

As pointed out in paper~I, there could be an initial onset of H$_2$ formation as interstitial H atoms, with binding energies of the order of $0.05-0.2$\,eV \citep{1984JAP....55..764S,1989ApPhL..54.1412S}, become mobile, combine to form H$_2$ and are lost from the solid or form internal CH bonds, a process that is rapidly quenched at higher temperatures \cite[{\it e.g.},][]{1996MNRAS.283..343D}. 

Ion and UV irradiation studies indicate that H atom loss from a-C:H stops at about $X_{\rm H}\simeq 0.05$ or $E_{\rm g} \simeq 0.22$\,eV \citep[{\it e.g.},][]{1989JAP....66.3248A,1996MCP...46...198M,2011A&A...529A.146G,2011A&A...528A..56G}, which appears to impose a minimum possible H atom content in processed a-C:H materials. If this same condition were to hold true under ISM conditions it could have important consequences for the evolution of carbonaceous dust in the ISM because it would impose a lower limit on the H atom content and hence on the band gap. In the laboratory it is possible to produce a-C materials with band gaps much smaller than this apparent ion and UV irradiation-imposed limit. Thus, in principle, the lower limit to the H atom content of hydrogenated amorphous carbons ought to be zero when, as in these laboratory experiments, the solids are made in the absence of hydrogen. However, in the ISM where H atoms are abundant the lower limit ought to apply, thus providing a natural `block' on the evolution of a-C(:H) to materials with less that $\simeq 5$\% atomic hydrogen content. Nevertheless, the hydrogen atom content could perhaps be driven lower than 5\% by dehydrogenation in intense UV photon fields that would remove most H atoms from the structure leading to further structural re-arrangement, `complete' aromatisation and to band gaps smaller than 0.2\,eV. However, such extreme processing could only occur to the very smallest a-C(:H) particles and in such environments they would be destroyed \cite[{\it e.g.},][]{1998ASPC..132...15B}. 

The formation of H$_2$, and other associated molecules, following a-C(:H) `decomposition' associated with the photo- or thermally-induced aliphatic to aromatic transformation (aromatisation) was discussed in detail in earlier work \citep[{\it e.g.}, see ][and \S~4.2 in paper ~I]{1984JAP....55..764S} and was also briefly mentioned as a possibility in the astrophysical context by \cite{2005A&A...432..895D}. It is apparent that the photo-darkening and thermal annealing time-scales, as calculated above, must be directly related to the H$_2$ formation time-scale because the inherent process is the same in each case, {\it i.e.}, the breaking of CH bonds, followed by H$_2$ loss and a structural re-arrangement to an olefinic-rich  material and eventually to a more aromatic-rich material, {\it e.g.}, 
\[
-\,{\rm CH}_{(n)}-{\rm CH}_{(m)}- \ \ \rightarrow \ \ -\,{\rm CH}_{(n-1)}={\rm CH}_{(m-1)}-  \ \ + \ \ {\rm H}_2
\]
\begin{equation}
\circ\{(-{\rm CH}_{(n)}={\rm CH}_{(m)}-)_3\} \ \ \rightarrow \ \ \circ\{-{\rm C}_6{\rm H}_{(3[n+m]-2)}-\} \ \ + \ \ {\rm H}_2
\label{eq_struct_rearrange}
\end{equation}
where the $\circ\{ S \}$ symbol is used to indicate that the enclosed cluster species, $S$, is cyclic; respectively, olefinic on the left and aromatic on the right in the lower equation.  

The H$_2$ formation rate, $R_{\rm H_2}$, arising from a-C(:H) aromatisation will be a first order reaction, {\it i.e.}, $k_f [{\rm CH}]$ where, in the absence of H atom re-incorporation into the structure, the rate coefficient, $k_f$, is the the photo-dissociation rate, $\Lambda_{\rm UV,pd}$ [s$^{-1}$]. $R_{\rm H_2}$ is then given by  
\begin{equation}
R_{\rm H_2} = k_f [{\rm CH}] = \Lambda_{\rm UV,pd} \ N_{\rm CH}(a,d_1) \ X_i(a) \ n_{\rm H} \ \ \ {\rm [cm^{-3} s^{-1}]}
\label{eq_H2_formation_rate}
\end{equation}
where $X_i(a)$ is the particle relative abundance and $n_{\rm H}$ is the proton density in the medium. In Eq.~(\ref{eq_H2_formation_rate}) we implicitly assume that every photo-dissociated CH bond leads to the formation of an H$_2$ molecule because of the resulting structural re-arrangement \citep{1989JAP....66.3248A,1996MCP...46...198M}. 
We then predict an H$_2$ formation rate arising, from a-C(:H) photo-decomposition, of the order of $R_{\rm H_2}/n_{\rm H} \approx 10^{-16}$\,s$^{-1}$ or $\approx10^{-8}$\,yr$^{-1}$. For the diffuse ISM ($n_H \simeq 10^2$\,cm$^{-3}$) this implies that an H$_2$ formation rate $R_{\rm DISM} \approx10^{-6}$\,H$_2$ molecules yr$^{-1}$\,cm$^{-3}$ would be sustainable for of the order of $\approx {\rm [C/H]} \, n_{\rm H} / R_{\rm DISM}  = 10^4$\,yr, {\it i.e.}, until all of the available C-H bonds in the a-C(:H) dust ($\approx 10^{-4}$\,$n_{\rm H}$) have been photo-dissociated, where we have assumed that about half of the available carbon (${\rm [C/H]} \simeq 2 \times 10^{-4}$) is in a-C(:H) dust and $X_{\rm H}=0.5$ ($\equiv$ C/H = 1). 

The two modes of H$_2$ formation from a-C:H particles, {\it i.e.}, interstitial H atom recombination and the photolysis of CH bonds in a-C:H, should lead to two zones of H$_2$ formation across PDRs, one closer to the molecular cloud (interstitial H atom recombination), where dust heating begins, and the other in the main PDR region where the dust is exposed to the more intense UV radiation (CH bond photolysis). 

Given that there is advection across the photodissociation fronts in PDRs, there will be a continual replenishment of the photolysable, H$_2$-producing, small carbonaceous grains as they are likely released from larger, ice-mantled aggregates.  Thus, it is apparent that H$_2$ formation via interstitial H atom recombination in a-C:H and small a-C:H grain photolysis ({\it i.e.}, via chemical decomposition) could be sustainable over rather long time-scales and thus explain the H$_2$ formation rate in regions where the dust is rather warm ($\sim 50$\,K) and gas-phase H atom physi-sorption, sticking, diffusion and surface recombination is inhibited.  It is likely that molecular hydrogen formation via a-C:H decomposition will lead to the ejection of H$_2$ in excited states, which could thus provide an additional source for the H$_2$ excitation expected and observed in PDRs and shocked regions of the ISM.  Given that small and `hot' a-C(:H) particles will likely be chemi-sputtered by atomic oxygen ({\it aka}, combustion), CO formation in excited states might be possible. 

As shown by \cite{2003ApJ...587..727M} and \cite{2011A&A...529A.146G} the effects of cosmic rays on a-C:H dust evolution is rather negligible when compared to UV photon irradiation from the ISRF. Thus, the assumed dominance of UV photo-processing on a-C(:H) dust evolution in the diffuse ISM, as discussed here, appears to be a rather reasonable approach.  

Implicit within this H$_2$ formation scenario is that there is no C-H bond regeneration in the a-C(:H) particles by H atom re-incorporation and that there is no `PAH-edge' catalytic formation of H$_2$ as has been proposed \citep[{\it e.g.},][]{2008ApJ...679..531R,2009ApJ...704..274L}. Clearly, however, if small a-C(:H) particles can catalyse H$_2$ formation, via a mechanism similar to that proposed for PAHs, then small carbonaceous grains in PDRs could provide an important source of H$_2$ here and might turn out to be the dominant source of H$_2$ if PAHs turn out to be rather transient species in PDRs \citep[as is perhaps indicated by the work of ][]{1996ApJ...472L.123S,2005A&A...435..885P,2009ASPC..414..473J}.

The effects of photo-darkening, the aromatisation $sp^3 \rightarrow sp^2$ transition, as a result of UV photo-processing, could be counterbalanced or perhaps even reversed by H atom addition (re-introduction) into the structure as a result of (energetic) collisions in the interstellar medium. Low energy (80\,K) H atom irradiation of ice-coated carbon particles shows that H atom addition leads to the formation of a 3.4\,$\mu$m band but with no evidence for the introduction of aromatic and aliphatic CH$_n$ bonds ($n = 2, 3$) into the structure until after sample warm-up \citep{2010ApJ...718..867M}. Such H atom insertion into the carbon backbone of the material must engender aromatic $C\simeq C$ bond breaking (bond energy $\simeq 5$\,eV). Using simple bond energy considerations, assuming the bond energies given in Table~\ref{table_bond_energies}, it can be shown that the direct chemical insertion of an H atom into the aromatic structure is likely to be endothermic by about 0.2\,eV, whereas insertion into the aromatic domain periphery, at a CH site (leading to $sp^3$ CH$_2$ formation), should by comparison be exothermic by about 1.3\,eV and therefore likely to be energetically favoured, but only if the appropriate energy is available for the breaking of the two adjacent aromatic C$\simeq$C bonds. This transformation appears to be at the heart of the PAH catalytic formation of H$_2$ \citep[{\it e.g.},][]{2008ApJ...679..531R,2009ApJ...704..274L}. Such hydrogenation has been widely seen in the laboratory but only where an appropriate catalyst is present. Thus, and under typical ISM conditions, it is not clear how and under what circumstances H atom insertion into aromatics, and therefore their hydrogenation and aliphatisation, {\it i.e.}, the reversal of the photo-darkening process, could occur without the presence of an appropriate catalyst. 

% TABLE
\begin{table}
\caption{Typical aromatic, olefinic and aliphatic hydrocarbon bond energies.}
\begin{center}
\begin{tabular}{lc}
                                         &                               \\[-0.35cm]
\hline
\hline
                                        &                          \\[-0.35cm]
  Bond                              &  Energy [eV]      \\[0.05cm]
\hline
                                        &                         \\[-0.35cm]
\hline
$sp^3$ aliphatic C-C                   &     4.0                \\
$sp^2$ aromatic C$\simeq$C     &     5.1                \\
$sp^2$ olefinic C=C                    &     6.2                \\
$sp^3$ or $sp^2$ C-H                &     4.3                \\
\hline
                                        &                          \\[-0.25cm]
\end{tabular}
\end{center}
\label{table_bond_energies}
\end{table}

%------------------------------------------------------------------
\subsection{a-C(:H) extinction and absorption}
%------------------------------------------------------------------

% ORIGINAL VERSION
%% FIGURE 16 *********************************************************
%\begin{figure} 
% %\resizebox{\hsize}{!}{\includegraphics{the_model/AAA_Qcalcs/ZZ_Qext-abs_vs_Eg_100nm_aCH_2011.ps}}
% %\resizebox{\hsize}{!}{\includegraphics{the_model/AAA_Qcalcs/ZZ_Qext-abs_vs_Eg_100nm_full_aCH_2011.ps}}
% \resizebox{\hsize}{!}{\includegraphics{ZZ_Qext-abs_vs_Eg_100nm_aCH_2011.ps}}
% \resizebox{\hsize}{!}{\includegraphics{ZZ_Qext-abs_vs_Eg_100nm_full_aCH_2011.ps}}
% \caption{The  optEC$_{(\rm s)}$ model extinction and absorption coefficient data, for clarity plotted as $\lambda Q_{\rm ext}/a$ (solid lines) and $\lambda Q_{\rm abs}/a$ (dashed lines), as a function of $E_{\rm g}$ and wavelength  for 100\,nm radius particles. From top to bottom on the right the dashed blue lines indicate wavelength dependencies, $\lambda^{-\beta}$, where $\beta = 1$, 1.5 and 2.}
% \label{fig_Qs_100_30nm}
%\end{figure}
%% *********************************************************
%
In Fig.~\ref{fig_Qs_100_30nm} we show the optEC$_{(\rm s)}$ model extinction and absorption efficiency factors, plotted as $\lambda Q_{\rm ext}/a$ and $\lambda Q_{\rm abs}/a$, respectively, as a function of wavelength  for 100\,nm radius particles. These data can be used to predict general evolutionary trends in the a-C(:H) optical properties that have observable consequences. We outline these changes, as a function of decreasing band gap (on going from 2.67 to -0.1\,eV), namely:
\begin{itemize}
  \item Decreasing UV extinction/absorption (peaking at $\sim 0.3\,\mu$m) gives way to increased visible absorption (peaking at about $0.7-0.8\,\mu$m). This is just a reflection of the trend that is clearly seen in Fig.~\ref{fig_e2_limited}, for example. 
  \item Longward of the mid-IR bands in the $\sim 3-10\,\mu$m region $Q_{\rm ext}$  can increase by up to $4-5$ orders of magnitude,  for the same grain size (mass). 
  \item At near- to mid-IR wavelengths both $Q_{\rm ext}$ and $Q_{\rm abs}$ increase, and $Q_{\rm ext}$ is increasingly dominated by absorption. 
  \item At visible to mid-IR wavelengths ($0.4 \leqslant \lambda \leqslant3\,\mu$m) $Q_{\rm ext}$ is dominated by scattering, $Q_{\rm sca}$, for H-rich materials. 
  \item The near- to mid-IR CC and CH bands decrease in intensity, and eventually `disappear'.
  \item At FIR-mm wavelengths ({\it e.g.}, $100\,\mu$m), where $Q_{\rm ext} = Q_{\rm abs} \propto \lambda^{-\beta}$,  $\beta \sim 2$ for $E_{\rm g} = 2.67 - 1.5$\,eV, a value determined by the long-wavelength wings of the Drude profiles adopted for the IR band profiles.  As the band gap further reduces ($E_{\rm g} \sim 1.5$\,eV $\rightarrow -0.1$\,eV) $\beta$ first increases to $\approx 3$ for $E_{\rm g} = 1-1.25$\,eV, drops to $\approx 1.3$ at $E_{\rm g} = 0.1$\,eV, and then rises to 1.7 for the smallest band gap material with $E_{\rm g} = -0.1$\,eV. 
\end{itemize} 
We  note that the evolution of a-C(:H) materials could be responsible for variations in the slope of the FIR-cm emissivity of dust in the ISM (see Figs.~\ref{fig_long_wave_Q} and \ref{fig_beta_Eg}).  The black line in Fig.~\ref{fig_beta_Eg} indicates the expected behaviour of $\beta$ based on the band gap dependence of $\gamma$, given by Eq.~(\ref{eq_k_lowE}), and its imprint on the long wavelength behaviour through $Q_{\rm abs} \propto \lambda^{-\beta}$ and  $Q_{\rm abs}  \propto k / \lambda$ \citep{2011A&A...528A..98J}, with $k \propto  \lambda^{-\gamma}$ we then have  $\beta = (\gamma + 1)$. The vertical dashed line indicates the limiting value for $X_{\rm H}=0.05$ ($\equiv E_{\rm g} = 0.22$\,eV) and the horizontal band indicates the typical diffuse ISM values $\beta = 1.8 \pm 0.1$ \citep[][see below]{2011A&A...536A..25P}. 

The observed variations in $\beta$ at FIR-mm  wavelengths, {\it e.g.}, as seen in studies using {\em IRAS}, {\em PRONAOS}, {\em Archaeops},  {\em Herschel} and {\em Planck} data,  could be due, in part, to an evolution of the ISM carbonaceous dust properties as it responds to changes in its environment (through the effects of UV photolysis and/or accretion on grain surfaces). Principally, we note that the above-described behaviours are only valid for particles larger than 100\,nm, {\it i.e.}, essentially only for the largest interstellar grains, or bulk materials, which ought then to be reflected in the recent {\em Herschel} and {\em Planck} space observatory observations.  In paper~I we showed that the observed diffuse ISM absorption spectra in the $3-4\,\mu$m region indicate H atom fractions $\geqslant 0.57$ ($\equiv E_{\rm g} \geqslant 2.5$\,eV), implying (see Fig.~\ref{fig_beta_Eg}) an emissivity $\beta \simeq 2.1$ at 100$\,\mu$m and $\beta \simeq 2.0$ for 300$\,\mu$m to 1\,mm wavelengths, which is close to that indicated for diffuse ISM dust ($\beta \simeq 1.8\pm0.1$) from analyses of the most recent Planck data coupled with {\em Herschel} and {\em IRAS} data \citep{2011A&A...536A..25P}, and also close to that for ``astronomical silicate'' \citep{1984ApJ...285...89D}.  A better fit to the observed value of $\beta = 1.8$ could perhaps be explained by a combination of a-C(:H) and amorphous silicate, a-Sil, materials, {\it i.e.}, 
\begin{equation}
\beta_{\rm obs.} \simeq f\beta_{\rm a-C:H} + (1-f)\beta_{\rm a-Sil}, 
\label{ }
\end{equation} 
where $f$ is the fractional contribution of a-C(:H) materials to $\beta$. 
If we assume that the carbonaceous component on the large amorphous silicate grains, with $\beta_{\rm a-Sil} \simeq 2$, is in the form of thin mantles of accreted a-C(:H) or coagulated a-C(:H) particles \citep[{\it e.g.},][]{2011A&A...528A..96K,2011A&A...Koehler_submitted} then $\beta_{\rm a-C(:H)} \simeq 1.4$ for UV-photolysed carbonaceous mantles a few tens of nm thick ({\it i.e.}, $E_{\rm g} \sim 0.2$\,eV, see Fig.~\ref{fig_beta_Eg}). In this case we have  
\begin{equation}
f \simeq \frac{(\beta_{\rm a-Sil} - \beta_{\rm obs.})}{(\beta_{\rm a-Sil} - \beta_{\rm a-C(:H)})} = \frac{(2.0-1.8)}{(2.0-1.4)} = \frac{1}{3},  
\label{eq_beta_fit}
\end{equation} 
which implies that large silicate grains make up the bulk of the FIR-mm dust emissivity in the diffuse ISM but that they must also contain a significant H-poor carbonaceous component.  

As indicated by ion and UV irradiation studies \citep{1989JAP....66.3248A,1996MCP...46...198M,2011A&A...529A.146G,2011A&A...528A..56G} there may, except in the case of very extreme radiation fields, be a natural `block' on the band gap of a-C(:H) materials in the ISM at about $0.2-0.25$\,eV ($X_{\rm H} \approx 0.04-0.06$), which would imply a lower limit to the carbon dust emissivity index, $\beta$, of about $1.4-1.5$. Only in the most extreme radiation fields, where dehydrogenation can be driven to `completion' would $E_{\rm g}$ evolve further and, as pointed out earlier, this can only occur to the smallest carbonaceous grains, which we do not consider here but leave  to a more detailed analysis in a follow-up paper.

However, it should be remembered that the observed slope of the ISM dust emissivity at these wavelengths {\em will be} determined by an appropriate mix of materials that emit significantly (carbonaceous, silicate, {\it etc.}). 
The size-dependence of the predicted and observable dust extinction and emission properties will be presented in follow-up papers.

For a-C:H materials with band gaps larger than $\sim 1$\,eV Fig.~\ref{fig_beta_Eg} now shows  a weaker effect and slightly lower absolute values of $\beta$ at FIR-mm wavelengths. The changes are most marked for $E_{\rm g} = 1-1.5$\,eV materials and for $\beta$ in the $\lambda \sim 100\,\mu$m region. Hence, the last sentence of the last itemized point should read:
\begin{itemize}
  \item \ldots As the band gap further reduces ($E_{\rm g} \sim 1.5$\,eV $\rightarrow -0.1$\,eV) $\beta$ first increases to $\approx 2.7$ for $E_{\rm g} = 1$\,eV, drops to $\approx 1.3$ at $E_{\rm g} = 0.1$\,eV, and then rises to 1.7 for the smallest band gap material with $E_{\rm g} = -0.1$\,eV.
\end{itemize}
In the upper plot in Fig.~\ref{fig_beta_Eg} the same $\beta(E_{\rm g})$ data is shown plotted in logarithmic form over a narrower range of $E_{\rm g}$ in order to allow a comparison with $\beta$ {\it vs.} $T_{\rm dust}$ plots, where $T_{\rm dust}$ could be considered a proxy for $E_{\rm g}$, implying that higher band gap materials are to be found in lower dust temperature-higher extinction regions, as might be expected by the accretion of increasingly hydrogen-rich a-C:H materials in denser regions of the ISM \citep[{\it e.g.},][]{1990QJRAS..31..567J}. 

% figure counter reset 
\setcounter{figure}{15}
% *********************************************************
% {\bf Paper~II, Fig.~16:} 
\begin{figure} 
 \resizebox{\hsize}{!}{\includegraphics{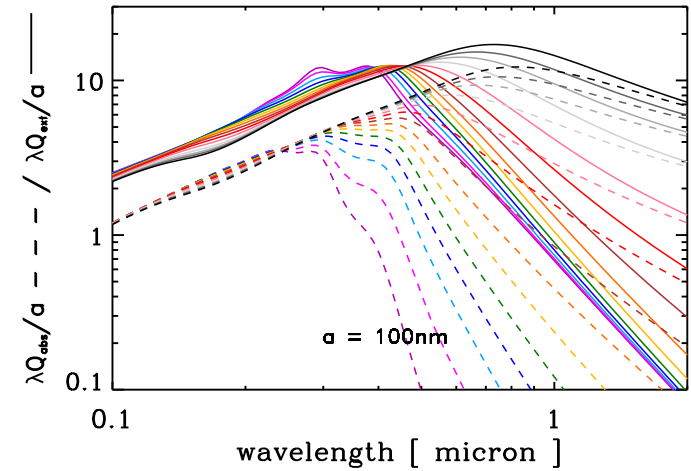}}
 \resizebox{\hsize}{!}{\includegraphics{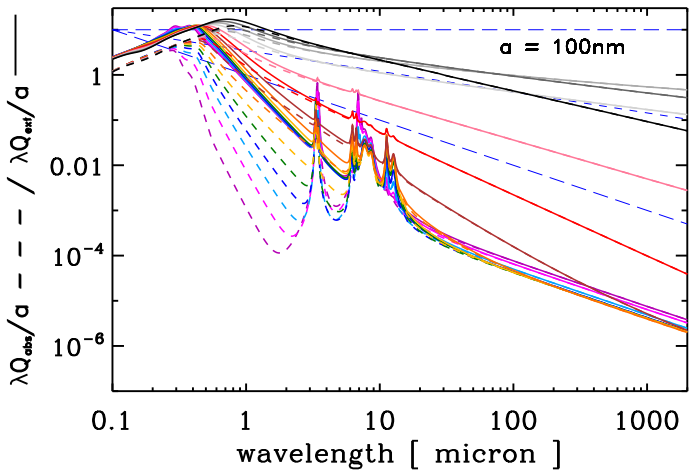}}
 \caption{The  optEC$_{(\rm s)}$ model extinction and absorption coefficient data, for clarity plotted as $\lambda Q_{\rm ext}/a$ (solid lines) and $\lambda Q_{\rm abs}/a$ (dashed lines), as a function of $E_{\rm g}$ and wavelength  for 100\,nm radius particles. From top to bottom on the right the dashed blue lines indicate wavelength dependencies, $\lambda^{-\beta}$, where $\beta = 1$, 1.5 and 2.}
 \label{fig_Qs_100_30nm}
\end{figure}
% *********************************************************

% *********************************************************
% {\bf Paper~II, Fig.~17:} 
\begin{figure} 
  \resizebox{\hsize}{!}{\includegraphics{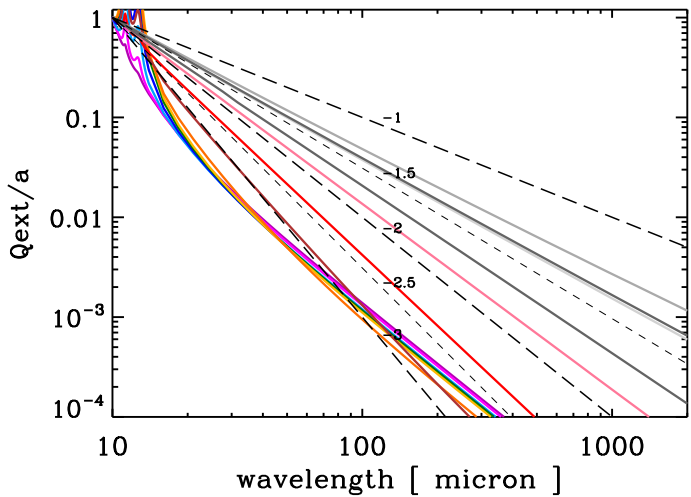}}
 \caption{The  optEC$_{(\rm s)}$ model extinction/absorption coefficient data at FIR-mm wavelengths, plotted as $Q_{\rm ext}/a$ and normalised at 10\,$\mu$m, as a function of $E_{\rm g}$  for 100\,nm radius particles. From top to bottom, the grey dashed lines indicate wavelength dependencies, $\lambda^{-\beta}$, where $\beta = 1$, 1.5, 2, 2.5  and 3. 
 %For reference the vertical grey line indicates the central position of the UV extinction bump at 217\,nm. 
 }
 \label{fig_long_wave_Q}
\end{figure}
% *********************************************************

% *********************************************************
% {\bf Paper~II, Fig.~18:} 
\begin{figure} 
 %\resizebox{\hsize}{!}{\includegraphics{the_model/AAA_Qcalcs/ZZ_FIRmm_slope_vs_Eg_aCH_2011.ps}}
 \resizebox{\hsize}{!}{\includegraphics{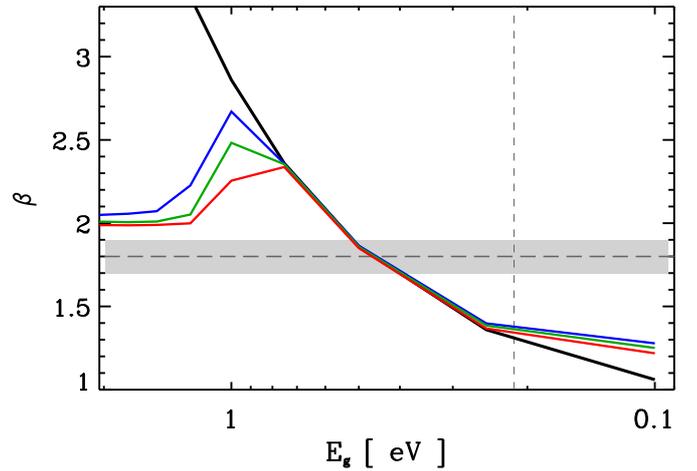}}
 \resizebox{\hsize}{!}{\includegraphics{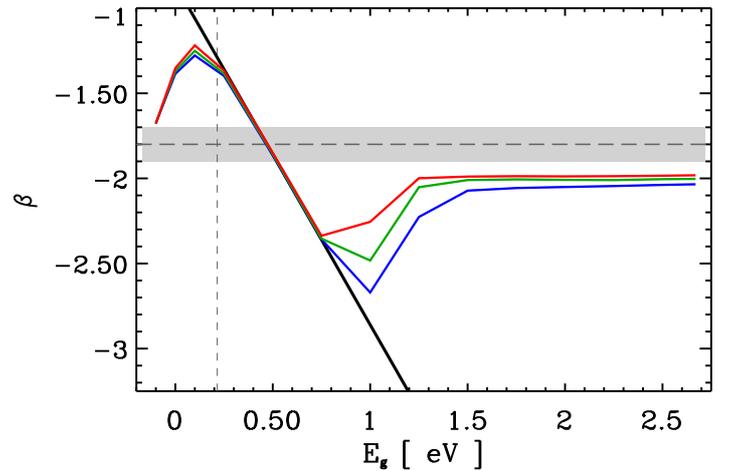}}
 \caption{The predicted emissivity slope, $\beta$, at FIR-mm wavelengths as a function of $E_{\rm g}$  for 100\,nm radius particles: blue (at 100\,$\mu$m), green (at 300\,$\mu$m) and red (at 1\,mm); lower, middle and upper lines, respectively.}
 \label{fig_beta_Eg}
\end{figure}
% *********************************************************

%------------------------------------------------------------------
\subsection{Scattering by a-C(:H) grains at mid-IR wavelengths}
%------------------------------------------------------------------

It has recently been proposed \citep{2010Sci...329.1622P,2010A&A...511A...9S} that the observed cloud ``emission'' seen at 3.6 and 4.5\,$\mu$m in the Spitzer IRAC bands can be explained by scattering from `larger' grains in the outer regions of molecular clouds and that this represents direct evidence for grain growth in these regions. 

\begin{figure} 
  \resizebox{\hsize}{!}{\includegraphics{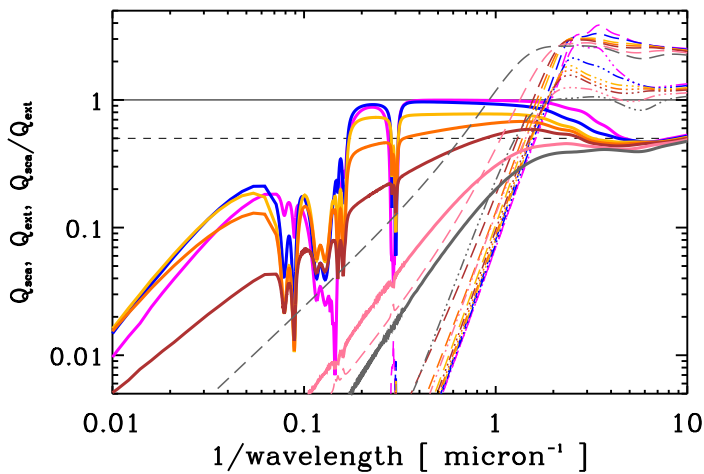}}
 \caption{$Q_{\rm ext}$ (dashed), $Q_{\rm sca}$ (triple-dotted-dashed) and $Q_{\rm sca}/Q_{\rm ext}$ (solid lines) as a function of $E_{\rm g}$ (0, 0.5, 1.0, 1.25, 1.5, 2.0 and 2.5\,eV, from bottom to top) for grains of radius $a = 100$\,nm.}
 \label{fig_QscaQext}
\end{figure}

% ORIGINAL VERSION
%In Fig.~\ref{fig_QscaQext} we show $Q_{\rm ext}$ and $Q_{\rm sca}$, and also the ratio of scattering to extinction $Q_{\rm sca}/Q_{\rm ext}$, as a function of $E_{\rm g}$ for 100\,nm grains. What this figure shows is, that for a fixed grain size,  large band gap a-C(:H) materials ($E_{\rm g} \geqslant 2$\,eV) exhibit almost pure scattering behaviour at wavelengths from 0.3 to 10\,$\mu$m ($0.1-3\,\mu$m$^{-1}$), with sharp downturns at the positions of the strong IR C$-$C and C$-$H resonances, and plausibly significant scattering ($Q_{\rm sca}/Q_{\rm ext} > 0.5$) out to wavelengths as short as $\simeq 0.2\,\mu$m and as long as $\simeq 30\,\mu$m. We note that the scattering fraction (see Fig.~\ref{fig_QscaQext}) decreases as the band gap of the material decreases but is still significant for materials with band gaps as low as 1.5\,eV. 
% NEW UPDATED CORRIGENDUM VERSION
In Fig.~\ref{fig_QscaQext} we show $Q_{\rm ext}$ and $Q_{\rm sca}$, and also the ratio of scattering to extinction $Q_{\rm sca}/Q_{\rm ext}$, as a function of $E_{\rm g}$ for 100\,nm grains. What this figure shows is, that for a fixed grain size,  large band gap a-C(:H) materials ($E_{\rm g} \geqslant 2$\,eV) exhibit almost pure scattering behaviour at wavelengths from 0.5 to 5\,$\mu$m ($0.2-2\,\mu$m$^{-1}$), with sharp downturns at the positions of the strong IR C$-$H resonances in the $3\,\mu$m region. We note that the scattering fraction (see Fig.~\ref{fig_QscaQext}) decreases as the band gap of the material decreases but is still significant for materials with band gaps as low as 1.5\,eV. 
Thus, it is possible that the ``coreshine effect'' could, in part, be explained by the presence of H-rich carbonaceous grains in the outer regions of molecular clouds. This would imply that the observed ``coreshine'' could be a manifestation of the material properties rather than due to the effects of grain growth.  If this is indeed an important contributor to the observed ``coreshine'', then this implies that the effect may have its origin in the addition of ``fresh'' H-rich, carbonaceous material to the grains, possibly via accretion in the interiors of molecular clouds. Thus, dust scattering at near-IR wavelengths could be a key tracer of the evolution of the dust optical properties resulting from the accretion of H-rich carbonaceous material onto grains in molecular clouds. 
 
% ORIGINAL
%% FIGURE 19 *********************************************************
%% plot of Qsca/Qext
%\begin{figure} 
% %\resizebox{\hsize}{!}{\includegraphics{the_model/AAA_Qcalcs/ZZ_Qext-abs_vs_Eg_100nm_Qexscab_aCH_2011.ps}}
% \resizebox{\hsize}{!}{\includegraphics{ZZ_Qext-abs_vs_Eg_100nm_Qexscab_aCH_2011.ps}}
% \caption{$Q_{\rm ext}$ (dashed), $Q_{\rm sca}$ (triple-dotted-dashed) and $Q_{\rm sca}/Q_{\rm ext}$ (solid lines) as a function of $E_{\rm g}$ (0, 0.5, 1.0, 1.25, 1.5, 2.0 and 2.5\,eV, from bottom to top) for grains of radius $a = 100$\,nm.}
% \label{fig_QscaQext}
%\end{figure}
%% *********************************************************
% NEW
% *********************************************************
% plot of Qsca/Qext - {\bf Paper~II, Fig.~19:} 
% *********************************************************

%------------------------------------------------------------------
\subsection{Applications to Solar System studies}
%------------------------------------------------------------------

\begin{figure} 
 %\resizebox{\hsize}{!}{\includegraphics{the_model/ZZ_reflectance_aCH_2011.ps}}
 \resizebox{\hsize}{!}{\includegraphics{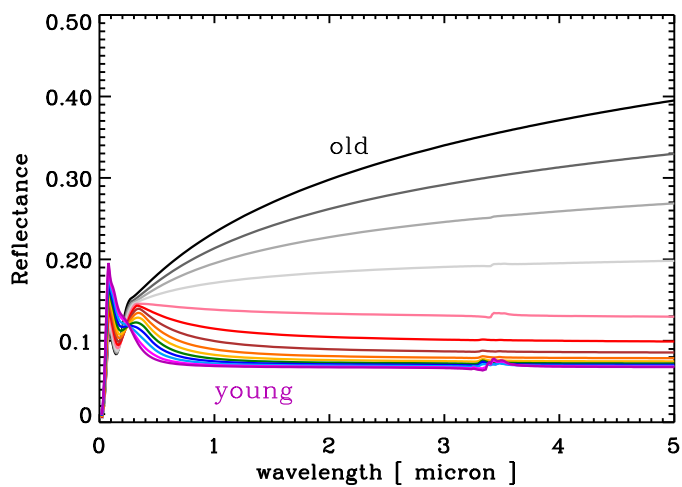}}
 \caption{The eRCN/DG model predicted reflectance spectrum as a function of $E_{\rm g}$ from `young' materials, assumed to be H-rich to `old' materials assumed to be H-poor.}
 \label{fig_reflectance}
\end{figure}

In a recent paper by \cite{2011A&A...533A..98D}  the nature of carbonaceous materials in the Solar System, and in particular the question as to whether their evolution is determined by nature or nurture, was studied and reviewed. In such studies it is clear that it is the albedo and/or reflectance, $R$, of the relevant analogue materials, generally in the $0.5-2.5\,\mu$m wavelength region, that is of primary importance. The reflectance of a material can be calculated from the complex refractive index via 
\begin{equation}
R =  \frac{\{ (n-1)^2 + k\}}{\{(n+1)^2 + k\}}. 
\label{eq_reflectance}
\end{equation} 
In Fig.~\ref{fig_reflectance} we therefore plot the reflectance for the optEC$_{(\rm s)}$ model data. These data indicate clear tendencies with material evolution. If we label wide band gap, H-rich a-C(:H) materials as `young' or little processed and small band gap, H-poor materials as evolved or `old' then the trend from young to old is for an increase in the slope or reddening of the material over the $1-3\,\mu$m region. This trend appears to be consistent with the  general evolution required to explain the history of solar system bodies. However, it must be remembered that in these observations the carbonaceous material will be mixed with at least silicate and ice components, which each in their own way affect the observed albedos for asteroid and cometary bodies. 

We note that the predicted effects of the C$-$C and C$-$H resonances in the $\approx 3.5\,\mu$m region are very weak in reflectance.  
 
% FIGURE 20 *********************************************************
% plot of a-C:H / a-C reflectivity
% *********************************************************

%------------------------------------------------------------------
\section{Predictions of the coupled eRCN/DG-optEC$_(s)$ model}
\label{sect_predictions}
%------------------------------------------------------------------

Here we summarise the major and specific predictions arising from our study of the evolution of hydrogenated amorphous carbons, as exemplified by our use of the derived optEC$_{(s)}$ model optical constants, they are: 
\begin{enumerate}
  \item  The electrical conductivity of a-C(:H) is expected to be low, and much lower than graphite, at the temperatures typical of interstellar grains, which implies that the thermal conductivity will also be low. 
  \item  The expected UV photo-processing rate for a-C(:H) particles indicates that the large grains ($a \sim 100$\,nm) should remain predominantly aliphatic- and H-rich, while their smaller brethren ($a \lesssim 10$\,nm) should be olefinic- and aromatic-rich.  
  \item The photo-processing of a-C(:H) particles should be an important source for (excited) H$_2$ formation and chemi-sputtering by O atoms (slow combustion) could be a source of (excited) CO formation. 
  \item Two zones of H$_2$ formation from a-C:H particles ought to be apparent across PDR regions: close to the molecular cloud (due interstitial H atom recombination) and in the main PDR region (due to CH bond photolysis). 
  \item As small a-C:H grains are progressively aromatised,  and the band gap decreases,  there will be a (per unit dust mass) concomitant:
  \begin{itemize}
  \item decrease in the UV extinction, 
  \item increase in the FIR-mm extinction and emission, 
  \item decrease in scattering at mid-IR wavelengths, 
  \item transition from CH- to CC-dominated IR modes, 
  \item increase in the FIR-mm emissivity for $E_{\rm g} = 0-0.5$\,eV ($\beta \simeq 1.3-1.8$) and
  \item decrease in the FIR-mm emissivity for $E_{\rm g} = 0.5-1.5$\,eV ($\beta \simeq 1.8-3.1$).
  \end{itemize} 
  \item The observed coreshine is due, at least in part, to the presence of H-rich carbonaceous matter accreted onto grains in molecular clouds, which is subsequently photo-processed to H-poorer material once in the diffuse ISM.  
  \item Carbonaceous materials on Solar System objects should show increased reddening in the $1-5\,\mu$m region as they evolve from `young', H-rich to `old', H-poor matter.  
\end{enumerate}

%------------------------------------------------------------------
\section{Limitations of the optEC$_{(s)}$ model}
\label{sect_limitations}
%------------------------------------------------------------------

In addition to the implicit limitations imported into the optEC$_{(s)}$ model from the eRCN/DG model, as pointed out in paper~I (\S\,4.4), we now consider the specific limitations to this version of the optEC$_{(s)}$ model. These include:
\begin{itemize}
  \item The possibly important contribution of as-yet uncharacterised long wavelength bands to the optical properties. Such bands are clearly indicated in the DDOP data used here. This could be a major drawback of the current data, which is unavoidable until extensive new laboratory data for carbonaceous materials at long wavelengths ($\lambda \geqslant 5\,\mu$m) become available. However, given the complexity of these materials this is not expected to be an easy task. 
  \item The possibly important effects of size on the determined optical properties, which will be treated in a follow-up paper. 
\end{itemize}

%------------------------------------------------------------------
\section{Concluding summary}
\label{sect_conclusion}
%------------------------------------------------------------------

In this paper we have further developed and extended the eRCN/DG model, presented in paper~I, to include the calculation of the complex refractive index of this suite of amorphous hydrocarbon solids as a function of the key defining parameter, the material band gap. A study and an anlaysis of the available laboratory and modelled data for a-C(:H) materials was made using a ``Tauc analysis'' to determined their behaviour as a function of the so-determined Tauc gap, $E_{\rm g}$, and the evolution of that gap and the material properties with photo- and/or thermal-annealing.

We present a model, {\bf o}ptical property {\bf p}rediction {\bf t}ool for the {\bf E}volution of {\bf C}arbonaceous {\bf (s)}olids, {\bf optEC$_{(s)}$}, that can be used to determine optical properties of carbonaceous materials, a-C(:H), from EUV to cm wavelengths as a function of $E_{\rm g}$. To this end we provide ASCII tables of the real and imaginary parts of the complex refractive index, as a function of $E_{\rm g}$, that can serve to study the full implications for the evolution of carbonaceous materials in the ISM. 

Clearly the optEC$_{(s)}$ model data is not intended to replace laboratory measured data but it should rather be seen to serve as a `place holder' until such time as the relevant data can be measured for the wide range of low-temperatre a-C(:H) materials and over the full energy range required for astrophysical dust modelling. Such work is essential to further our understanding of the long-wavelength modes that are so critical in determining the FIR-mm emission behaviour of amorphous carbon materials.

The particular strength of optEC$_{(s)}$ model is that it provides a test-able, and readily modifiable, data platform for the determination of the optical properties of amorphous hydrocarbon materials where data is currently rather incomplete. The model is a predictor of how the structural, spectral and optical properties of carbonaceous dust evolve in the ISM and in the Solar System as they respond to their local environment (see \S\,\ref{sect_predictions}).  An initial analysis of the applicability of the optEC$_{(s)}$ model shows its utility in the interpretation of many astronomical observations, namely:
\begin{itemize}
  \item a continuity in the optical properties of a-C(:H) solids from large to small band gap and from the largest interstellar grains down to particles with radii of the order of a fraction of a nm (to be fully presented in a follow-up paper), 
  \item large carbonaceous grains are H-rich and predominantly aliphatic, while their smaller cousins are H-poor and aromatic-rich, 
  \item carbonaceous grain decomposition/aromatisation can contribute to H$_2$ formation in the ISM, albeit perhaps on limited time-scales, and to the formation of ``free flying'' small hydrocarbon molecules and ``PAHs'', 
  \item carbonaceous dust evolution could be playing a key role in the observed variations in interstellar extinction and emission by dust, 
  \item scattering by H-rich carbonaceous grains at mid-IR wavelengths might be an alternative to the dust-growth explanation of the observed ``coreshine'' seen in the outer regions of molecular clouds, and
  \item carbonaceous dust evolution is consistent with the major trends seen in ``organic matter'' evolution in Solar System studies. 
\end{itemize}
Further the optEC$_{(s)}$ model makes a number of predictions that, depend principally upon the material band gap, $E_{\rm g}$, and will hopefully prove test-able with laboratory data and astrophysical observations in the near future, these include:
\begin{itemize}
  \item the IR spectral behaviour of carbonaceous materials with band intensities decreasing with decreasing band gap, 
  \item a ``rapid'' aromatisation of small carbonaceous particles in PDR regions; with a time-scale of the order of a few million years for nm-sized particles in the diffuse ISM, decreasing to of the order of hundreds of years in PDRs. 
  \item a ``transient'' steepening ($\beta \simeq 1.8-3.1$ for $E_{\rm g} = 0.5$ to $1.5$\,eV), and then a flattening of the carbonaceous dust emissivity at FIR-mm wavelengths ($\beta \simeq 1.4-1.8$ for $E_{\rm g} = -0.5$ to 0.5\,eV) as H-rich material ($\beta \simeq 2$) 
is dehydrogenated and aromatised, which may have a natural `block' at a value of $1.4-1.5$ due to a lower limit to the band gap ($E_{\rm g} \simeq 0.2$\,eV) imposed by the results of ion and UV irradiation studies, and  \item enhanced scattering at mid-IR wavelengths for H-rich carbonaceous materials ($E_{\rm g} \gtrsim 1$\,eV).
\end{itemize}
We leave a full elucidation of the optEC$_{(s)}$ predictions for a follow-up paper, where the crucial effects of particle size on the physical properties are evaluated.

It is hoped that the tabulated refractive index data provided here, and those to follow for the size-dependent behaviours, will be of use to the community in providing and adaptable tool that can be further tuned as and when better laboratory data become available.  However, it is clear that new laboratory data on a-C(:H) solids is urgently needed in order to assess and quantify the likely very important contribution of long wavelength transitions to the mm-cm emissivity of carbonaceous dust in the ISM.

%%%%%%%%%%%%%%%%%%%%%%%%%%%%%%%%%%%%%%%%%

\begin{acknowledgements}
I greatly indebted to Laurent Verstraete for many stimulating discussions on the optical and electronic properties of carbonaceous particles. I would also like to thank the referee, Walt Duley, for many valuable suggestions. 
This research was, in part, made possible through the financial support of the Agence National de la Recherche (ANR) through the program {\it Cold Dust} (ANR-07-BLAN-0364-01).
\end{acknowledgements}

%%%%%%%%%%%%%%%%%%%%%%%%%%%%%%%%%%%%%%%%%

\bibliographystyle{aa} 
\bibliography{biblio_HAC.bib} 

\Online

%%%%%%%%%%%%%%%%%%%%%%%%%%%%%%%%%%%%%%%%%

% APPENDICES

\appendix
\clearpage

\section{Tauc analysis of the avaialble data}
\label{app_full_Tauc_analysis} 

In \S~\ref{opt_props} we showed how optical data can be fitted, and the band gap (Tauc gap) determined, using the linear portion of a Tauc plot, {\it i.e.}. $(\alpha E)^{0.5}$ {\it vs.} $E$. At higher and lower energies the behaviour deviates from this linear trend but can be extended to lower energies with the addition of an exponential Urbach tail contribution (see \S\,\ref{sect_Tauc_fits}). At higher energies the data can be fitted, for aesthetic purposes only (see \S\,\ref{sect_Tauc_fits}), with 
\begin{equation}
(\alpha E)^{1/2}  = \surd B(E-E_{\rm g}) - \left( B\ E\ {\rm exp} \left[ \frac{-(E-E_{\rm UV})^2}{2 \sigma_{\rm UV}^2} \right] \right)^{1/2},
\label{eq_gaussian_tail}
\end{equation}
where the left-hand term is the linear portion and the right hand is an ``inversed gaussian tail'' fit to the data.
Table~\ref{table_Tauc_params_0} lists the adopted values for the parameters $E_{\rm UV}$ and $\sigma_{\rm UV}$.

% TABLE
\begin{table}
\caption{Parameters used in Eq.~(\ref{eq_gaussian_tail}) for the ``aesthetic'' fit to the UV portion of the Tauc plots for optical data for the given sources.}
\begin{center}
\begin{tabular}{lcccccc}
                                         &       &       &        &       &                   &                    \\[-0.35cm]
\hline
\hline                                         &                             &                    &            &       &        &             \\[-0.25cm]

\multicolumn{7}{l}{\cite{1984JAP....55..764S}}  \\[0.1cm]
$T_{\rm c}$ [$^\circ$C]        &   250       &   350       &   450       &   550       &   650       &   750     \\
$E_{\rm UV}$ [eV]                           &   10.0     &    10.0       &   10.0     &   10.0      &   10.0      &   10.0     \\
$\sigma_{\rm UV}$ [eV]                  &   1.0     &    1.0       &   1.7     &    1.7     &    1.6     &   1.4      \\[0.1cm]

\hline
                                         &                             &                    &            &       &        &             \\[-0.25cm]
\multicolumn{7}{l}{\cite{1991ApJ...377..526R}}  \\[0.1cm]       
Material                               &    AC1   &      BE1     &    FC21PS  &  HAPS     &        &         \\
$E_{\rm UV}$ [eV]                           &    9.8    &    8.8    &    7.8     &   7.8          &        &         \\
$\sigma_{\rm UV}$ [eV]                   &     1.5   &     1.3    &    1.15    &    0.9         &        &         \\[0.1cm]
        
\hline
                                         &                             &                    &            &       &        &             \\[-0.25cm]
\multicolumn{7}{l}{\cite{1995ApJS..100..149M}}  \\[0.1cm]   
$T_{\rm c}$ [$^\circ$C]        &       20      &    415     &    600     &   800    &        &         \\
$E_{\rm UV}$ [eV]                           &     9.8       &     8.8    &         7.8        &         7.8 &        &         \\
$\sigma_{\rm UV}$ [eV]                   &     1.5      &    1.3        &         1.15         &        0.9  &        &         \\[0.1cm]
           
\hline
                                         &                             &                    &            &       &        &             \\[-0.25cm]
\multicolumn{7}{l}{Jena DDOP}  \\[0.1cm]  
$T_{\rm c}$ [$^\circ$C]        &       400    &    600    &   800     &   1000    &        &         \\
$E_{\rm UV}$ [eV]                           &     9.8      &       9.0       &      9.0       &      9.0  &        &         \\
$\sigma_{\rm UV}$ [eV]                   &      1.9      &       1.5       &      1.5       &      1.5  &        &         \\[0.1cm]
             
\hline
                                         &                             &                    &            &       &        &             \\[-0.25cm]
\multicolumn{7}{l}{\cite{1996MNRAS.282.1321Z}}  \\[0.1cm]  
Material                                &  BE   &     ACH2     &   ACAR     &       &        &         \\
$E_{\rm UV}$ [eV]                           &   7.0      &       10.5      &       7.0  &   &       &         \\
$\sigma_{\rm UV}$ [eV]                   &   1.44     &        1.7      &       0.75  &   &       &         \\[0.1cm]
             
\hline
\hline
                       &                  &       &        &                 &                   &                    \\[-0.25cm]
\end{tabular}
\begin{list}{}{}
\item[] Notes: - 
The characteristic temperature parameter, $T_{\rm c}$ indicates the relevant material deposition or annealing temperature. 
\end{list}
\end{center}
\label{table_Tauc_params_0}
\end{table}

\begin{figure}
 %\resizebox{\hsize}{!}{\includegraphics{the_model/ZZ_Tauc_plots_Mennella_energy_2011.ps}}
 %\resizebox{\hsize}{!}{\includegraphics{the_model/ZZ_Tauc_plots_Mennella_wavelength_2011.ps}}
 \resizebox{\hsize}{!}{\includegraphics{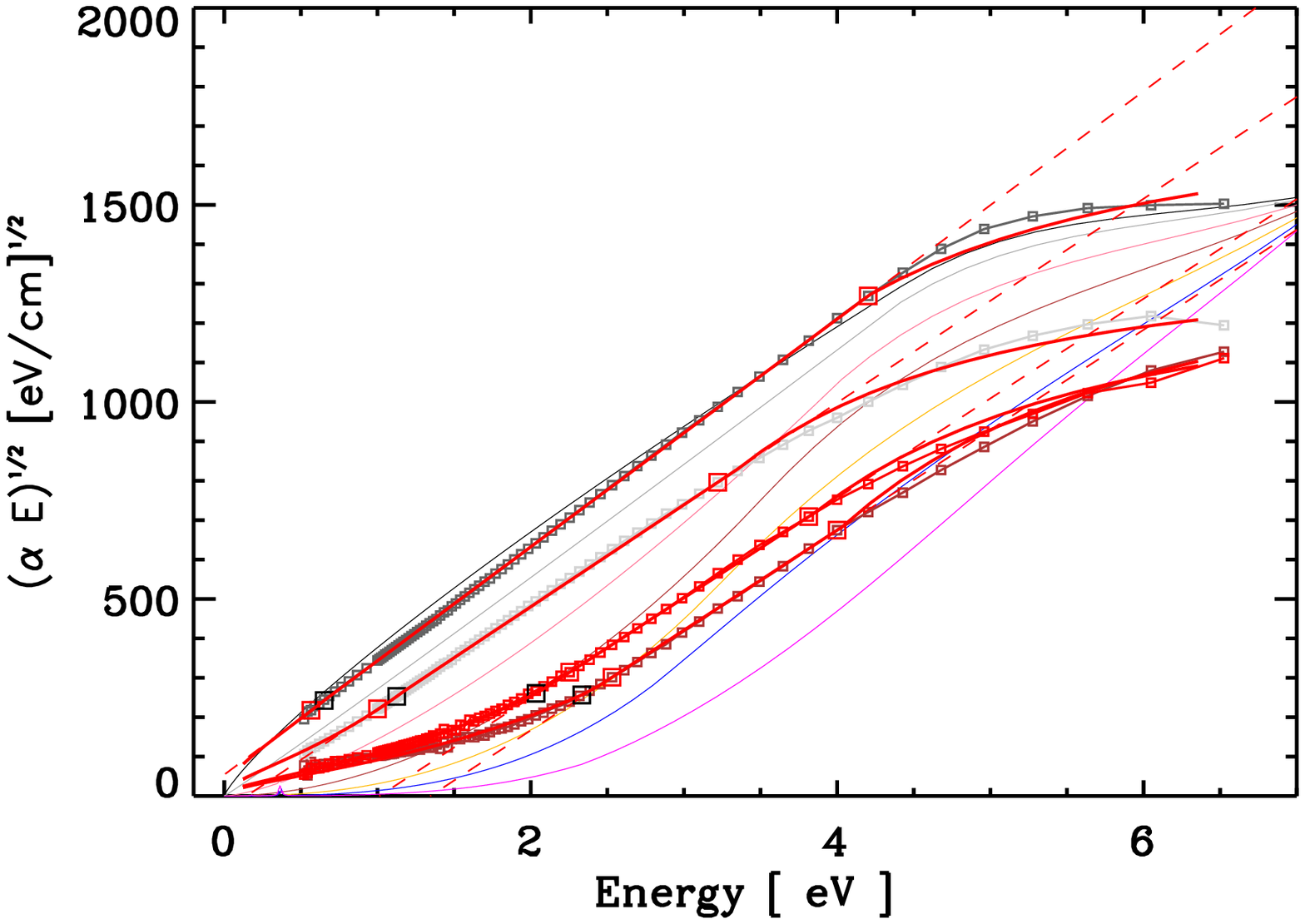}}
 \resizebox{\hsize}{!}{\includegraphics{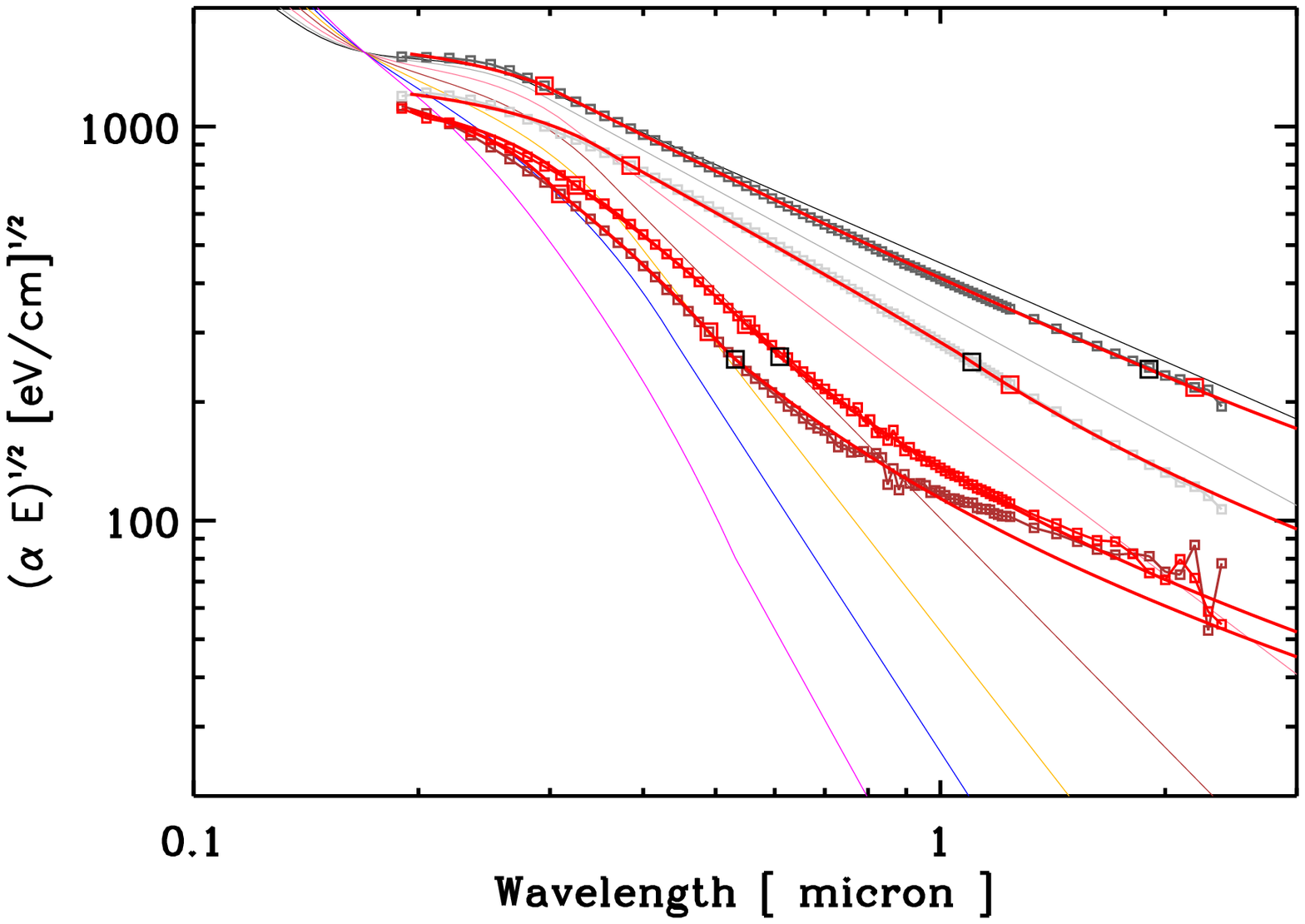}}
 \caption{The laboratory data from \cite[][(coloured curves with small data points)]{1995ApJS..100..149M} plotted as $(\alpha E)^{0.5}$ {\it vs.} energy, Tauc plot (upper), and also versus wavelength (lower). The red lines show the modelled fits to the data using Eqs.~(\ref{alphaE_fit_0}) and (\ref{eq_Urbach}), with the addition of an ``inverse Gaussian tail'' at high energies. The red squares delineate the portion used to determine $B$ and the extrapolation to $E_{\rm g}$ (dashed red lines). The black squares indicate the transition energies to the Urbach tail, $E_1$. The lines without data points are the optEC$_{(s)}$ model data.}
 \label{fig_Mennella_Tauc}
\end{figure}

\begin{figure}
 %\resizebox{\hsize}{!}{\includegraphics{the_model/ZZ_Tauc_plots_Rouleau_energy_2011.ps}}
 %\resizebox{\hsize}{!}{\includegraphics{the_model/ZZ_Tauc_plots_Rouleau_wavelength_2011.ps}}
 \resizebox{\hsize}{!}{\includegraphics{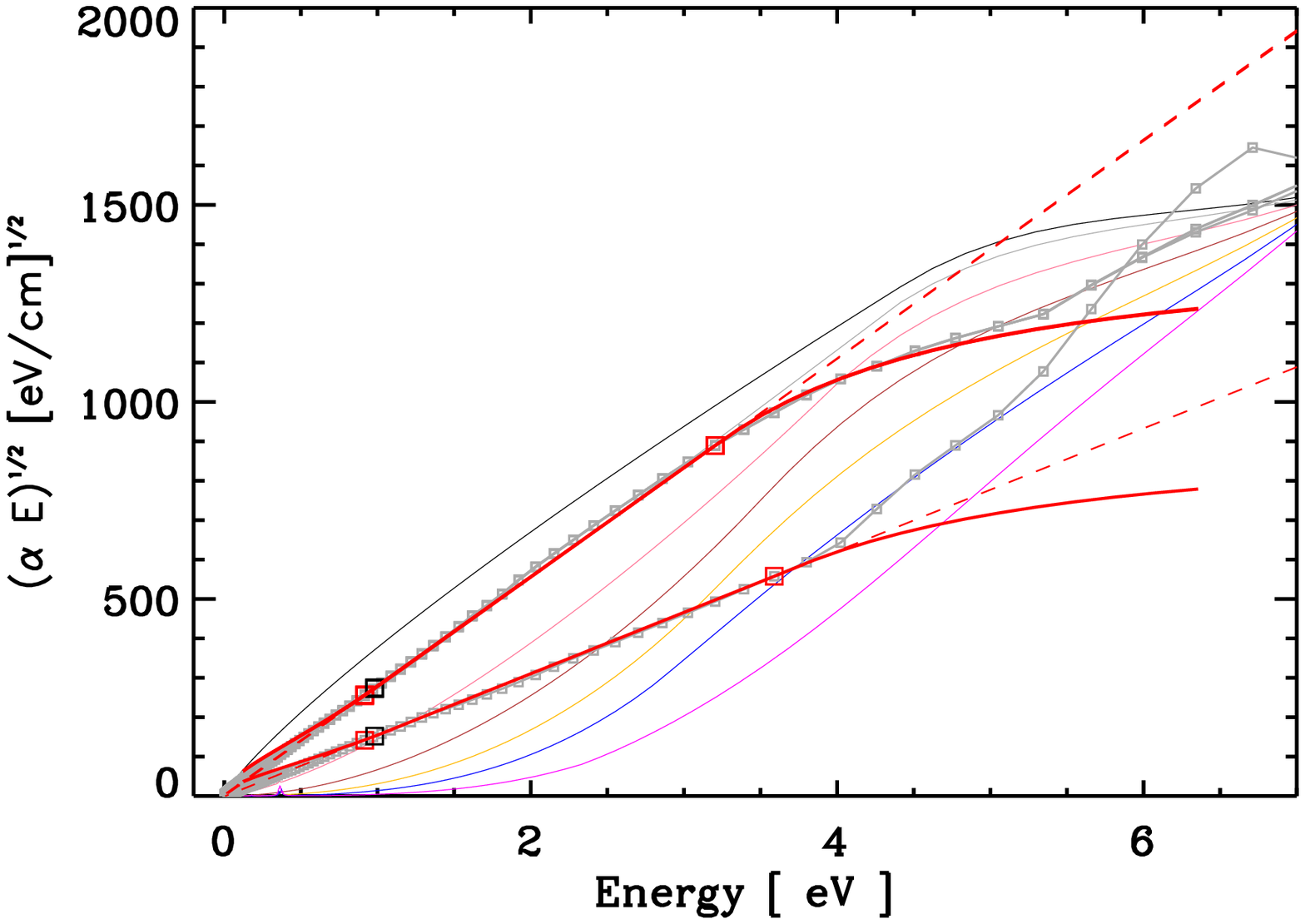}}
 \resizebox{\hsize}{!}{\includegraphics{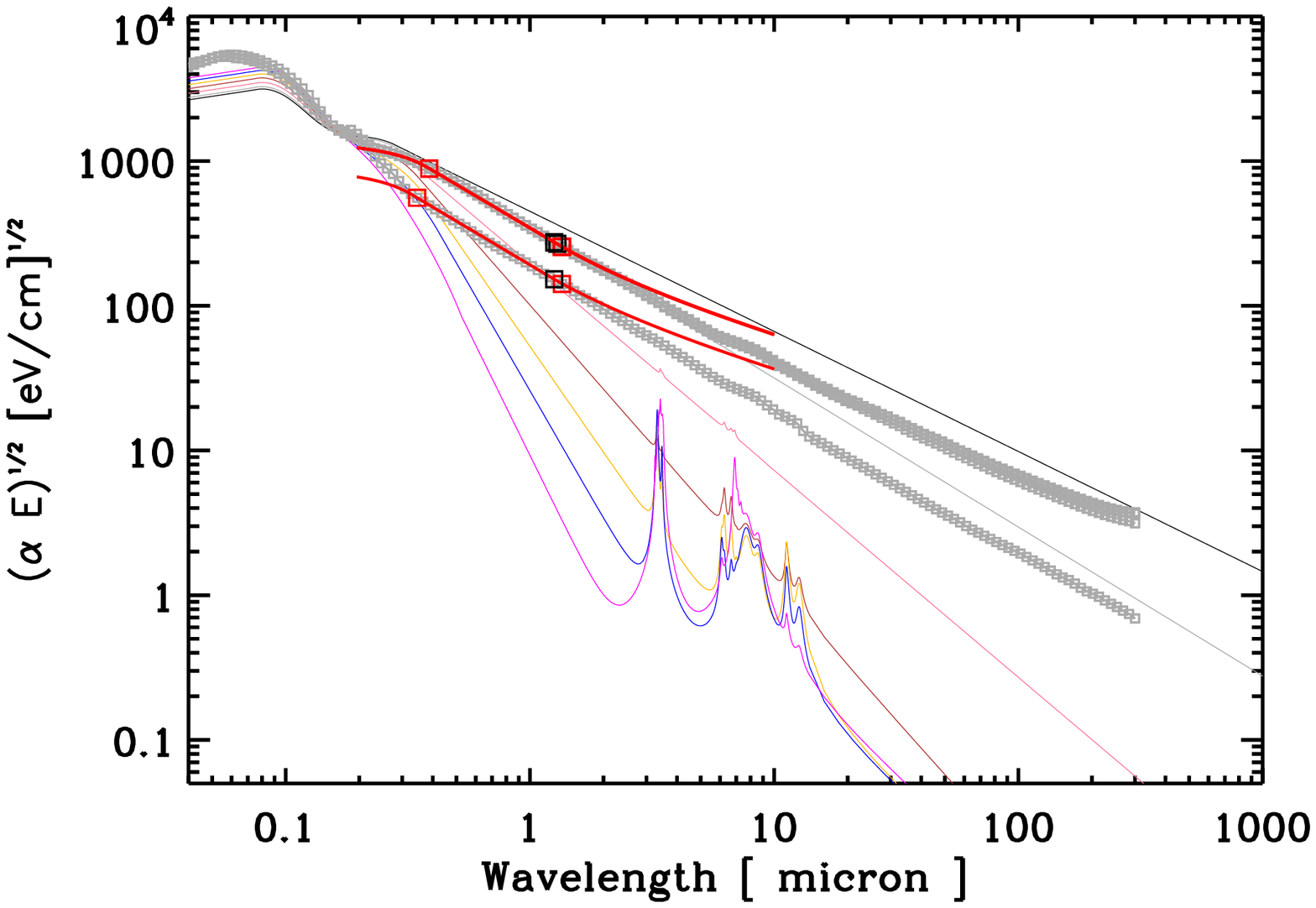}}
 \caption{As per Figs.~\ref{fig_Smith_Tauc} and \ref{fig_Mennella_Tauc} but for the \cite{1991ApJ...377..526R} data.}
 \label{fig_Rouleau_Tauc}
\end{figure}

\begin{figure}
 %\resizebox{\hsize}{!}{\includegraphics{the_model/ZZ_Tauc_plots_Jena_energy_2011.ps}}
 %\resizebox{\hsize}{!}{\includegraphics{the_model/ZZ_Tauc_plots_Jena_wavelength_2011.ps}}
 \resizebox{\hsize}{!}{\includegraphics{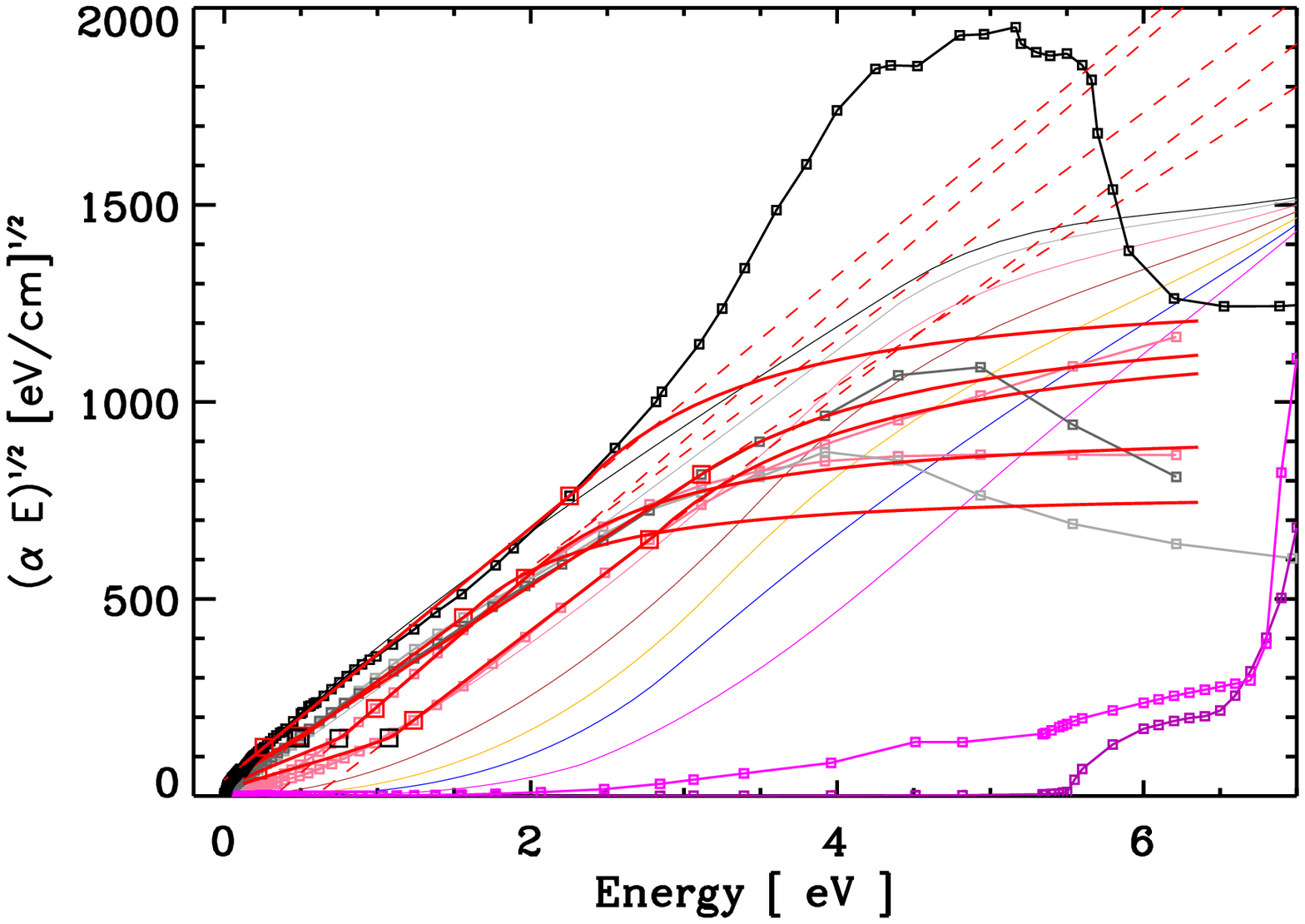}}
 \resizebox{\hsize}{!}{\includegraphics{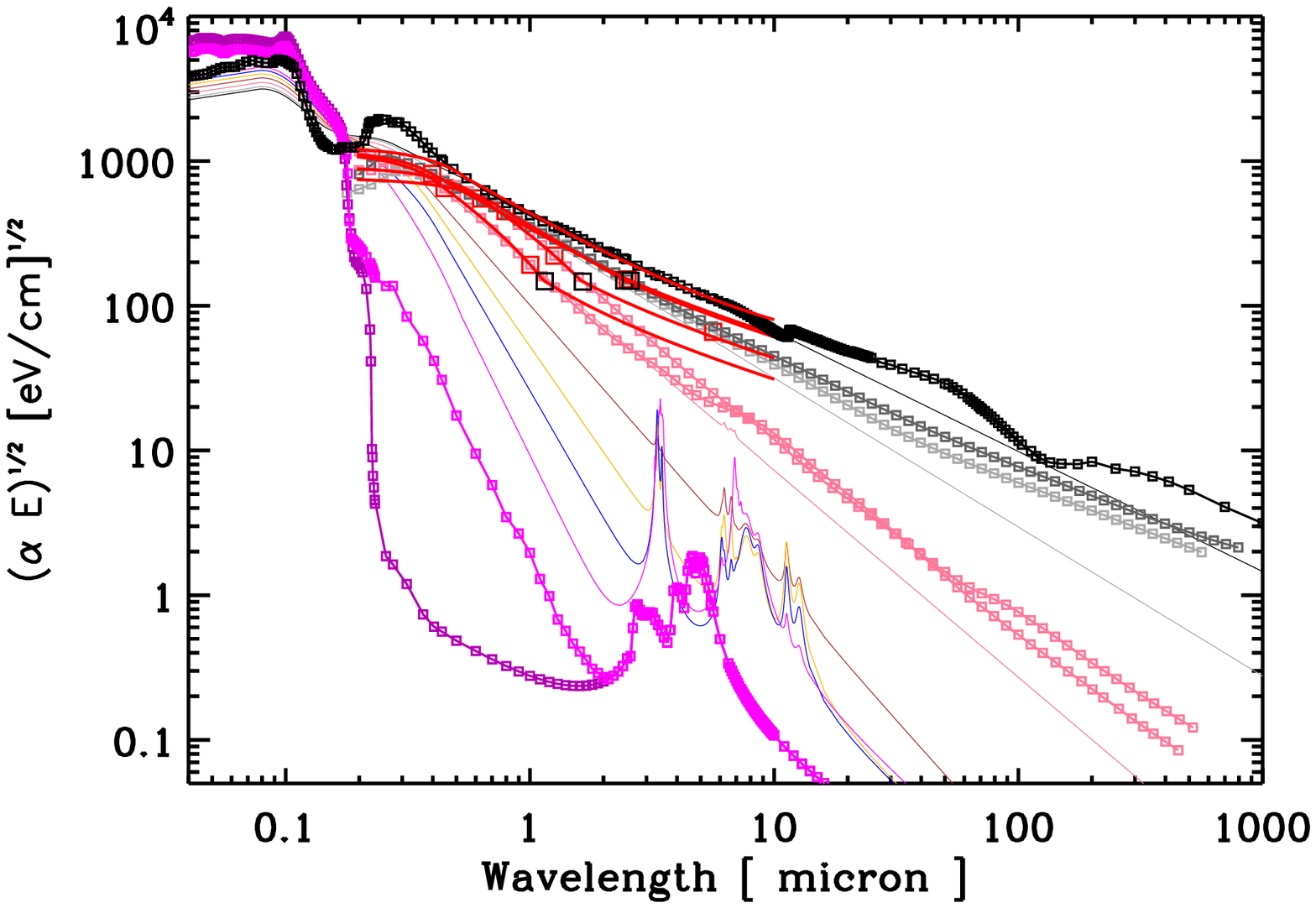}}
 \caption{As per Fig.~\ref{fig_Mennella_Tauc} but for the Jena DDOP amorphous carbon data. Also shown for comparison are the data for graphite using the usual $\frac{1}{3}\epsilon_\bot + \frac{2}{3}\epsilon_\|$ approximation \citep[black lines,][]{1984ApJ...285...89D} and diamond \citep[purple and violet lines,][respectively]{1985HandbookOptConst...665,1989Natur.339..117L}.}
 \label{fig_Jena_Tauc}
\end{figure}

\begin{figure}
 %\resizebox{\hsize}{!}{\includegraphics{the_model/ZZ_Tauc_plots_Zubko_energy_2011.ps}}
 %\resizebox{\hsize}{!}{\includegraphics{the_model/ZZ_Tauc_plots_Zubko_wavelength_2011.ps}}
 \resizebox{\hsize}{!}{\includegraphics{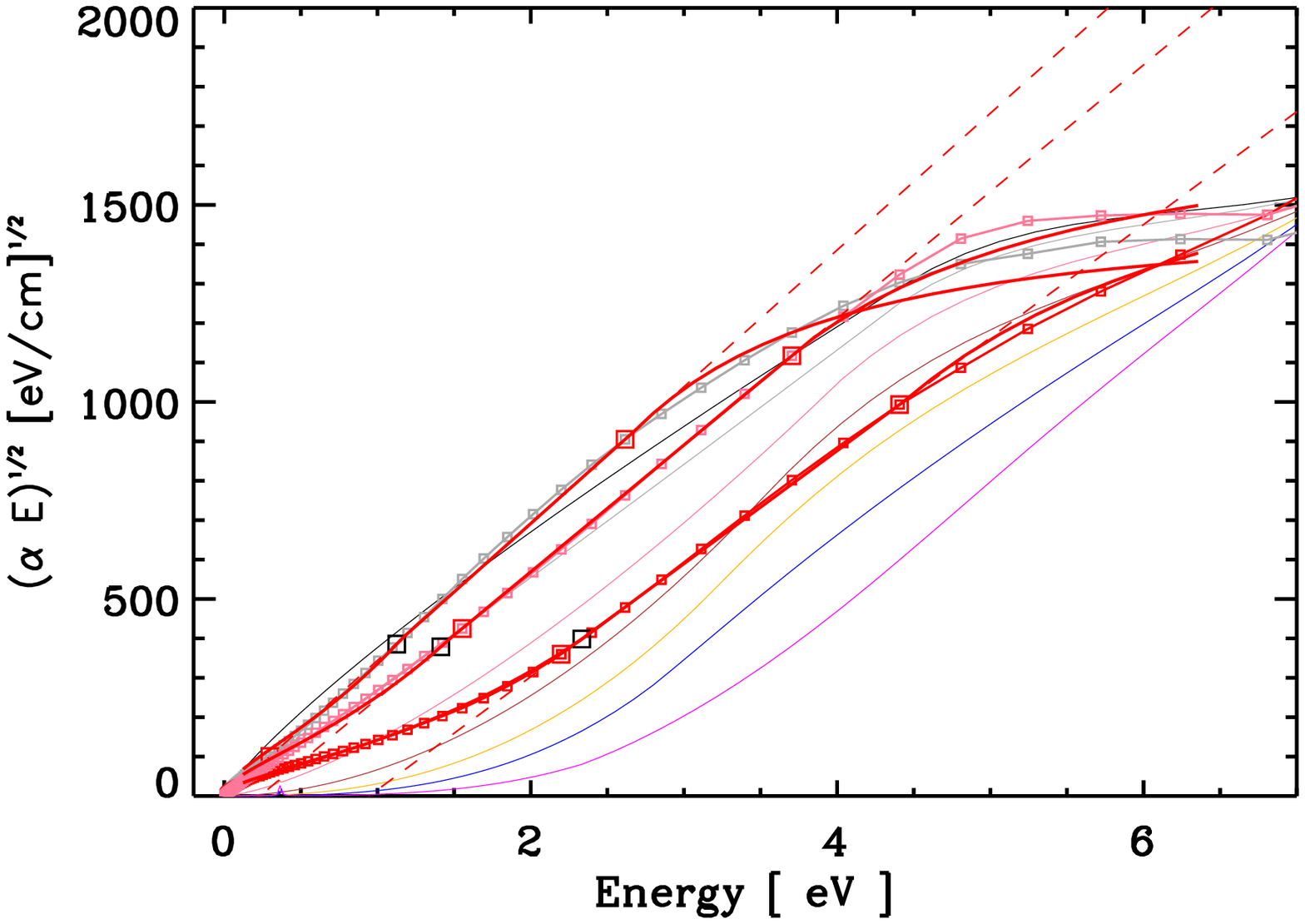}}
 \resizebox{\hsize}{!}{\includegraphics{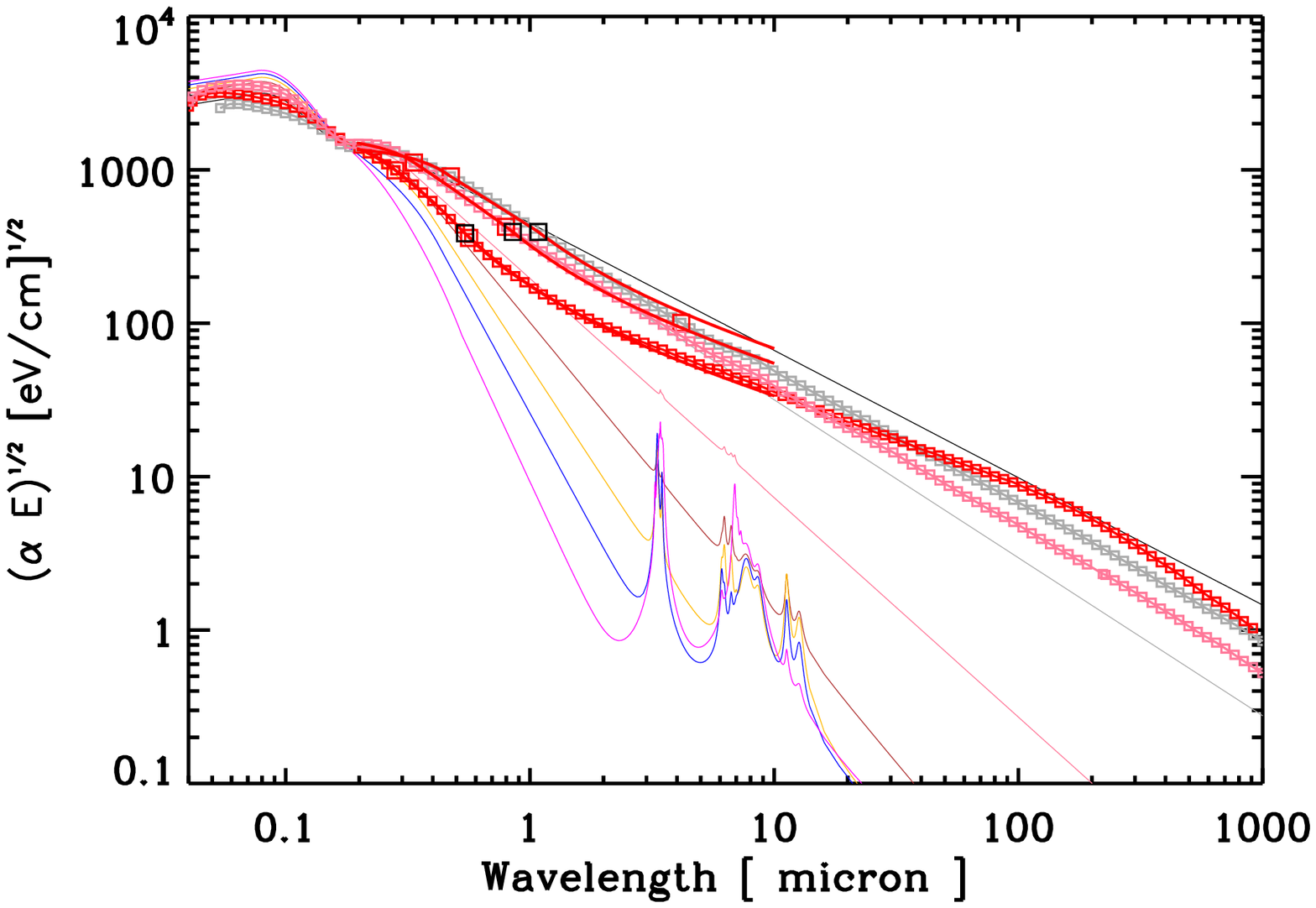}}
 \caption{As per Fig.~\ref{fig_Mennella_Tauc} but for the \cite{1996MNRAS.282.1321Z} data.}
 \label{fig_Zubko_Tauc}
\end{figure}

We have appled this ``Tauc analysis'' to the available laboratory-measured and modelled optical data \citep[][and DDOP]{1984JAP....55..764S,1991ApJ...377..526R,1995ApJS..100..149M,1996MNRAS.282.1321Z} and have derived the parameters for the Tauc fits, these are given in Table~\ref{table_Tauc_params}. The derived values for $B$ shown in Table~\ref{table_Tauc_params} are in good agreement with those expected for these materials, {\it i.e.}, $\simeq 4 \times 10^4$\,eV$^{-1}$ cm$^{-1}$ for a-C and $\simeq 6 \times 10^4$\,eV$^{-1}$ cm$^{-1}$ for a-C:H  \citep{1986AdPhy..35..317R}. Additionally, and as \cite{1986AdPhy..35..317R} points out, $E_{\rm g}$ is a measure of the largest significant aromatic cluster size and $B$ gives the range of cluster sizes. In the accompanying figures (Fig.~\ref{fig_Smith_Tauc} and Figs.~\ref{fig_Mennella_Tauc} to \ref{fig_Zubko_Tauc}) we show these fits along with the optEC$_{(s)}$ model data. We recall here that we have colour-coded the data (see Table~\ref{table_colour_code}) in order to facilitate the comparison between the model and laboratory data. 

As can be seen, by comparing these colour-coded data, the optEC$_{(s)}$ model, qualitatively at least, follows all of the major trends in these data over the entire visible-UV wavelengths region. However, certain laboratory and modelled data appear to deviate from the general trends when plotted in this way. We do not here wish to enter into any speculations as to why this might be the case, other than by referring to the comments about possible electrical conductivity effects mentioned in the following appendix.  

% TABLE
\begin{table}
\caption{The Tauc analysis-derived parameters for the optical data in the given references.}
\begin{center}
\begin{tabular}{lcccccc}
                                         &       &       &        &       &                   &                    \\[-0.35cm]
\hline
\hline                                         &                             &                    &            &       &        &             \\[-0.25cm]

\multicolumn{7}{l}{\cite{1984JAP....55..764S}}  \\[0.1cm]
$T_{\rm c}$ [$^\circ$C]        &   250       &   350       &   450       &   550       &   650       &   750     \\
$E_{\rm g}$ [eV]                  &   2.12      &   2.10       &   1.54     &   0.78      &   0.24      &   0.02     \\
$B$                                      &   7.99      &   10.8       &   12.9     &   9.89      &   8.44       &   8.13 \\
$E_1$ [eV]                           &   3.10     &    2.94       &   2.30     &   1.66      &   1.19      &   0.98     \\
$\sigma_{\rm U}$ [eV]         &   0.40     &    0.43       &   0.60     &    0.90     &    1.00     &   1.00      \\
$X_{\rm H}$                         &    0.49    &    0.49       &  0.36      &    0.18     &    0.06     &   0.00     \\
$N_{\rm R}$                         &    7         &    8            &   14        &    55        &   586       &   1e4      \\
$a_{\rm R}$  [nm]                &     0.4   &     0.4      &   0.5     &    1.0      &   3.1     &   40      \\
$L_{\rm a}$  [nm]                &     0.4   &     0.4      &   0.5     &    1.0      &   3.2     &   42      \\[0.1cm]

\hline
                                         &                             &                    &            &       &        &             \\[-0.25cm]
\multicolumn{7}{l}{\cite{1991ApJ...377..526R}}  \\[0.1cm]       
Material                               &    AC1   &      BE1     &    FC21PS  &  HAPS     &        &         \\
$E_{\rm g}$ [eV]                  &    0.01   &     0.00    &    -0.01        &   -0.01     &        &         \\
$B$                                      &    2.42   &     7.71     &    7.64        &   7.66       &        &         \\
$E_1$ [eV]                           &    1.01    &    1.00    &    0.99     &   0.99          &        &         \\
$\sigma_{\rm U}$ [eV]         &     1.10   &     1.00    &    1.00    &    1.00         &        &         \\
$X_{\rm H}$                         &      0       &   0          &   0           &   0                &        &         \\
$N_{\rm R}$                         &      7e5   &  7e7       &  2e5       &  4e5             &        &         \\
$a_{\rm R}$  [nm]                &    106     &  1093       &  61       &  80               &        &         \\
$L_{\rm a}$  [nm]                &     110     &  1140      &   63       &   83                 &        &         \\[0.1cm]
        
\hline
                                         &                             &                    &            &       &        &             \\[-0.25cm]
\multicolumn{7}{l}{\cite{1995ApJS..100..149M}}  \\[0.1cm]   
$T_{\rm c}$ [$^\circ$C]        &       20      &    415     &    600     &   800    &        &         \\
$E_{\rm g}$ [eV]                  &      1.35     &   1.01     &    0.15   &   -0.19  &        &         \\
$B$                                      &    6.46       &     6.40     &       6.70      &      8.35 &        &         \\
$E_1$ [eV]                           &     2.40       &     2.07    &         1.17        &         0.67 &        &         \\
$\sigma_{\rm U}$ [eV]          &     1.10      &    1.00        &         0.75         &        1.00  &        &         \\
$X_{\rm H}$                         &      0.31      &   0.24       &          0.03        &        0  &        &         \\
$N_{\rm R}$                         &     19         &     33         &        1595       &         958  &        &         \\
$a_{\rm R}$  [nm]                &      0.6        &     0.8         &        5        &         4  &        &         \\
$L_{\rm a}$  [nm]                &      0.6        &     0.8         &        5        &        4  &        &         \\[0.1cm]
           
\hline
                                         &                             &                    &            &       &        &             \\[-0.25cm]
\multicolumn{7}{l}{Jena DDOP}  \\[0.1cm]  
$T_{\rm c}$ [$^\circ$C]        &       400    &    600    &   800     &   1000    &        &         \\
$E_{\rm g}$ [eV]                  &     0.59      &      0.33     &      -0.01     &      -0.11      &        &         \\
$B$                                      &     8.87    &    11.4   &     8.33    &    6.41    &        &         \\
$E_1$ [eV]                           &     1.09      &       0.77       &      0.51       &      0.48  &        &         \\
$\sigma_{\rm U}$ [eV]         &      1.00      &       1.00       &      1.00       &      1.00  &        &         \\
$X_{\rm H}$                         &      0.14       &      0.08      &      0       &     0 &        &         \\
$N_{\rm R}$                         &      96      &      317     &    5e5     &      2898 &        &         \\
$a_{\rm R}$  [nm]                &       1.3      &       2.3      &      87      &       6.8    &        &         \\
$L_{\rm a}$  [nm]                &       1.3      &       2.4      &     91      &      7.2      &        &         \\[0.1cm]
             
\hline
                                         &                             &                    &            &       &        &             \\[-0.25cm]
\multicolumn{7}{l}{\cite{1996MNRAS.282.1321Z}}  \\[0.1cm]  
Material                                &  BE   &     ACH2     &   ACAR     &       &        &         \\
$E_{\rm g}$ [eV]                  &   0.01      &      0.95     &       0.23  &  &        &         \\
$B$                                      &   12.1   &     8.23    &    10.3  &  &        &         \\
$E_1$ [eV]                           &   1.16      &       2.34      &       1.48  &   &       &         \\
$\sigma_{\rm U}$ [eV]         &   0.80     &        1.10      &       0.90  &   &       &         \\
$X_{\rm H}$                         &   0      &       0.22       &      0.05   &    &      &         \\
$N_{\rm R}$                         &   5e5      &       38     &       628   &    &      &         \\
$a_{\rm R}$  [nm]                &   88      &       0.81     &        3.2  &    &      &         \\
$L_{\rm a}$  [nm]                &   92      &       0.81      &       3.3   &    &      &         \\[0.1cm]
             
\hline
\hline
                       &                  &       &        &                 &                   &                    \\[-0.25cm]
\end{tabular}
\begin{list}{}{}
\item[] Notes: - 
The characteristic temperature parameter, $T_{\rm c}$ indicates the relevant material deposition or annealing temperature. 
The slope parameter $B$ is in units of $10^4$ cm$^{-1}$ eV$^{-1}$. 
\end{list}
\end{center}
\label{table_Tauc_params}
\end{table}

% FIGURE A.1 *********************************************************
% *********************************************************

% FIGURE A.2 *********************************************************
% *********************************************************

% FIGURE A.3 *********************************************************
% *********************************************************

% FIGURE A.4 *********************************************************
% *********************************************************

\section{The derived optical constants compared with the available data}
\label{app_derived_nk} 

\begin{figure}
 %\resizebox{\hsize}{!}{\includegraphics{the_model/ZZ_k_vs_wavelength_Smith_2011.ps}} 
 %\resizebox{\hsize}{!}{\includegraphics{the_model/ZZ_n_vs_wavelength_Smith_2011.ps}}
 \resizebox{\hsize}{!}{\includegraphics{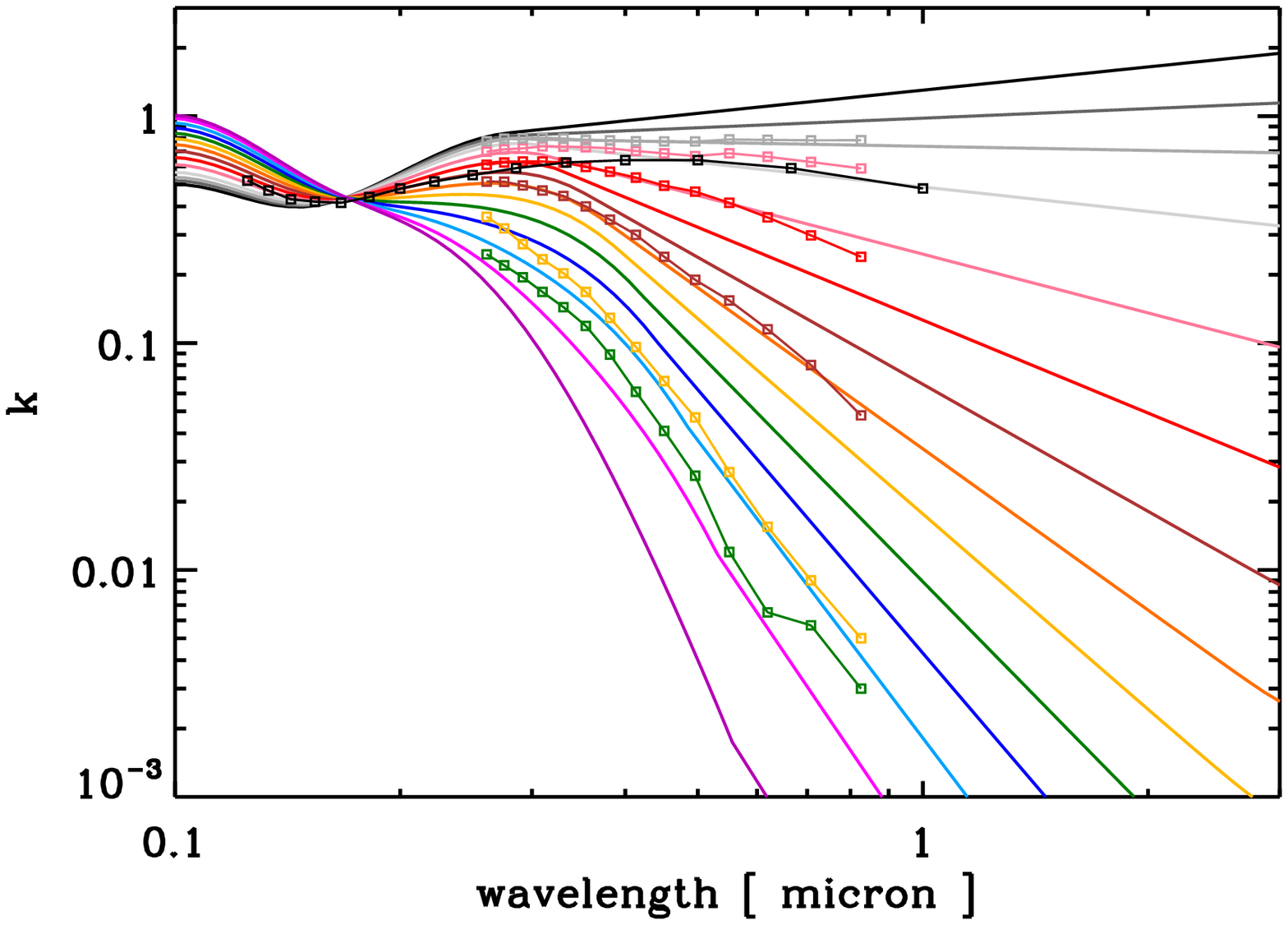}} 
 \resizebox{\hsize}{!}{\includegraphics{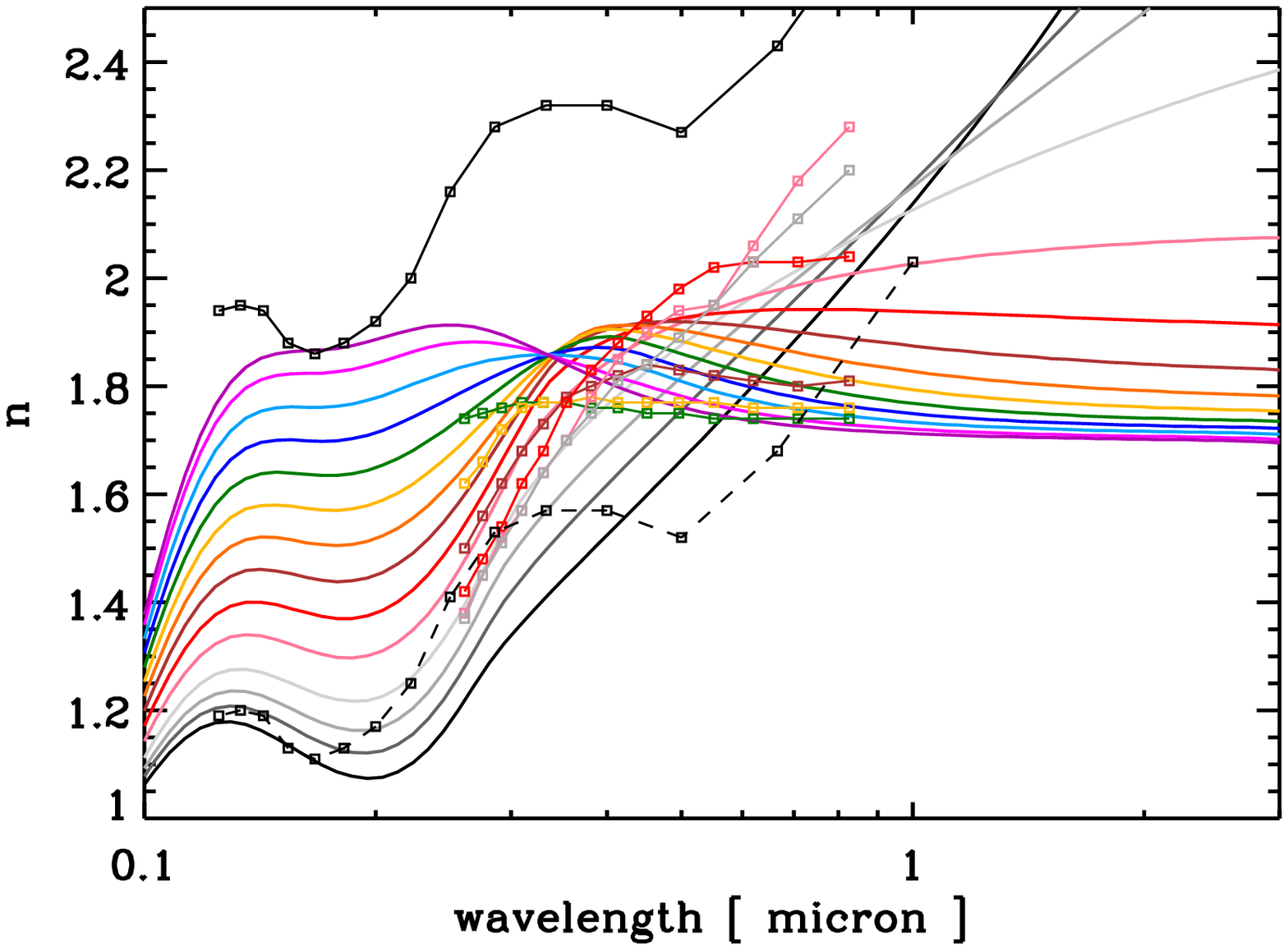}}
 \caption{The imaginary (upper) and real (lower) parts of the refractive index for the \cite{1984JAP....55..764S} data (lines with data points) compared to the optEC$_{(s)}$ model data (lines without data points). The solid black lines with data points show the \cite{1984ApJ...287..694D} a-C $k$ (upper) and $n$ data (lower); for comparison purposes only, the dashed line shows the $n$ data shifted down by 0.75.  }
 \label{fig_Smith_k}
\end{figure}

\begin{figure}
 %\resizebox{\hsize}{!}{\includegraphics{the_model/ZZ_k_vs_wavelength_Mennella_2011.ps}}
 \resizebox{\hsize}{!}{\includegraphics{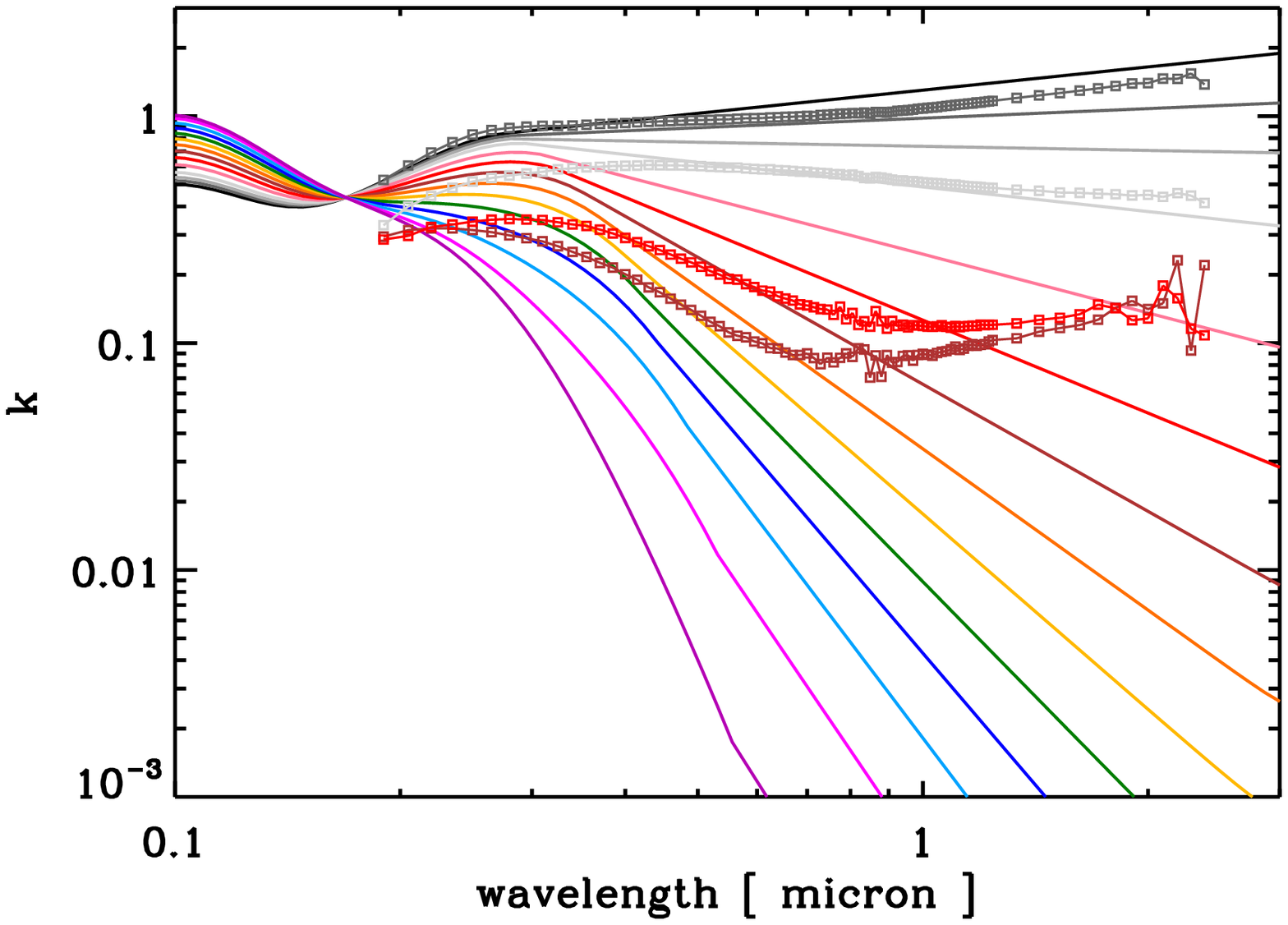}}
 \caption{Same as Fig.~\ref{fig_Smith_k} but for the \cite{1995ApJS..100..149M} data.}
 \label{fig_Mennella_k}
\end{figure}

\begin{figure}
 %\resizebox{\hsize}{!}{\includegraphics{the_model/ZZ_k_vs_wavelength_Rouleau_2011.ps}} 
 %\resizebox{\hsize}{!}{\includegraphics{the_model/ZZ_n_vs_wavelength_Rouleau_2011.ps}}
 \resizebox{\hsize}{!}{\includegraphics{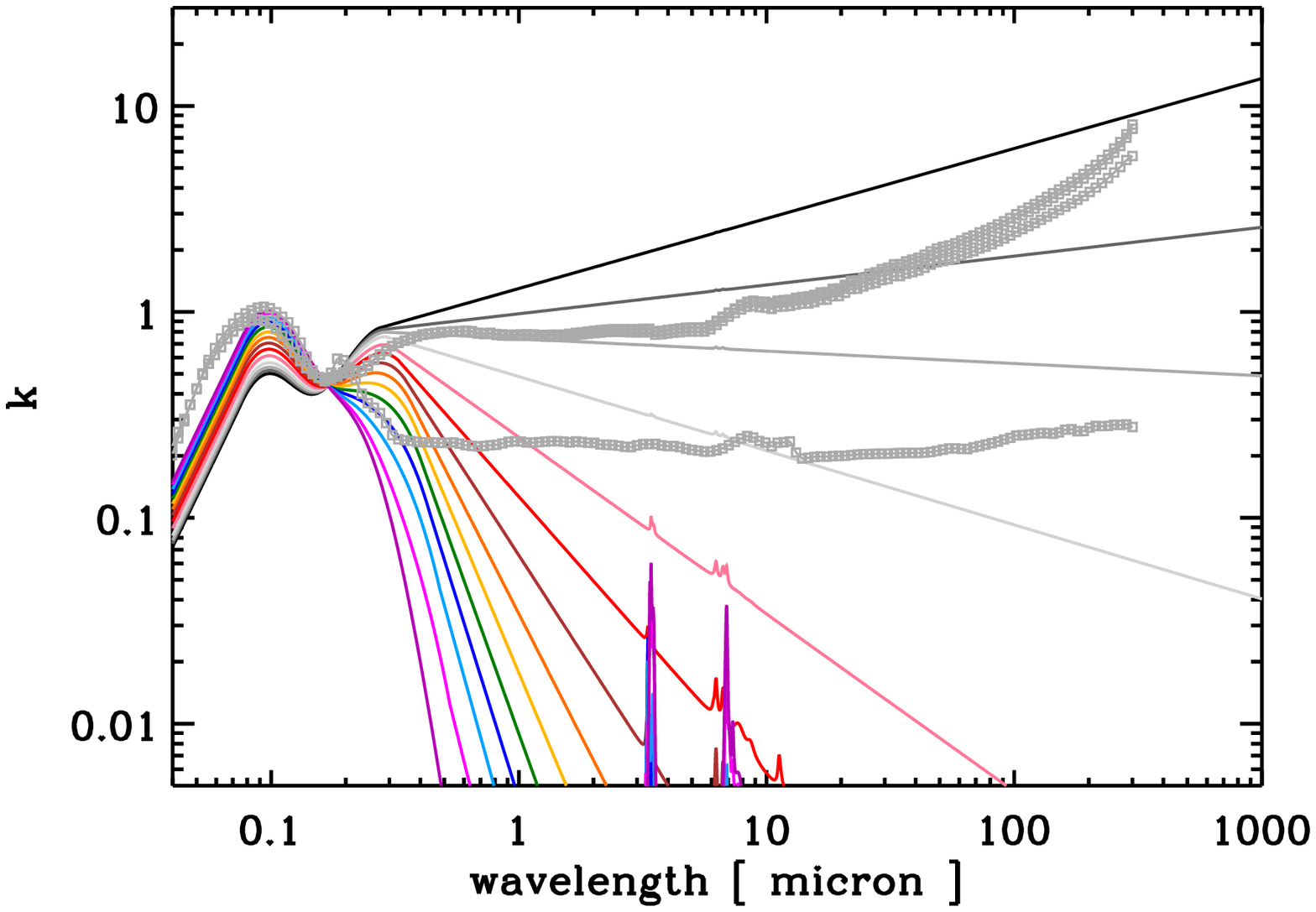}} 
 \resizebox{\hsize}{!}{\includegraphics{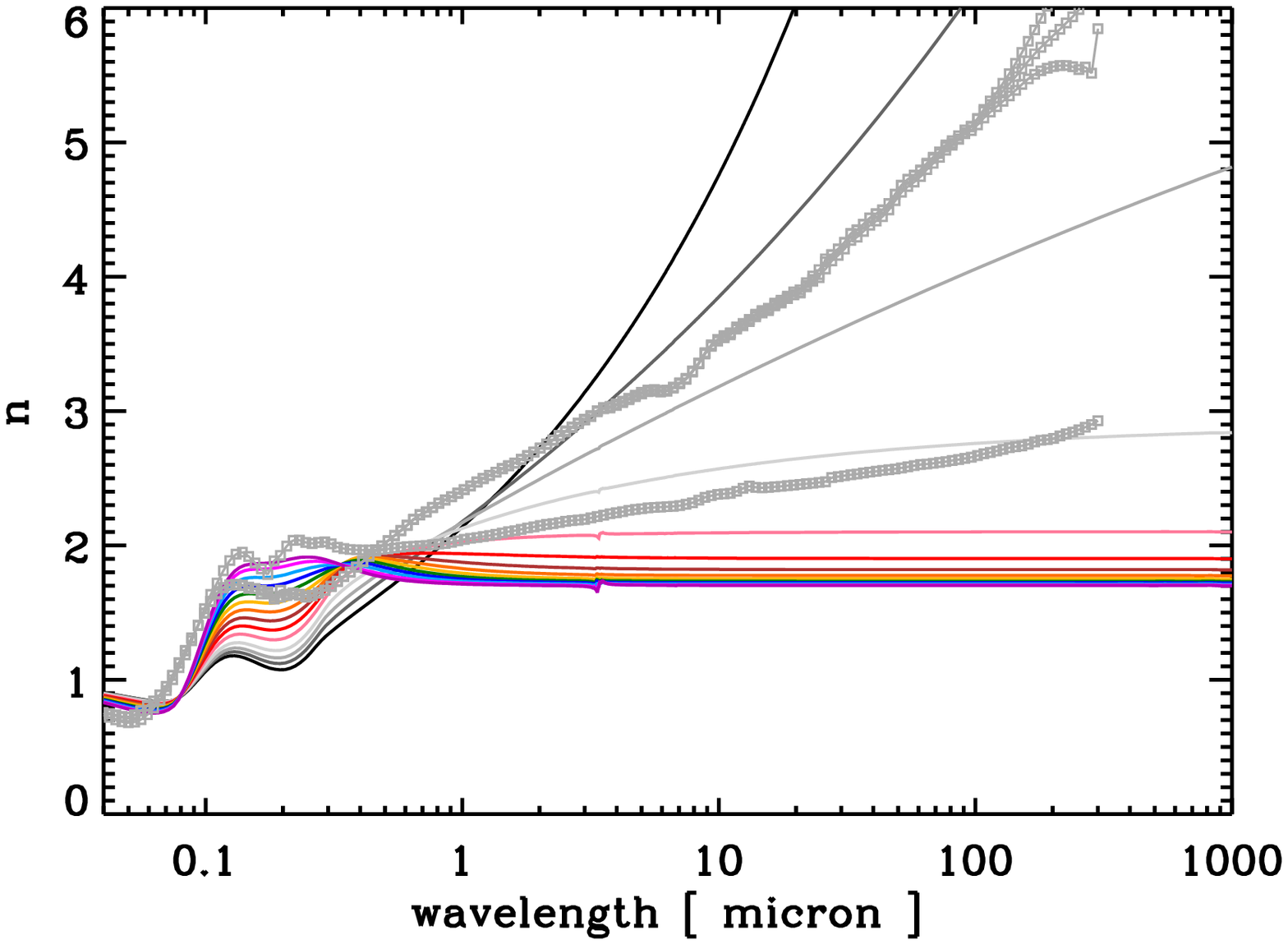}}
 \caption{Same as Fig.~\ref{fig_Smith_k} but for the \cite{1991ApJ...377..526R} data.}
 \label{fig_Rouleau_k}
\end{figure}

\begin{figure}
 %\resizebox{\hsize}{!}{\includegraphics{the_model/ZZ_k_vs_wavelength_Jena_2011.ps}} 
 %\resizebox{\hsize}{!}{\includegraphics{the_model/ZZ_n_vs_wavelength_Jena_2011.ps}}
 \resizebox{\hsize}{!}{\includegraphics{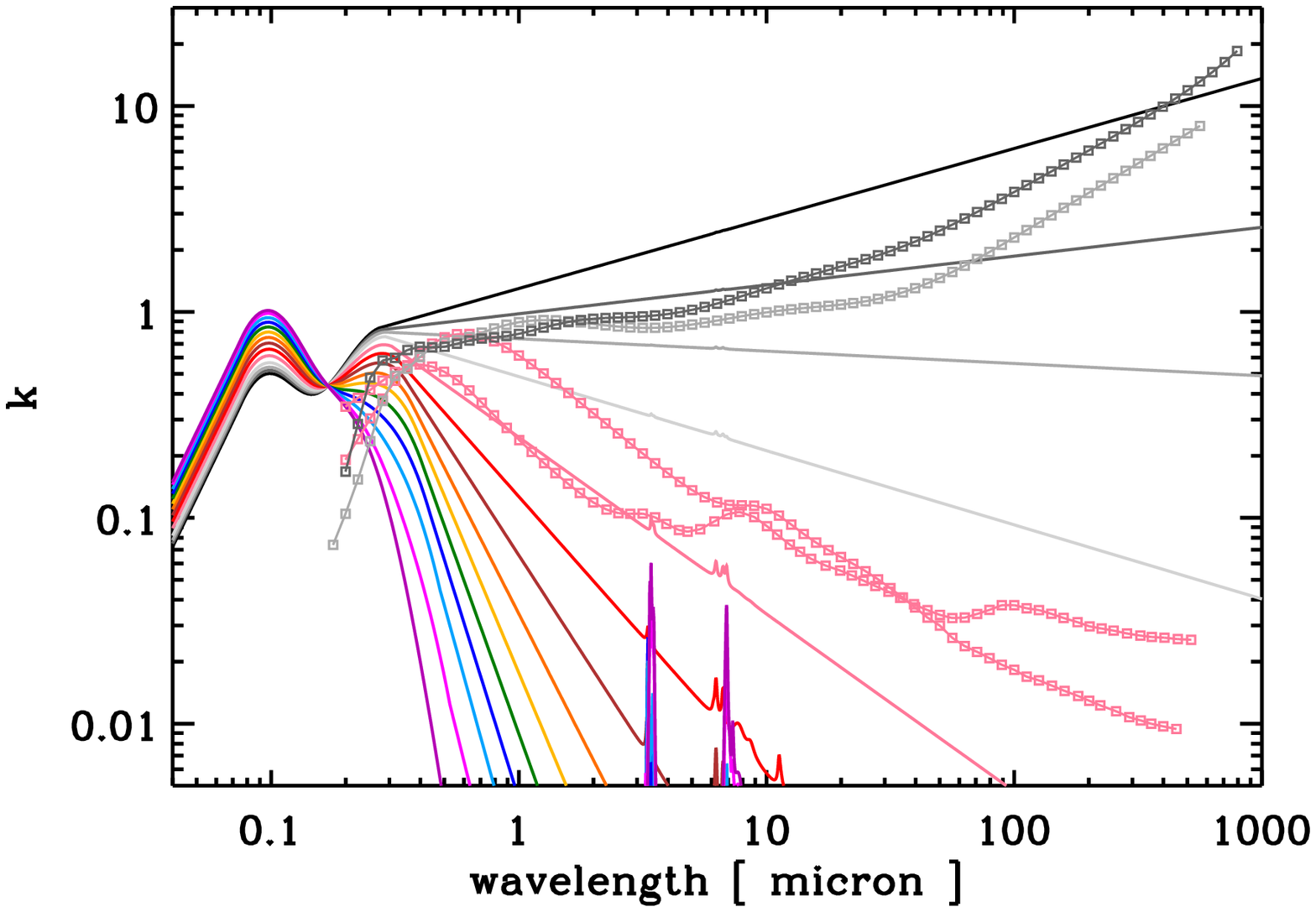}} 
 \resizebox{\hsize}{!}{\includegraphics{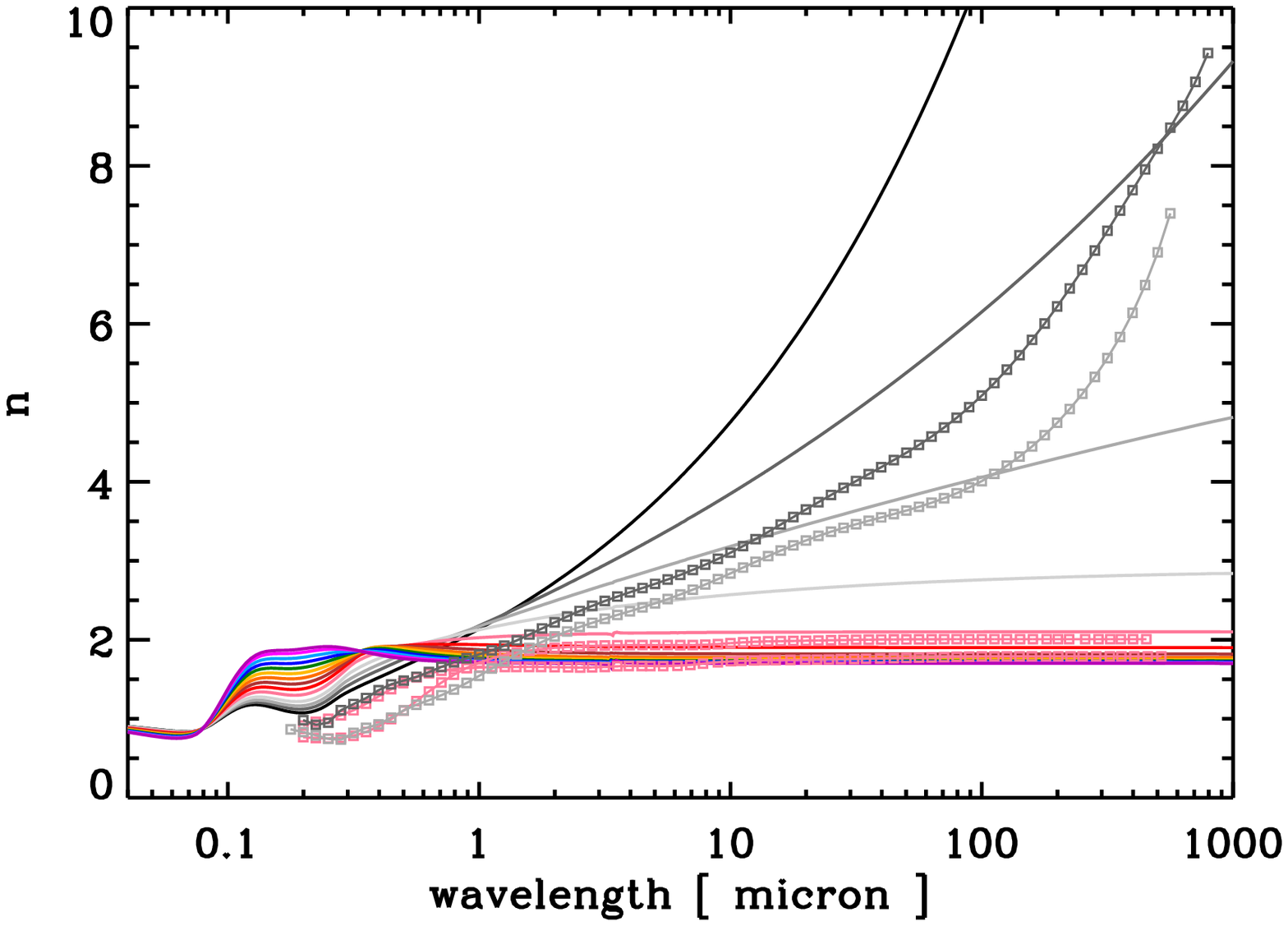}}
 \caption{Same as Fig.~\ref{fig_Smith_k} but for the Jena DDOP amorphous carbon data.}
 \label{fig_Jena_k}
\end{figure}

\begin{figure}
 %\resizebox{\hsize}{!}{\includegraphics{the_model/ZZ_k_vs_wavelength_Zubko_2011.ps}}
 %\resizebox{\hsize}{!}{\includegraphics{the_model/ZZ_n_vs_wavelength_Zubko_2011.ps}}
 \resizebox{\hsize}{!}{\includegraphics{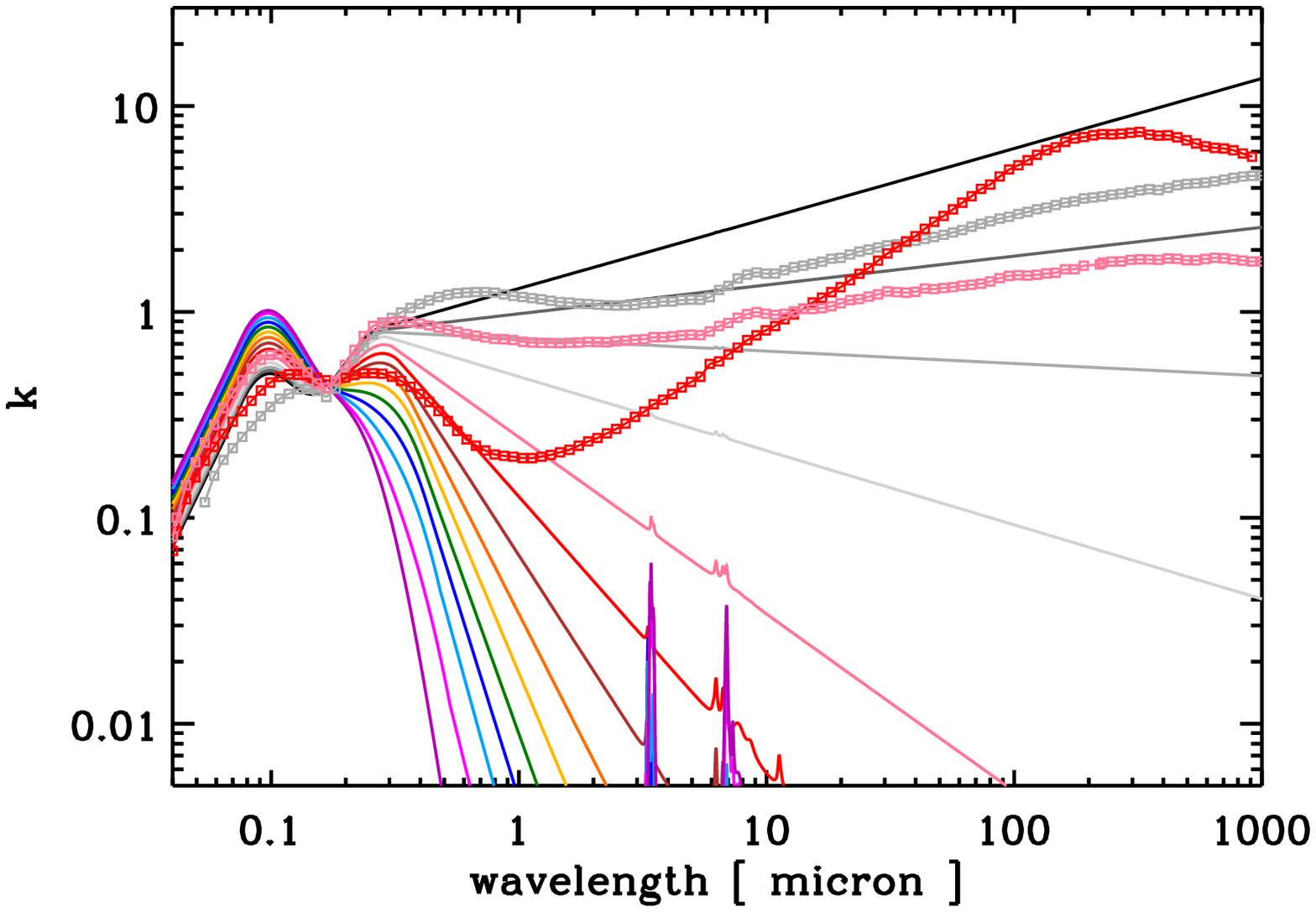}}
 \resizebox{\hsize}{!}{\includegraphics{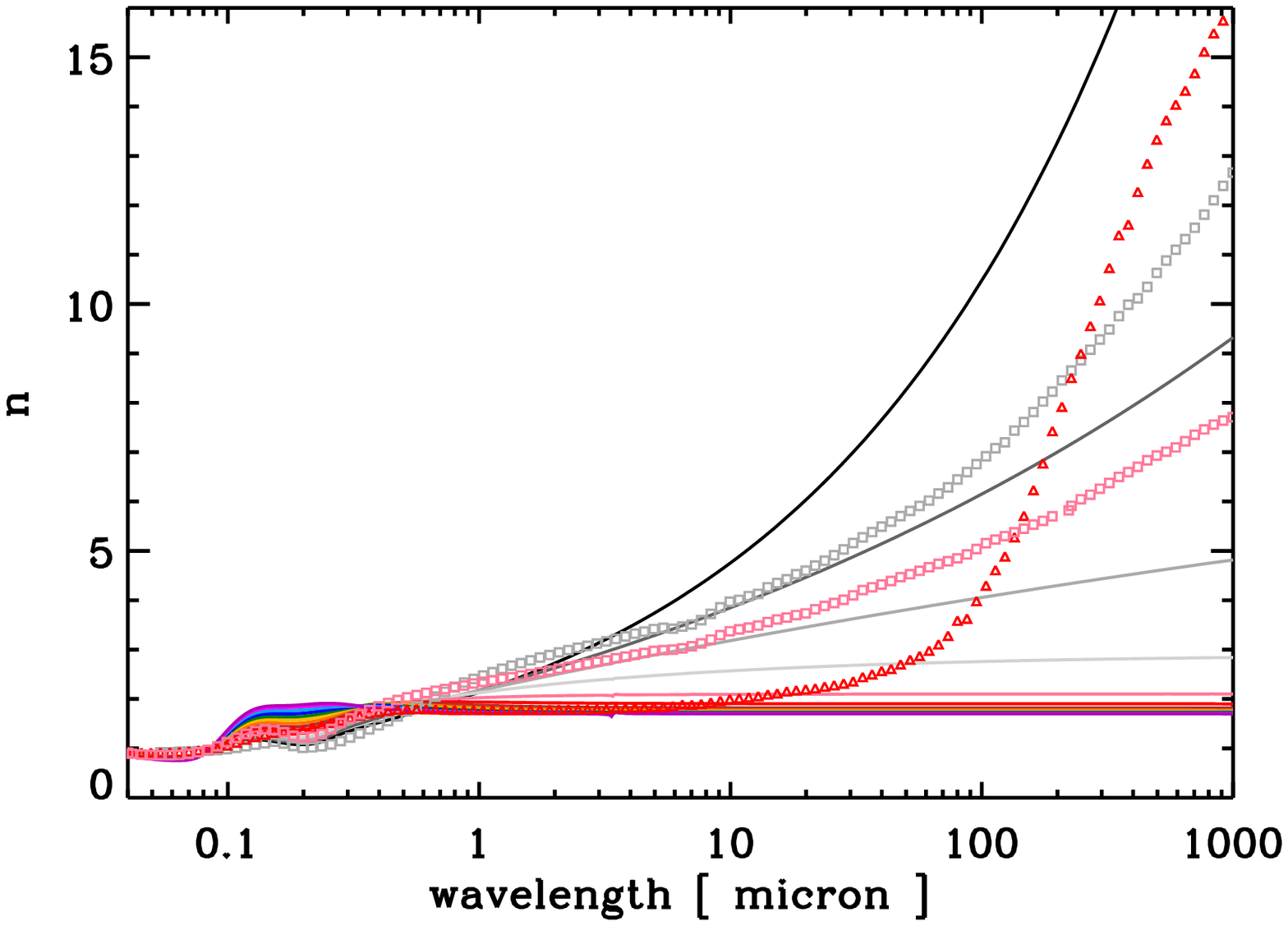}}
 \caption{Same as Fig.~\ref{fig_Smith_k} but for the \cite{1996MNRAS.282.1321Z} data.}
 \label{fig_Zubko_k}
\end{figure}

In Figs.~\ref{fig_Smith_k} to \ref{fig_Zubko_k} we present the real and imaginary parts of the complex refractive index for the available laboratory and modelled data \citep[][and DDOP]{1984ApJ...287..694D,1984JAP....55..764S,1991ApJ...377..526R,1995ApJS..100..149M,1996MNRAS.282.1321Z} compared to the optEC$_{(s)}$ model-derived  values. 

Clearly the optEC$_{(s)}$ model reproduces all of the observed trends in the laboratory and the other modelled data extremely well, however, the data do fall ``short'' in explaining {\em all} of the long wavelength optical property behaviour. For example, looking at the DDOP $k$ data presented in Fig.~\ref{fig_Jena_k}, in particular in the 100\,$\mu$m region, it is clear that there appear to be other bands that contribute to the long wavelength behaviour in these samples. However, it should be pointed out that the presence of structure at $\approx 3, 10$ and 100\,$\mu$m, and perhaps some other underlying broad structures, is rather reminiscent of water ice bands at $\approx 3, 6, 12, 44$ and 100\,$\mu$m. Thus, it is possible that some of the features present in these data could be due to adsorbed water in the samples but this possibility needs to be eliminated. We have therefore not tried to fit the $\approx 100\,\mu$m band, or indeed any others that might contribute at long wavelengths (in these and the other data), and advise that until such time as the exact origin of these bands is understood and quantified it is rather premature to do so.

We also note that there is present, to a greater or lesser extent, an `upward' curvature in some of the laboratory $k$ data. This kind of behaviour, especially evident in the \cite{1995ApJS..100..149M} and \cite{1991ApJ...377..526R} data  \citep[the latter being derived from an earlier version of the data in][]{1995ApJS..100..149M}, is reminiscent of an electrical conductivity contribution, such as seen in the graphite $k$ data, and we therefore suspect that inter-particle electrical conduction may be playing a role here. Conduction between particles is notoriously difficult to eliminate in these demanding experimental measurements. Thus, it may not be too surprising to see residual electrical conductivity effects in these data arising from un-wanted particle clustering in the laboratory samples. 

\cite{1996MNRAS.282.1321Z} took careful account of the clustering effects in their data analysis and modelling and this is evident in their AC1 and ACAR data, which show clear linear tendencies from FIR to mm wavelengths with no indication of an `upturn' in the $k$ data. Hence, it appears that particle clustering (and the associated conductivity effect) could indeed be the origin of the FIR-mm `upturn'. \cite{1991ApJ...377..526R} also made allowance for particle clustering effects but arrived at rather different results. 

In the DDOP data this `upward' curvature is apparent in the $n$ data, which perhaps indicates that in these data, at least, the curvature in the $k$ data is due to the presence of the long wavelength bands. That the long wavelength wings of, as-yet uncharacterised, FIR bands can contribute to the behaviour in the FIR-cm region will have major implications for the observed dust emissivity, as it would lead to a flattening of the carbonaceous dust emissivity at these wavelengths. Thus, and until such time as these bands are well characterised by laboratory measurements, some caution should be exercised in the use of these optEC$_{(s)}$ model data at mm wavelengths and beyond. 

% FIGURE B.1 *********************************************************
% *********************************************************

% FIGURE B.2 *********************************************************
% *********************************************************

% FIGURE B.3 *********************************************************
% *********************************************************

% FIGURE B.4 *********************************************************
% *********************************************************

% FIGURE B.5 *********************************************************
% *********************************************************

\listofobjects

\end{document}